\documentclass[journal]{IEEEtran}
\usepackage{geometry}
\usepackage{amsbsy,amsmath,amssymb,epsfig,bbm,mathrsfs,multirow,amsthm}
\usepackage{float}
\usepackage{nomencl}
\usepackage{tabularx}
\usepackage{makecell}
\usepackage{booktabs}
\usepackage{pdflscape} 
\usepackage{algpseudocode}
\usepackage{longtable}
\usepackage{array, graphicx}
\usepackage{graphics}
\usepackage{enumitem}
\usepackage{subcaption}
\usepackage[utf8]{inputenc}
\usepackage[utf8]{vietnam} 
\usepackage[justification=centering]{caption}
\usepackage{graphicx}
\usepackage{silence}
\usepackage[english]{babel}
\usepackage{epstopdf}
\usepackage{nomencl}
\usepackage{cite}
 
\usepackage{soul,color}
\usepackage{amssymb}
\usepackage{optidef}
\usepackage{hyperref}
\usepackage{cleveref}
\usepackage{subcaption}
\usepackage{lipsum}
\usepackage{epsfig} 
\usepackage{balance}
\usepackage{textgreek}
\usepackage{bbm}
\usepackage{cite}
\usepackage{algpseudocode}
\usepackage{mathtools}
\usepackage{pdfpages}
\usepackage{setspace}
\usepackage{makecell}
\usepackage[linesnumbered,ruled,vlined]{algorithm2e}
\usepackage{subcaption}
\usepackage{graphicx}
\usepackage{subcaption}

\usepackage[acronym,nonumberlist]{glossaries}

\makeglossaries
\setglossarystyle{super}

\usepackage{supertabular} 
\usepackage{pifont}
\usepackage{makeidx}

\usepackage[table]{xcolor} 
\usepackage{xltabular}
\renewcommand{\arraystretch}{1.2}
\usepackage{comment}

\newacronym{5g}{5G}{Fifth-Generation}
\newacronym{6g}{6G}{Sixth-Generation}
\newacronym{abc}{ABC}{Artificial Bee Colony}
\newacronym{aco}{ACO}{Ant Colony Optimization}
\newacronym{acs}{ACS}{Ant Colony System}
\newacronym{ai}{AI}{Artificial Intelligence}
\newacronym{ann}{ANN}{Artificial Neural Network}
\newacronym{ar}{AR}{Augmented Reality}
\newacronym{auv}{AUV}{Autonomous Underwater Vehicle}
\newacronym{awgn}{AWGN}{Additive White Gaussian Noise}
\newacronym{bs}{BS}{Base Stations}
\newacronym{cceas}{CCEAs}{Cooperative Co-Evolutionary Algorithms}
\newacronym{crns}{CRNs}{Cognitive Radio Networks}
\newacronym{de}{DE}{Differential Evolution}
\newacronym{eas}{EAs}{Evolutionary Algorithms}
\newacronym{embb}{eMBB}{Enhanced Mobile Broadband}
\newacronym{es}{ES}{Evolutionary Strategies}
\newacronym{flc}{FLC}{Fuzzy Logic Controller}
\newacronym{ga}{GA}{Genetic Algorithms}
\newacronym{geo}{GEO}{Geostationary Orbit}
\newacronym{gp}{GP}{Genetic Programming}
\newacronym{hap}{HAP}{High Altitude Platform}
\newacronym{hetnet}{HetNet}{Heterogeneous Network}
\newacronym{hflga}{H-FLGA}{Hybrid-Fuzzy Logic Guided GA}
\newacronym{iot}{IoT}{Internet of Things}
\newacronym{ledma}{LEDMA}{LLM-Enabled Decomposition-based Multi-Objective Algorithm}
\newacronym{leo}{LEO}{Low Earth Orbit}
\newacronym{llms}{LLMs}{Large Language Models}
\newacronym{llm}{LLM}{Large Language Model}
\newacronym{los}{LoS}{Line-of-Sight}
\newacronym{meo}{MEO}{Medium Earth Orbit}
\newacronym{mr}{MR}{Mixed Reality}
\newacronym{ml}{ML}{Machine Learning}
\newacronym{mlp}{MLP}{Multi-Layer Perceptron}
\newacronym{mimo}{MU-MIMO}{Multi-User Multi-Input Multi-Output}
\newacronym{moea}{MOEA}{Multi-Objective Evolutionary Algorithm}
\newacronym{moead}{MOEA/D}{Multi-Objective Evolutionary Algorithm based on Decomposition}
\newacronym{mmtc}{mMTC}{Massive Machine-Type Communication}
\newacronym{mop}{MOP}{Multi-Objective Optimization Problem}
\newacronym{gan}{GAN}{Generative Adversarial Network}
\newacronym{vae}{VAE}{Variational Autoencoder}

\newacronym{mmwave}{mmWave}{Millimeter Wave}
\newacronym{nfv}{NFV}{Network Function Virtualization}
\newacronym{ng}{NG}{Next Generation}
\newacronym{nsgaii}{NSGA-II}{Non-dominated Sorting Genetic Algorithm II}
\newacronym{nsgaiii}{NSGA-III}{Non-dominated Sorting Genetic Algorithm III}
\newacronym{pso}{PSO}{Particle Swarm Optimization}
\newacronym{qoas}{QOAs}{Quantum Optimization Algorithms}
\newacronym{qoe}{QoE}{Quality of Experience}
\newacronym{qos}{QoS}{Quality of Service}
\newacronym{ris}{RIS}{Reconfigurable Intelligent Surfaces}
\newacronym{si}{SI}{Swarm Intelligence}
\newacronym{sinr}{SINR}{Signal-to-Interference-plus-Noise Ratio}
\newacronym{sc}{SC}{Sub-Carrier}
\newacronym{sdn}{SDN}{Software-Defined Networking}
\newacronym{soea}{SOEA}{Single-Objective Evolutionary Algorithm}
\newacronym{spea2}{SPEA2}{Strength Pareto Evolutionary Algorithm II}
\newacronym{spso}{SPSO}{Spherical vector-based PSO}
\newacronym{tbps}{Tbps}{Terabits per second}
\newacronym{ts}{TS}{Tabu Search}
\newacronym{uavs}{UAVs}{Unmanned Aerial Vehicles}
\newacronym{uav}{UAV}{Unmanned Aerial Vehicle}
\newacronym{ub}{UB}{UAVs with Backscatter Devices}
\newacronym{uebs}{UE-BS}{User-to-Base Station}
\newacronym{ueh}{UEH}{Untrusted Relaying Energy Harvesting}
\newacronym{urllc}{URLLC}{Ultra-Reliable Low-Latency Communication}
\newacronym{uuv}{UUV}{Unmanned Underwater Vehicle}
\newacronym{vr}{VR}{Virtual Eeality}
\newacronym{wsns}{WSNs}{Wireless Sensor Networks}
\newacronym{zf}{ZF}{Zero-Forcing}
\newacronym{gwo}{GWO}{Grey Wolf Optimizer}
\newacronym{so}{SO}{Stochastic Optimization}
\newacronym{rl}{RL}{Reinforcement Learning}
\newacronym{sa}{SA}{Simulated Annealing}
\newacronym{psd}{PSD}{Power Spectral Density}
\newacronym{snr}{SNR}{Signal-to-Noise Ratio}
\newacronym{kpis}{KPIs}{Key Performance Indicators}
\newacronym{dqn}{DQN}{Deep Q-Network}
\newacronym{ep}{EP}{Evolutionary Programming}
\newacronym{QA}{QA}{Quantum Annealing}
\newacronym{NISQ}{NISQ}{Noisy Intermediate-Scale Quantum}
\newacronym{QUBO}{QUBO}{Quadratic Unconstrained Binary Optimization}
\newacronym{PQCs}{PQCs}{Parameterized Quantum Circuits}
\newacronym{MARL}{MARL}{Multi-Agent Reinforcement Learning}

\newacronym{MEC}{MEC}{Mobile Edge Computing}
\newacronym{DDPG}{DDPG}{Deep Deterministic Policy Gradient}
\newacronym{MADDPG}{MADDPG}{Multi-Agent Deep Deterministic Policy Gradient}
\newacronym{PPO}{PPO}{Proximal Policy Optimization}
\newacronym{DQN}{DQN}{Deep Q-Network}
\newacronym{SARSA}{SARSA}{State–action–reward–state–action}
\newacronym{LSTM}{LSTM}{Long Short-term Memory}
\newacronym{DNN}{DNN}{Deep Neural Network}
\newacronym{MODE}{MODE}{Multi-objective Differential Evolution}

\newacronym{IoT}{IoT}{Internet of Thing}
\newacronym{BCD}{BCD}{Block Coordinate Descent}
\newacronym{LS}{LS}{Local Search}
\newacronym{MOPSO}{MOPSO}{Multi-objective PSO}
\newacronym{MDP}{MDP}{Markov Decision Process}
\newacronym{DRL}{DRL}{Deep Reinforcement Learning}
\newacronym{DDQN}{DDQN}{Double Deep Q-Network}
\newacronym{RE}{RE}{Resource Efficiency}
\newacronym{SE}{SE}{Spectral Efficiency}
\newacronym{EE}{EE}{Energy Efficiency}
\newacronym{NOMA}{NOMA}{Non-Orthogonal Multiple Access}
\newacronym{HT-PSO}{HT-PSO}{Hyperbolic Tangent PSO}
\newacronym{AoI}{AoI}{Age of Information}
\newacronym{ISAC}{ISAC}{Integrated Sensing and Communication}
\newacronym{UASNs}{UASNs}{Underwater Acoustic Sensor Networks}
\newacronym{FL}{FL}{Federated Learning}
\newacronym{SGD}{SGD}{Stochastic Gradient Descent}

\newacronym{QC}{QC}{Quantum Computing}
\newacronym{bo}{BO}{Bayesian Optimization}
\newacronym{qrl}{QRL}{Quantum Reinforcement Learning}
\newacronym{ran}{RAN}{Radio Access Network}
\newacronym{gpu}{GPU}{Graphics Processing Unit}
\newacronym{cpu}{CPU}{Central Processing Unit}
\newacronym{sac}{SAC}{Soft Actor-Critic}
\newacronym{QoE}{QoE}{Quality of Experience}

\newacronym{LMEA}{LMEA}{LLM-driven Evolutionary Algorithm}
\newacronym{QIM}{QIM}{Quantum-inspired Metaheuristic}
\newacronym{QIGA}{QIGA}{Quantum-inspired Genetic Algorithms}
\newacronym{VQCs}{VQCs}{Variational Quantum Circuits}
\newacronym{GSP}{GSP}{Genetic Simulated-Annealing-based Particle Swarm Optimization}
\newacronym{swipt}{SWIPT}{Simultaneous Wireless Information and Power Transfer}
\newacronym{soa}{SoA}{State-of-the-Art}
\newacronym{ea}{EA}{Evolutionary Algorithm}
\newacronym{ser}{SER}{Symbol Error Rate}

\newacronym{td3}{TD3}{Twin Delayed Deep Deterministic Policy Gradient}
\newacronym{aps}{APs}{Access Points}
\newacronym{gls}{GLS}{Guided Local Search}
\newacronym{ils}{ILS}{Iterated Local Search}
\newacronym{trpo}{TRPO}{Trust Region Policy Optimization}
\newacronym{a3c}{A3C}{Asynchronous Advantage Actor Critic}
\newacronym{a2c}{A2C}{Advantage Actor-Critic}

\newtheorem{definition}{Definition}
\newcolumntype{P}[1]{>{\raggedright\arraybackslash}p{#1}}
\usepackage{adjustbox}
\makeindex 
\usepackage{nomencl}
\makenomenclature
\begin{document}
\title{\huge Single- and Multi-Objective Stochastic Optimization for Next-Generation Networks in the
Generative AI and Quantum Computing Era}
\author{\normalsize
\IEEEauthorblockN{$\text{Trinh Van Chien}, \textit{Member, IEEE}$, $\text{Bui Trong Duc}$, $\text{Nguyen Xuan Tung}$,  \text{Van Duc Nguyen}, $\text{Waqas Khalid}, \textit{Member, IEEE}$, $\text{Symeon Chatzinotas}, \textit{Fellow, IEEE}$, and $\text{Lajos Hanzo}, \textit{Life Fellow, IEEE}$
\vspace*{-0.9cm}
}
\thanks{Trinh Van Chien and Bui Trong Duc are with the School of Information and Communications Technology, Hanoi University of Science and Technology, Hanoi 100000, Vietnam. (e-mail: chientv@soict.hust.edu.vn, duc.buitrong@hust.edu.vn).}
\thanks{Nguyen Xuan Tung is with Faculty of Interdisciplinary Digital Technology,
PHENIKAA University, Yen Nghia, Ha Dong, Hanoi 12116, Viet Nam (email: tung.nguyenxuan@phenikaa-uni.edu.vn).
}
\thanks{Van Duc Nguyen is with the School of Electrical and Electronic Engineering, Hanoi University of Science and Technology, Hanoi 100000, Vietnam (e-mail: duc.nguyenvan1@hust.edu.vn).}
\thanks{Waqas Khalid is with the Institute of Industrial Technology, Korea University, Sejong 30019, South Korea. (e-mail: waqas283@gmail.com).}
\thanks{Symeon Chatzinotas is with the Interdisciplinary Centre for Security, Reliability and Trust (SnT), University of Luxembourg, L-1855 Luxembourg, Luxembourg. (e-mail: symeon.chatzinotas@uni.lu).}
\thanks{Lajos Hanzo is with the School of Electronics and Computer Science, University of Southampton, Southampton, SO17 1BJ, U.K. (e-mail: lh@ecs.soton.ac.uk).}
}

\maketitle
\begin{abstract}
\gls{ng} networks move beyond simply connecting devices to creating an ecosystem of connected intelligence, especially with the support of generative \gls{ai} and quantum computation. These systems are expected to handle large-scale deployments and high-density networks with diverse functionalities. As a result, there is an increasing demand for efficient and intelligent algorithms that can operate under uncertainty from both propagation environments and networking systems. Traditional optimization methods often depend on accurate theoretical models of data transmission, but in real-world \gls{ng}  scenarios, they suffer from high computational complexity in large-scale settings. \gls{so} algorithms, which are designed to accommodate extremely high density and extensive network scalability, have emerged as a powerful solution for optimizing wireless networks. This includes various categories that range from model-based approaches to learning-based approaches. These techniques are capable of converging within a feasible time frame while addressing complex, large-scale optimization problems. However, there is currently limited research on \gls{so} applied for \gls{ng} networks, especially the upcoming  \gls{6g}. In this survey, we emphasize the relationship between \gls{ng} systems and \gls{so} by eight open questions involving the background, key features, and lesson learned. Overall, our study starts by providing a detailed overview of both areas, covering fundamental and widely used \gls{so} techniques, spanning from single to multi-objective signal processing. Next, we explore how different algorithms can solve \gls{ng} challenges, such as load balancing, optimizing energy efficiency, improving spectral efficiency, or handling multiple performance trade-offs. Lastly, we highlight the challenges in the current research and propose new directions for future studies.
\end{abstract}

\IEEEpeerreviewmaketitle
\begin{small}
\makenomenclature
\printnomenclature
\nomenclature[5g]{5G}{Fifth-Generation}
\nomenclature[6g]{6G}{Sixth-Generation}
\nomenclature[abc]{ABC}{Artificial Bee Colony}
\nomenclature[aco]{ACO}{Ant Colony Optimization}
\nomenclature[ai]{AI}{Artificial Intelligence}
\nomenclature[ann]{ANN}{Artificial Neural Network}
\nomenclature[auv]{AUV}{Autonomous Underwater Vehicle}
\nomenclature[awgn]{AWGN}{Additive White Gaussian Noise}
\nomenclature[bss]{BSs}{Base Stations}
\nomenclature[cpu]{CPU}{Central Processing Unit}
\nomenclature[de]{DE}{Differential Evolution}
\nomenclature[ea]{EA}{Evolutionary Algorithm}
\nomenclature[ep]{EP}{Evolutionary Programming}
\nomenclature[es]{ES}{Evolutionary Strategy}
\nomenclature[ga]{GA}{Genetic Algorithms}
\nomenclature[gp]{GP}{Genetic Programming}
\nomenclature[gpu]{GPU}{Graphics Processing Unit}
\nomenclature[gwo]{GWO}{Grey Wolf Optimizer}
\nomenclature[KPIs]{KPIs}{Key Performance Indicators}
\nomenclature[LLM]{LLM}{Large Language Model} 
\nomenclature[ml]{ML}{Machine Learning}
\nomenclature[mmWave]{mmWave}{Millimeter Wave}
\nomenclature[NFV]{NFV}{Network Function Virtualization}
\nomenclature[MU-MIMO]{MU-MIMO}{Multi-User Multi-Input Multi-Output}
\nomenclature[NG]{NG}{Next Generation}
\nomenclature[NSGA-II]{NSGA-II}{Non-Dominated Sorting Genetic Algorithm II}
\nomenclature[NSGA-III]{NSGA-III}{Non-Dominated Sorting Genetic Algorithm III}
\nomenclature[PSD]{PSD}{Power Spectral Density}
\nomenclature[PSO]{PSO}{Particle Swarm Optimization} 
\nomenclature[QA]{QA}{Quantum Annealing}
\nomenclature[QRL]{QRL}{Quantum Reinforcement Learning}
\nomenclature[QoE]{QoE}{Quality of Experience}
\nomenclature[QoS]{QoS}{Quality of Service}
\nomenclature[RIS]{RIS}{Reconfigurable Intelligent Surfaces} 
\nomenclature[RL]{RL}{Reinforcement Learning}
\nomenclature[SA]{SA}{Simulated Annealing}
\nomenclature[sac]{SAC}{Soft Actor-Critic}
\nomenclature[SDN]{SDN}{Software-Defined Network}
\nomenclature[SI]{SI}{Swarm Intelligence} 
\nomenclature[SINR]{SINR}{Signal-to-Interference-plus-Noise Ratio} 
\nomenclature[SNR]{SNR}{Signal-to-Noise Ratio}
\nomenclature[SO]{SO}{Stochastic Optimization}
\nomenclature[SOA]{SoA}{State-of-the-Art}
\nomenclature[SPSO]{SPSO}{Sherical Vector-based PSO}
\nomenclature[SWIPT]{SWIPT}{Simultaneous Wireless Information and Power Transfer}
\nomenclature[Tbps]{Tbps}{Terabits per Second}
\nomenclature[TS]{TS}{Tabu Search} 
\nomenclature[UAV]{UAV}{Unmanned Aerial Vehicle} 
\nomenclature[URLLC]{URLLC}{Ultra-Reliable Low-Latency Communication} 
\nomenclature[UUV]{UUV}{Unmanned Underwater Vehicle} 
\nomenclature[WSNs]{WSNs}{Wireless Sensor Networks} 
\nomenclature[NISQ]{NISQ}{Noisy Intermediate-Scale Quantum}
\nomenclature[QUBO]{QUBO}{Quadratic Unconstrained Binary Optimization}
\nomenclature[PQCs]{PQCs}{Parameterized Quantum Circuits}

\nomenclature[MARL]{MARL}{Multi-Agent Reinforcement Learning}

\nomenclature[MEC]{MEC}{Mobile Edge Computing}
\nomenclature[DDPG]{DDPG}{Deep Deterministic Policy Gradient}
\nomenclature[MADDPG]{MADDPG}{Multi-Agent Deep Deterministic Policy Gradient}
\nomenclature[PPO]{PPO}{Proximal Policy Optimization}
\nomenclature[DQN]{DQN}{Deep Q-Network}
\nomenclature[SARSA]{SARSA}{State–action–reward–state–action}
\nomenclature[LSTM]{LSTM}{Long Short-term Memory}
\nomenclature[DNN]{DNN}{Deep Neural Network}
\nomenclature[MODE]{MODE}{Multi-objective Differential Evolution}

\nomenclature[IoT]{IoT}{Internet of Thing}
\nomenclature[BCD]{BCD}{Block Coordinate Descent}
\nomenclature[LS]{LS}{Local Search}
\nomenclature[MOPSO]{MOPSO}{Multi-objective PSO}
\nomenclature[MDP]{MDP}{Markov Decision Process}
\nomenclature[DRL]{DRL}{Deep Reinforcement Learning}
\nomenclature[DDQN]{DDQN}{Double Deep Q-Network}
\nomenclature[RAN]{RAN}{Radio Access Network}
\nomenclature[RE]{RE}{Resource Efficiency}
\nomenclature[SE]{SE}{Spectral Efficiency}
\nomenclature[EE]{EE}{Energy Efficiency}
\nomenclature[NOMA]{NOMA}{Non-Orthogonal Multiple Access}
\nomenclature[HT-PSO]{HT-PSO}{Hyperbolic Tangent PSO}
\nomenclature[AoI]{AoI}{Age of Information}
\nomenclature[ISAC]{ISAC}{Integrated Sensing and Communication}
\nomenclature[UASNs]{UASNs}{Underwater Acoustic Sensor Networks}
\nomenclature[FL]{FL}{Federated Learning}
\nomenclature[SGD]{SGD}{Stochastic Gradient Descent}

\nomenclature[QC]{QC}{Quantum Computing}

\nomenclature[LMEA]{LMEA}{LLM-driven Evolutionary Algorithm}
\nomenclature[QIM]{QIM}{Quantum-inspired Metaheuristic}
\nomenclature[QIGA]{QIGA}{Quantum-inspired Genetic Algorithms}
\nomenclature[VQCs]{VQCs}{Variational Quantum Circuits}
\nomenclature[GSP]{GSP}{Genetic Simulated-Annealing-based PSO}
\end{small}
\vspace{-0.25cm}
\section{Introduction}
Given the rise of revolutionary technologies and the expanding reliance on multimedia content and data traffic, there is a growing demand to enhance contemporary communication systems \cite{wang2020thirty,letaief2021edge}. \gls{ng} network deployments may face challenges due to this sharp rise in data demand, including the need for an increased system capacity,  and improved mobility while still being able to keep the energy consumption at a low level \cite{chen20235g}. Compared to previous generations, \gls{6g} is expected to offer improvements in transmission speed, data capacity, reliability, latency, and energy efficiency with the added features of connected intelligence \cite{letaief2021edge}. In \gls{ng} communication, the peak data rate is expected to reach the level of \gls{tbps} \cite{kiasaraei2023towards}. \gls{ng} networks are also projected to provide significantly increased capacity, enabling more connections for radio devices and applications \cite{zhang20196g, chen2020vision}. In terms of latency, \gls{6g} networks are expected to outperform \gls{5g} by reducing transmission delays below one millisecond \cite{chen20235g}, which enables new applications requiring near-real-time responses and high precision in self-driving cars, remote surgery, and smart-city technologies \cite{guo2023five}.
\begin{figure}[t]
\centering
\includegraphics[trim=2cm 0.1cm 0.3cm 1.8cm, clip=true, width=0.9\linewidth]{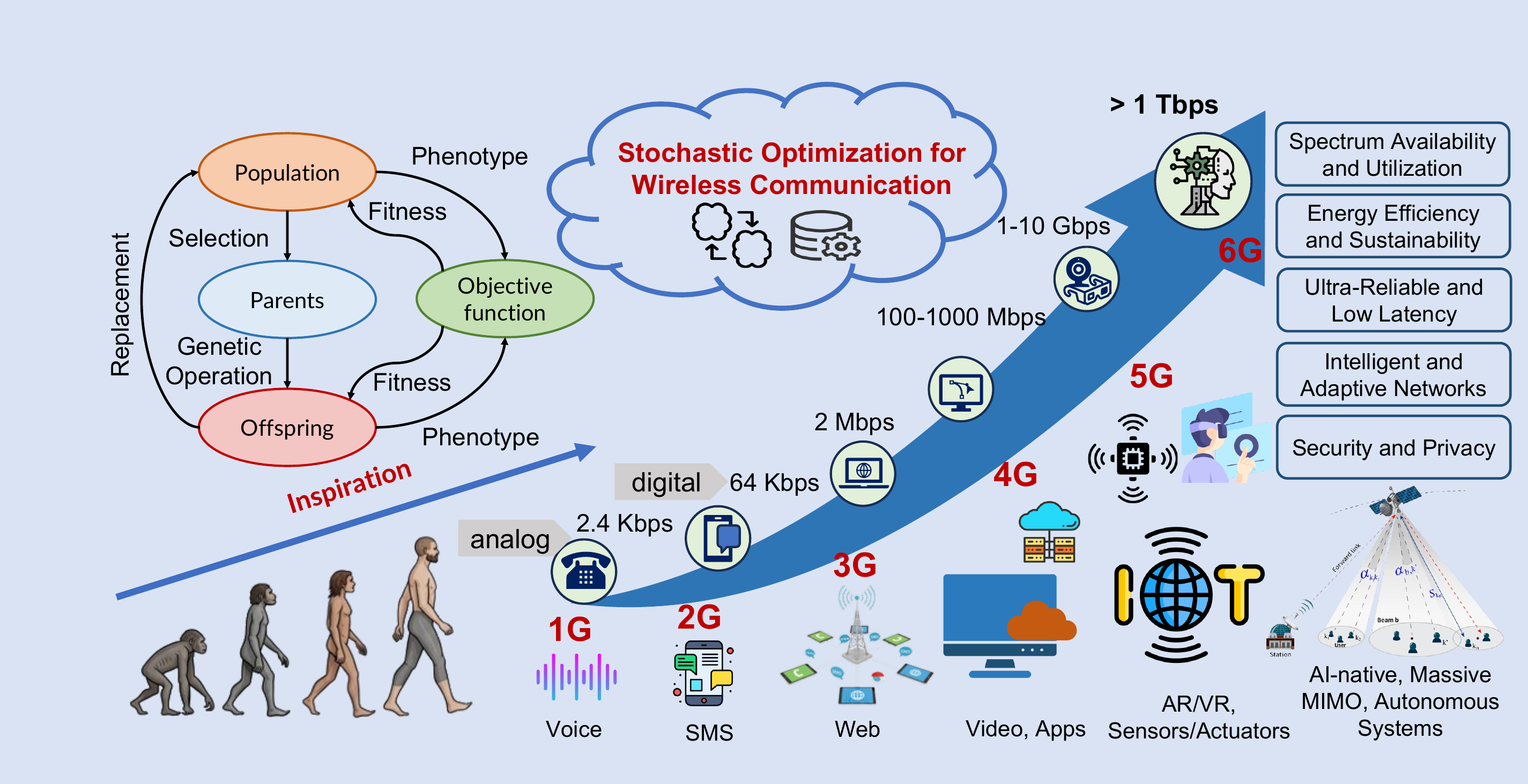}
    \caption{Evolution of \gls{ng} wireless networks and \gls{ea}.}
    \label{fig:evoEA}
\vspace{-0.5cm}
\end{figure}
The \gls{ng} network architecture is expected to foster advanced technologies, including integrated space-air-ground-underwater communication and cloud-based solutions. In a \gls{ng} maritime network, underwater devices, such as \gls{uuv}, buoys and ships, may use acoustic links to connect with surface nodes and aerial relays. This integration enables new services like real-time oceanic monitoring, adaptive \gls{auv}
 swarms and extended coverage across the sea, air, and space.
This network will create highly scalable, seamless and widespread connectivity. Additionally, by incorporating the Terahertz, visible light and unlicensed spectral bands, the architecture will support blue a full range of frequencies for attaining improved performance. Other key innovations, including quantum communication and  massive \gls{mimo} communications are also pushing the evolution of \gls{ng} networks \cite{van2024active}.

Future  advances in mobile technology are expected to support more reliable connections and improved coverage in remote areas,   despite hostile propagation conditions. This improved reliability will be crucial for critical emergency services, industrial automation, and healthcare. Additionally,  \gls{ng} systems are expected to be more energy-efficient than previous generations, aiming for reducing the carbon footprint \cite{he2023reconfigurable}, in the face of high demand for wireless access \cite{wedage2023climate}.   Numerous challenges need exploring, such as resource allocation, and spectrum management, and network security. Thus, effective optimization techniques are essential for \gls{ng} networks. These techniques include mathematical modeling, powerful algorithms, and specific measures to find the optimal solution set. A convex
optimization problem has mathematical properties allowing for well-established methods and existing solvers to reach the optimal solution. However, real-world optimization problems often exhibit non-convexity, making it difficult to guarantee
the global optimum.

\vspace{-0.5cm}
\subsection{\gls{so} in \gls{ng} networks}
Several potent learning-based approaches, such as machine learning, deep learning, and game theory \cite{shi2023machine}, have been demonstrated to be capable of tackling these nonconvex challenges. In many scenarios, such as large-scale \gls{IoT} networks \cite{jiang2020energy},  massive \gls{mimo} systems, or \gls{ris}-aided communications, the problem becomes highly complex, exhibiting multiple local optima and non-differentiable objective functions. Additionally, real-world environments are dynamic, implying that optimal solutions must promptly adapt to changes in traffic patterns, interference levels, and resource availability. Numerous extensive optimization algorithms exist that do not require such data. However, these optimization solvers tend to become excessively complex upon handling large-scale multi-objective problems. This makes real-time decision-making more challenging and less suitable for time-sensitive scenarios. In these cases, researchers often favor heuristic algorithms due to their low computational complexity, which can find sub-optimal solutions with modest complexity. 

\gls{so} algorithms, which are part of the \gls{ai} family, have gained popularity in research for their ability to provide high-quality solutions that are computationally feasible, while maintaining robustness and ensuring convergence. They offer several benefits over traditional methods and match well with the evolution of wireless network generations, as shown in Figure~\ref{fig:evoEA}. They are easy to implement, facilitate global optimization, solve a diverse variety of problems, handle constraints, and support efficient parallel processing. Many algorithms are inspired by natural processes and are based on the broad principles of \gls{eas} and \gls{si}. These algorithms are often inspired by nature, drawing on fields such as biology, physics, or animal behavior. They are stochastic, involving random elements, and typically including parameters that have to be tailored to the specific problem. 
\gls{eas} simulate natural selection using genetic techniques like crossover and mutation  for gradually improving the solutions. Swarm-based approaches, on the other hand, draw their inspiration from the collective behavior of social animals to address optimization challenges. \gls{so} methods typically require a trial-and-error phase to select appropriate algorithmic parameters such as the population size, mutation rate, inertia weights, and the process of tuning these parameters may significantly impact performance. However, unlike many \gls{ai} techniques, which tend to rely on large training datasets and extensive training time, stochastic methods are generally more lightweight and data-independent, making them more practical for resource-constrained or real-time optimization scenarios. 
\begin{table*}[t]
\centering
\caption{Comparison of systems and stochastic optimization topics across different recent surveys.}
\label{tab:comparison}
\scriptsize
\renewcommand{\arraystretch}{1.1}
\setlength{\tabcolsep}{2pt}
\resizebox{\textwidth}{!}{
\begin{tabular}{|l|l|c|c|c|c|c|c|c|c|c|c|c|c|c|c|}
\hline
\textbf{References} &  &
\cite{asim2020review} &
\cite{ji2021survey} &
\cite{pham2021swarm} &
\cite{zeb2022industrial} &
\cite{zhou2023survey} &
\cite{abasi2023survey} &
\cite{singh2024hybrid} &
\cite{abasi2024metaheuristic} &
\cite{butt2024quantum} &
\cite{survey2025_hassan2025quantum} &
\cite{survey2025_sun2025comprehensive} &
\cite{survey2025_wang2025survey} &
\textbf{Our} \\
\hline

\textbf{Years} &  & 
2020 & 2021 & 2021 & 2022 & 2023 & 2023 & 2024 & 2024 & 2024 & 2025 & 2025 & 2025 & 2026 \\
\hline

\multirow{2}{*}{\textbf{Objective}} 
& Single   &  &  & \checkmark & \checkmark & \checkmark & \checkmark &  & \checkmark & \checkmark & \checkmark & \checkmark & \checkmark & \checkmark \\
& Multiple & \checkmark & \checkmark &  & \checkmark & \checkmark & \checkmark & \checkmark & \checkmark & \checkmark & \checkmark & \checkmark & \checkmark & \checkmark \\
\hline
\multirow{6}{*}{\textbf{Optimization}}
& Local-search based algorithms & \checkmark &  & \checkmark &  & \checkmark & \checkmark &  & \checkmark & \checkmark &  &  &  & \checkmark \\
& \gls{eas}        & \checkmark & \checkmark &  & \checkmark & \checkmark & \checkmark & \checkmark & \checkmark & \checkmark &  &  &  & \checkmark \\
& \gls{si}                       & \checkmark & \checkmark & \checkmark & \checkmark & \checkmark & \checkmark & \checkmark & \checkmark & \checkmark & & & \checkmark & \checkmark \\
& Cooperative Co-\gls{eas}       & &  & \checkmark & \checkmark & & & &  &  &  &  &  & \checkmark \\
& \gls{rl}                       & \checkmark & \checkmark & \checkmark & \checkmark & \checkmark & \checkmark & & \checkmark & \checkmark & \checkmark & \checkmark & \checkmark & \checkmark \\
& Hybrid algorithms              &  & \checkmark & \checkmark & \checkmark &  & \checkmark & \checkmark & \checkmark & \checkmark & & \checkmark &  & \checkmark \\
\hline

\multirow{12}{*}{\textbf{Use cases}}
& \gls{ris}                  &  &  & \checkmark &  & \checkmark & \checkmark &  & \checkmark & \checkmark & \checkmark & \checkmark & & \checkmark \\
& Massive \gls{mimo}         &  &  & \checkmark & \checkmark & \checkmark & \checkmark &  & \checkmark & \checkmark & \checkmark & \checkmark &  & \checkmark \\
& Satellite                  &  &  & \checkmark & \checkmark & \checkmark &  &  & \checkmark & \checkmark & \checkmark & \checkmark & \checkmark & \checkmark \\
& \gls{uav}                  & \checkmark &  & \checkmark & \checkmark & \checkmark & \checkmark &  & \checkmark & \checkmark & \checkmark & \checkmark & \checkmark & \checkmark \\
& Sensing and communication  & \checkmark & \checkmark & \checkmark & \checkmark & \checkmark & \checkmark & \checkmark & \checkmark & \checkmark & \checkmark & \checkmark & \checkmark & \checkmark \\
& Edge computing             & \checkmark & \checkmark & \checkmark & \checkmark & \checkmark & \checkmark & \checkmark & \checkmark & \checkmark & \checkmark & \checkmark & \checkmark & \checkmark \\
& Underwater communication   &  &  & \checkmark &  &  &  &  &  & \checkmark & \checkmark & \checkmark & & \checkmark \\
& \gls{QC}                   &  &  & \checkmark & \checkmark & &  &  & \checkmark & \checkmark & \checkmark &  &  & \checkmark \\
& \gls{llm}                   &  &  &   &  &  & & &  &   & &\checkmark  & \checkmark & \checkmark\\
\hline

\end{tabular}}
  \vspace{-0.5cm}
\end{table*}

\vspace{-0.5em}
\subsection{Our key contributions}
\begin{figure}[t] 
    \centering
    \includegraphics[width=0.4\textwidth]{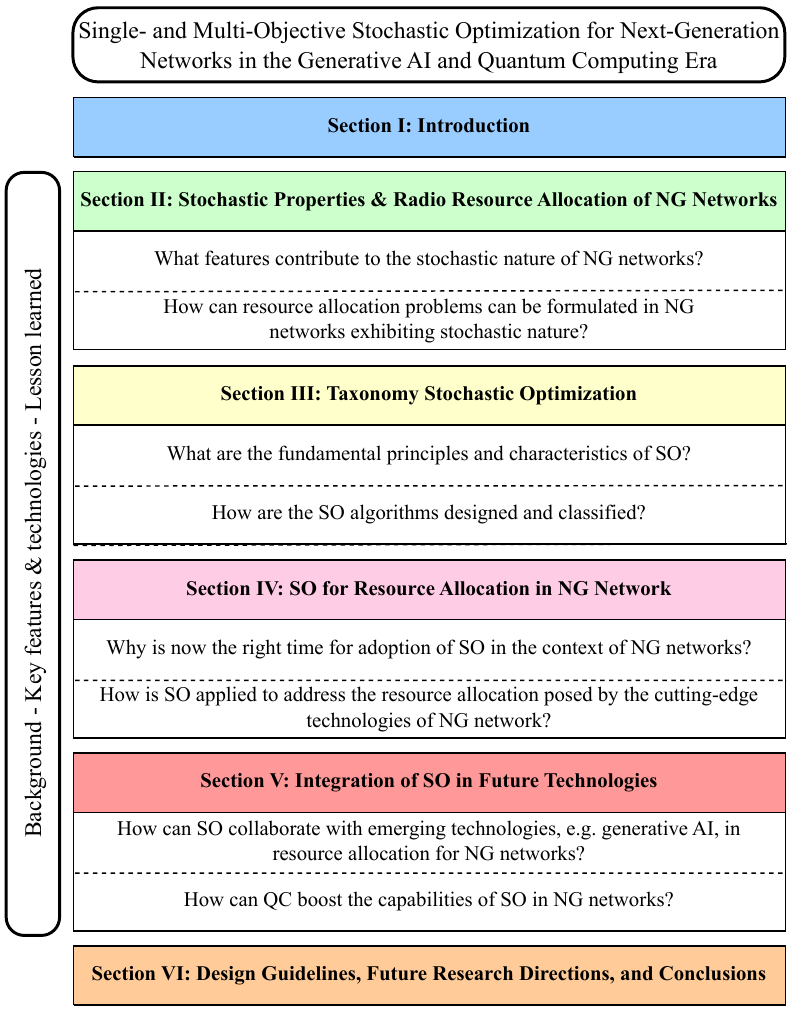} 
    \caption{Main content of the tutorials and surveys}
    \label{fig:6G_paper}
    \vspace{-0.5cm}
\end{figure}

The rise of \gls{ng} wireless technologies, generative \gls{ai}, and quantum computing brings about both opportunities and significant challenges. Conducting an in-depth review of recent advances in \gls{so} for resource allocation in \gls{ng} networks and addressing both the security and privacy concerns are of pivotal importance. We offer a critical appraisal of the latest stochastic optimization techniques, their roles in \gls{ng} communications, and their applications in solving real-world problems. We characterize a suite of potent  algorithms,  and practical use cases. Additionally, we highlight existing research challenges and suggest directions for future studies,  hence providing a resource for researchers focusing on \gls{so} of  \gls{ng} systems. Our key contributions are based on eight open questions and future search directions, as highlighted below: 
\begin{itemize}
    \item  We offer an explicit explanation of the basics, definitions, classifications, and applications of stochastic features of \gls{ng} networks as well as stochastic algorithms, and examine several notable ones. 
    \item We provide design guidelines on how stochastic optimization techniques may be applied to address the challenges of \gls{ng} resource allocation, security, and privacy issues.
    \item  We highlight key challenges and suggest potential directions for future research and development in this area. In addition, we critically contrast stochastic optimization algorithms to traditional deterministic optimization techniques such as \gls{BCD}, for example, for highlighting the associated pros and cons.
    \item Finally, we present Table~\ref{tab:comparison}, which provides a detailed comparison between our study and the latest research in the field. We discuss a suite of open issues, challenges, and the design guidelines of stochastic optimization conceived for resource allocation in \gls{ng} networks.
\end{itemize}
The structure of this paper is illustrated in Fig.~\ref{fig:6G_paper}, providing a comprehensive overview of current research on \gls{so} algorithms in \gls{ng} wireless systems. Section~\ref{Sec:StoPro} presents the stochastic properties of \gls{ng} networks and offers the background necessary for implementing optimization algorithms adopted for resource allocation. A detailed overview of the family of \gls{so} algorithms is given in Section~\ref{sec:Overview}. Section~\ref{sec:soa6G} presents the applications of \gls{so} in \gls{ng} resource allocation. The intrinsic integration of \gls{so} with \gls{ng} networks in the generative \gls{ai} and quantum computation era is discussed in Section~\ref{sec:Integrated}. Finally, our design guidelines take away messages and main conclusions are offered in Section~\ref{sec:conclusions}.

\vspace{-0.35cm}
\section{Stochastic Properties \& Radio Resource Allocation of \gls{ng} Networks}\label{Sec:StoPro}
This section presents the inherent stochasticity of \gls{ng} wireless networks.
Resource allocation problems under stochastic conditions should be formulated and solved in ways that account for random factors influencing the system performance. 

\vspace{-0.5em}
\subsection*{\textbf{Open question one: What features contribute to the stochastic nature of \gls{ng} networks?}}

\begin{figure}[t]
\centering
\includegraphics[trim=0.5cm 0.5cm 0.8cm 0.5cm, clip=true, width=2.8in]{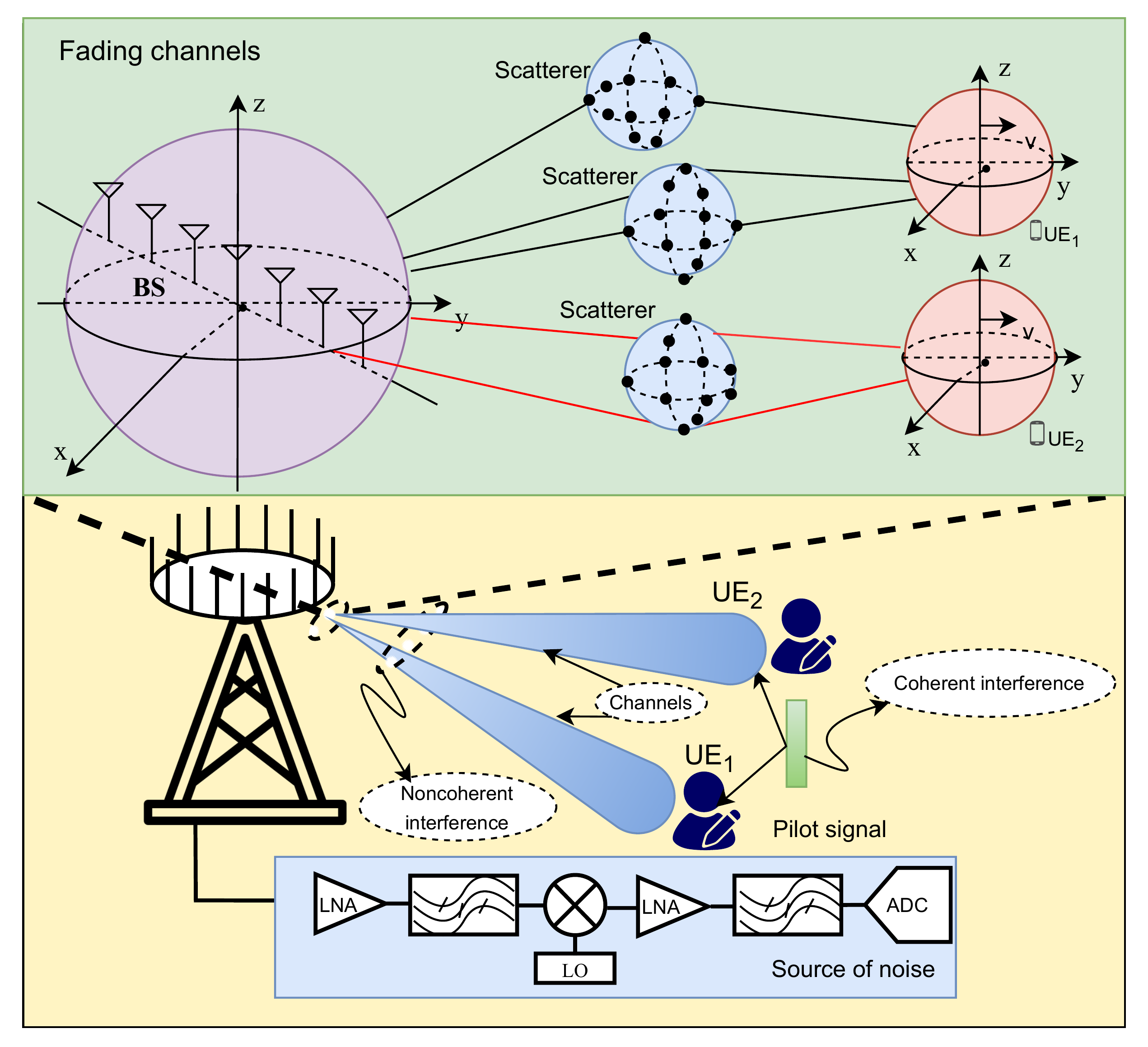}
    \caption{Source of stochastic features in \gls{ng} wireless networks: Channel fading, noise, and mutual interference. }
    \label{fig:StochasticNG}
    \vspace{-0.5cm}
\end{figure}

\subsubsection{Background}
In \gls{ng} systems, stochastic properties indicate that network performance is influenced by sources of uncertainty originating from random propagation conditions, noise and interference, and dynamic user mobility \cite{khan2024beyond}. These uncertainties may not be completely eliminated in typical propagation environments. Consequently, the same input conditions can lead to different solutions and strategies when designing and managing radio resources  \cite{khan2024percentile}. This stochastic characteristic challenges the conventional, optimistic assumptions of deterministic networks and calls for probabilistic modeling  \cite{wang2025terahertz}. Note that the uncertainty may obey diverse probability distributions \cite{zhang2024new}. User mobility introduces temporal variability in the channel impulse response, which affects system performance in unpredictable ways \cite{fan2025dynamic, biagi2025invisible}. Noise and mutual interference arriving from nearby users, as well as the violent received signal fluctuation, further exacerbate the nondeterministic features \cite{willhammar2025achieving}. 

\subsubsection{Key features and considerations}
We provide the principal reasons why \gls{ng} wireless communications are stochastic and  the related mathematical modeling, as shown in Fig.~\ref{fig:StochasticNG}:
\begin{itemize}[leftmargin=*]
\item[$i)$] \textit{Noise} is fundamentally owing to the Brownian motion of electrons in the receiver as highlighted in Fig.~\ref{fig:StochasticNG}. This poses a multifaceted challenge across various communication environments, including space, ground, and underwater \cite{nguyen2024semantic}. The impairments in satellite communications are exacerbated by cosmic radiation, atmospheric conditions, and hardware impairments \cite{wang2024safeguarding}. \gls{uav}-based communications introduce additional noise due to amplifiers, mixers, vibration, and electromagnetic emissions from motors and propellers \cite{gu2025ris}. Terrestrial \gls{ng} systems may operate in the millimeter-wave and terahertz bands, experiencing severe thermal noise due to the extremely high bandwidth \cite{jiang2024terahertz}. Those sources of noise may be modeled by the \gls{awgn} distribution having a flat \gls{psd} \cite{petrov2024wavefront}. 
Involving underwater acoustic communications in \gls{ng} networks introduces non-white (colored) noise, which is caused by marine life, water movement, and shipping. The \gls{psd} of underwater noise is a fluctuation of the frequency \cite{trucco2024predicting}. Note that low-noise amplifiers, digital filters, and robust noise-mitigation algorithms must be  designed and optimized to mitigate the effects of noise and hence improve the \gls{snr} \cite{mendes2025fully}. However, mitigating colored noise requires extra effort due to its frequency-dependent power, and its success depends on accurately estimating the spectral shape \cite{trucco2024predicting}. 
\item[$ii)$] \textit{Fading channels} are influenced by random physical phenomena in both the time and frequency \cite{xiao2024rethinking}. These include multipath propagation, where signals are reflected by buildings, terrain, or other objects (modeled by the scatterers in  Fig.~\ref{fig:StochasticNG}), causing time-varying delays and phases \cite{ju2024statistical}.  Since the exact positions and properties of the propagation environment are unknown or constantly changing, the variations in the transmitted signal are modeled statistically \cite{seid2023multiagent}. The movement of the transmitter, receiver, or nearby objects introduces Doppler shifts and rapid changes in the channel profile \cite{ke2024localization}. Scattering and reflection from irregular surfaces add further randomness to the received signal, as shown in Fig.~\ref{fig:StochasticNG}. This phenomenon is discussed in technical detail relying on the geometry-based stochastic
model proposed in \cite{zhang2023non}. Environmental factors, including weather, temperature, and the presence of obstacles, also evolve unpredictably, reinforcing the need for probabilistic modeling. As a result, the so-called frequency-flat fading is typically described using quasi-static models \cite{ghareeb2024statistical}. In satellite communication, tropospheric scintillation is more prominent at higher frequencies, such as Ka-band, and inflicts further degradations along with turbulent eddies caused by temperature and pressure fluctuations \cite{chakraborty2021tropospheric}. Space communications are also increasingly affected by dynamic frequency shifts in \gls{leo} orbits, self-shadowing from satellite structures, and intensified rain fade due to climate variability as well as dense constellations  \cite{he2024doppler,kisseleff2025massive}. In underwater environments, temperature and salinity fluctuations, surface bubble clouds, and motion from autonomous vehicles create complex, time-varying multipath and Doppler-induced fading \cite{yi2025unified}. Furthermore, man-made underwater structures and biological activities contribute to frequency-selective fading.
\item[$iv)$] \textit{Mutual interference}  refers to the unwanted interaction between transmitters operating in the same or adjacent frequency bands,  especially when multiple devices transmit simultaneously, causing extra degradation in signal quality and reliability \cite{zhang2023interdependent}. Mutual interference is especially critical in dense \gls{ng} networks such as cell-free and machine-type communication having limited co-use bandwidth \cite{mao2024communication}. Coherent interference occurs when the interfering signal has a phase relationship consistent with the desired signal, often resulting from synchronized sources or reflections of the same signal, and significantly impacts signal integrity. In \gls{ng} networks, pilot contamination is a key challenge, where identical or strongly correlated non-orthogonal pilot signals are reused across cells, leading to highly coherent interference during channel estimation and thereby degrading system performance \cite{pawar2024impact}. Noncoherent interference, on the other hand, arises from sources that are almost completely uncorrelated in phase or modulation with the desired signal. These typically behave like random noise, hence they are often modeled as \gls{awgn}. The cutting-edge \gls{ng} technologies  typically aim for mitigating both coherent and noncoherent interference by dynamically controlling the occurrence and disappearance of propagation paths, smart pilot assignments, and distributed antennas close to users \cite{van2021reconfigurable}. Additionally, learning-based interference prediction and reconfigurable spectrum access have been developed to manage interference in high-dynamic network environments \cite{deng2025csi}.
\end{itemize}

\subsubsection{Lessons learned}
The inherent stochasticity of \gls{ng} networks, imposed by dynamic device mobility and time-frequency varying channel conditions, presents significant challenges in evaluating system efficiency and effective resource management. The unpredictable nature of multiple sources motivates us to observe the average system performance over multiple consecutive coherence intervals and over many realizations of small-scale fading coefficients, as well as device locations. Stochastic features also complicate the optimization of key resources such as spectrum, power, and computational capability. Conventional deterministic allocation schemes are often inadequate for addressing these uncertainties, leading to suboptimal system performance and eroded \gls{qos} guarantees. Furthermore, the need to support heterogeneous applications having diverse conditions exacerbates the complexity of resource allocation under stochastic conditions.

\vspace{-1.2em}
\subsection*{\textbf{Open question two:  How can resource allocation problems be formulated in \gls{ng} networks exhibiting stochastic nature?}}

\setcounter{subsubsection}{0}
\subsubsection{Background}
Resource allocation is known as a key methodology for enhancing the efficiency of \gls{ng} systems operating in the face of limited radio resources and stochastic behaviors, \cite{prabhashana2025machine}. This involves the judicious allocation of resources among different users and their devices within a network. The main objective is to maximize performance, while satisfying the diverse requirements of devices, such as their throughput, spectrum, and energy efficiency, just to name a few. Moreover, high-performance, yet low-complexity and power-efficient algorithms are required for real-time implementation in the face of demanding network conditions and user requirements. The integration of optimization techniques into \gls{ng} wireless systems aims for agile resource allocation, leading to smarter and more adaptive management for satisfying the \gls{kpis}  \cite{liu2024survey}. 
\subsubsection{Key features and considerations}
An optimization problem is a mathematical formulation constructed to find the best solution from a set of feasible alternatives, according to a specific criterion \cite{yu2024optimization}. It involves either the maximization or minimization of one or multiple objective functions subject to a set of constraints that define the feasible region.

\textit{Preliminaries:} To provide a critical appraisal of the various algorithmic categories
and their applications, we 
commence with a general single-objective optimization problem formulated in its maximization form as
\begin{equation}
\begin{array}{rl}
    \underset{\mathbf{s} \in \mathcal{D}}{\text{maximization}} & f_0(\mathbf{s}) \\
    \text{subject to} & f_i(\mathbf{s}) \leq 0, \quad i = 1,2,\dots,m, \\
                      & h_j(\mathbf{s}) = 0, \quad j = 1,2,\dots,r,
\end{array}
\end{equation}
where $\mathbf{s} = [s_1, \ldots, s_N]^T$ includes the $N$ optimization variables; the set \( \mathcal{D} \) defines the feasible region of the optimization problem; \( f_0(\mathbf{s}) \) is known as the objective function, also known as the cost or fitness function; and the sets of \( \{ f_i (\mathbf{s}) \}_{i=1}^{m} \) and \( \{ h_j (\mathbf{s}) \}_{j=1}^{r} \) represent inequality and equality constraints. We stress that the objective and constraints include the stochastic features of \gls{ng} wireless networks as the large-scale and small-scale fading effects \cite{pereira2023mobility}. Hence, the feasible set \( \mathcal{D} \) may not always be explicitly defined in practice, but can be determined using an oracle, such as a user-provided software tool. The optimization variables in \( \mathbf{s} \) are considered feasible if \(  \mathbf{s} \in \mathcal{D} \) and satisfy all the given inequality and equality constraints. A solution \( \mathbf{s}_{\mathrm{best}} \) is globally optimal if it holds that
$f_0(\mathbf{s}_{\mathrm{best}}) \leq f_0(s), \forall \mathbf{s} \in \mathcal{D}$.
Conversely, a feasible vector \( \bar{\mathbf{s}} \) is locally optimal if there exists a small positive number \( \epsilon > 0 \) such that
$f_0(\bar{\mathbf{s}}) \leq f_0(\mathbf{s}), \forall \mathbf{s} \in \mathcal{D} \text{ satisfying } \| \mathbf{s} - \bar{\mathbf{s}} \| \leq \epsilon$.
The optimization problems may generally be categorized into convex and nonconvex problems, each requiring distinct solution approaches
\cite{liu2019stochastic}. Convex optimization involves maximizing a convex objective function while satisfying convex constraints. These techniques play a crucial role in engineering applications because in a convex problem any local optimum is also a global optimum \cite{survey2025_wang2025survey}. 

\begin{figure}[t]
    \centering
    \includegraphics[width=\linewidth]{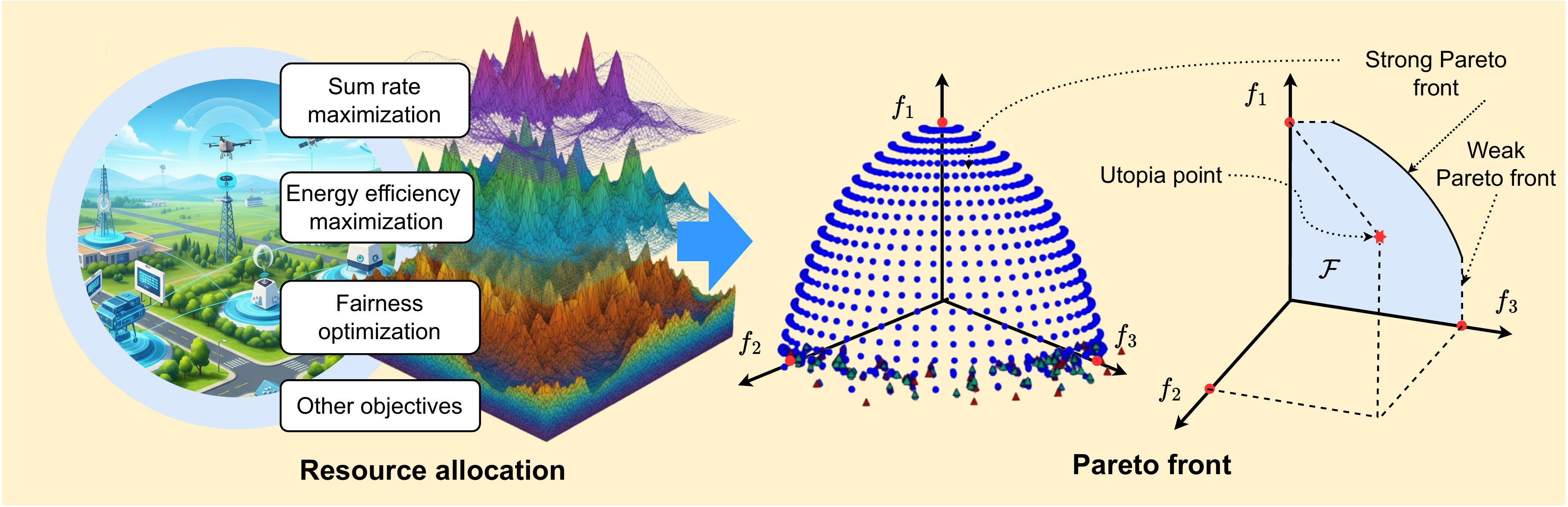}
    \caption{Conflicting objective functions leading to drive multiobjective optimization problems.}
    \label{fig:paretoFront}
    \vspace{-0.5cm}
\end{figure}

In \gls{ng} networks, we typically have multiple objectives $f_{01}(\mathbf{s}), \ldots, f_{0M}(\mathbf{s})$, with $M$ being the number of objectives. A fundamental assumption is that the $M$ objectives are unordered and thus analyzed without bias or predetermined preferences, allowing all possible outcomes to be considered. One might aim for maximizing the $M$ objectives simultaneously as 
\begin{equation} \label{Prob:MultiObj}
\begin{array}{rl}
    \underset{\mathbf{s} \in \mathcal{D}}{\text{maximization}} & \mathbf{f}_0(\mathbf{s}) = [f_{01}(\mathbf{s}),  \ldots, f_{0M}(\mathbf{s})]^T \\
    \text{subject to} & f_i(\mathbf{s}) \leq 0, \quad i = 1,2,\dots,m, \\
                      & h_j(\mathbf{s}) = 0, \quad j = 1,2,\dots,r,
\end{array}
\end{equation}
where the bold faced notation $\mathbf{f}_0(\mathbf{s})$ emphasized the vector-valued nature of the objective function. The $M$ objectives in \eqref{Prob:MultiObj} are inherently conflicting, and due to the absence of ordering among the vector-valued objectives, a global optimum generally does not exist. As only subjectively optimal solutions exist,  the multi-objective problem \eqref{Prob:MultiObj} cannot be solved in a globally optimal sense. Let us denote the achievable objective set as $\mathcal{F}= \{ \mathbf{f}_0(\mathbf{s}): \mathbf{s} \in \mathcal{D} \}$,  which involves all the possibilities of objective values. The set $\mathcal{F}$ is usually compact in the sense that $\mathbf{f}_0 \in \mathcal{F}$ leads to $\alpha \mathbf{f}_0 \in \mathcal{F}, \forall \alpha \in [0,1]$. Most points within the set $\mathcal{F}$ are strictly suboptimal; specifically, any point located in the interior of $\mathcal{F}$ can be excluded, as there are other points in 
$\mathcal{F}$ that are superior across all the $M$ objectives. The set of points that cannot be outperformed in terms of all objectives constitutes the Pareto boundary, see Fig.~\ref{fig:paretoFront} for its illustration. None of the solutions on the objectives of the Pareto boundary can be improved without degrading at least one of the others.

\begin{definition}
The strong Pareto boundary comprises all points in $\mathbf{f}_0 \in \mathcal{F}$ where there is no feasible point $\tilde{\mathbf{f}} \in \mathcal{F} \setminus \{ \mathbf{f}_0 \}$ with $\tilde{f}_{0m} \geq f_{0m}, \forall m \in \{1, \ldots, M\}$.
\end{definition}
The strong Pareto boundary comprises all the feasible points that cannot be objectively rejected because improving any one of the objectives would necessarily degrades at least one of the others. Clearly, any point not lying on the strong Pareto boundary is suboptimal, since there exist other points that are superior or equally effective across all the objectives. In the context of multi-objective optimization, the strong Pareto boundary represents the closest possible approximation to the global optimality. The strong Pareto boundary is a subset of the upper boundary of $\mathcal{F}$. The full upper boundary, known as the weak Pareto boundary, also includes points where some - but not all - objectives can be improved without eroding others. The utopia point is defined as
$\mathbf{s}_{\text{utopia}} = \left[\max\nolimits_{\mathbf{s} \in \mathcal{D}} f_{01} (\mathbf{s}), \ldots,  \max\nolimits_{\mathbf{s} \in \mathcal{D}} f_{0M} (\mathbf{s})\right]^T$, 
which concurrently maximizes all the $M$ objectives and a representative toy example is shown in Fig.~\ref{fig:paretoFront}. If we have $\mathbf{s}_{\text{utopia}} \in \mathcal{F}$, the multi-objective optimization problem is trivial, since the strong Pareto boundary is the utopia point, which is the global optimum. In \gls{ng} networks, the multi-objective optimization problems having multiple conflicting objectives are nontrivial and the global optimum does not exist as we have $\mathbf{s}_{\mathrm{utopia}} \notin \mathcal{F}$. Because the Pareto boundary includes all potentially optimal operating points,  the network should be optimized to represent these points.
\begin{definition}
A feasible point $\mathbf{s}^\ast \in \mathcal{D}$ is the Pareto optimal point if $\mathbf{f}_0 (\mathbf{s}^\ast)$ belongs to the strong Pareto boundary.
\end{definition}
We emphasize that  multiple feasible points in $\mathcal{D}$ might give exactly the
same objective point and therefore the multi-objective function is not bi-objective. The design of \gls{ng} networks is expected to tackle the nonconvex-nonsmooth optimization problems in the face of stochasticity.

\textit{Key features \& contemporary issues}: Convex
optimization may, for example, be harnessed for minimizing the total power consumption by ensuring convergence to a unique global optimum. This defines
the optimal power allocation strategy, while meeting specific
\gls{sinr} thresholds for
each receiver \cite{van2020joint}. Advalnces in modern systems have
significantly transformed the mathematical optimization problems underlying system design. These changes have introduced substantial challenges in understanding, analyzing, and
solving these optimization problems. Many of the single- and multi-objective optimization problems
are complex, often being non-convex and non-deterministic \cite{tang2023energy}. Additionally, the design variables may vary widely, from
continuous to integer or even mixed types. These emerging complexities have driven the development of
advanced optimization theories, algorithms, and techniques to improve the suboptimal solutions, guiding them toward the global optimum \cite{eichfelder2024solver}.

Non-convex problems present additional difficulties as they may not
converge to a single global optimum. Instead, they often
exhibit multiple local optima due to the inherent non-linear
dynamics of wireless networks \cite{ahmed2024active}. This further complicates
the optimization process, requiring specialized techniques for effectively
navigating the complex solution space. Traditional
algorithms conceived for solving non-convex problems are
often resource-intensive and computationally expensive. They typically rely on iterative approximations through convex
formulations \cite{han2021sparse}. Alternatively, one can decompose the original
problem into smaller, more manageable subproblems such as exploiting the block-coordinate descent \cite{xiu2024robust}.
Despite their widespread use, these methods exhibit inherent
limitations, including slow convergence, high computational
complexity, and the inability to guarantee global optimality.
Consequently, approximate approaches have emerged as a
practical alternative, providing high-quality solutions within
a reasonable time frame \cite{azarbahram2024waveform}. Non-convex problems often arise in optimization tasks like pilot assignment \cite{nguyen2025hybrid} or beamforming design \cite{zhu2024robust}, where the objective function is a complex nonlinear function of its parameters.

In the realm of small-scale optimization problems, exhaustive (or brute-force) search-based algorithms are capable of solving non-convex and deterministic problems due to their ability to systematically explore all possible solutions. These methods guarantee the finding of the optimal solution by evaluating every feasible option, making them particularly suitable for optimization problems where the global optimum can be reliably identified \cite{byeon2025next}. For instance, brute force algorithms can be used in combination with general-purpose
solvers to address deterministic optimization problems, delivering accurate results despite their high computational costs, expanding exponentially with the network scale
\cite{jia2020leo}. More explicitly,  the efficiency of
these algorithms is constrained by the problem size, rendering
them impractical for complex large-scale scenarios. This fact makes traditional exhaustive search methods impractical due to their excessive
computational costs and time requirements. Additionally, many
 optimization problems of \gls{ng} wireless networks are NP-hard, as noted in
authoritative studies \cite{butt2024quantum} and operate in uncertain environments \cite{kavehmadavani2024empowering,li2023heterogeneous,10264814}. Consequently, exhaustive search algorithms face significant limitations, necessitating the use of
stochastic algorithms. These stochastic methods are crucial
for efficiently handling the dynamic and fluctuating heterogeneous nature of \gls{ng} systems, which are expected to support large-scale ultrahigh density networks. As exemplified by
the framework in \cite{nguyen2023network}, the \gls{ng} network performance may be readily enhanced by stochastic algorithms conceived
for coordinated management, 
congestion control, and scheduling. This approach aims for
improving network efficiency and minimizing latency by integrating multi-layer optimization strategies that maximize network
utility and employ \gls{so} methods. These
features support rapid convergence and long-term optimization
in dynamic networks. On a similar note, distributed
machine learning techniques, which often rely on \gls{so}, are also popular
in the context of large-scale optimization, offering near-optimal performance and high computational efficiency \cite{shi2023machine}. 

\subsubsection{Lessons learned}
Resource allocation operating in the face of stochastic features imposes significant uncertainty on the network performance. Single-objective optimization techniques typically aim for maximizing a specific metric, possibly subject to some constraints. While effective in well-defined scenarios, single-objective methods often fail to capture the trade-offs that must be met in complex environments, where multiple performance criteria must be satisfied simultaneously. Multi-objective optimization, on the other hand, provides a more flexible framework by considering conflicting objectives. However, the stochastic nature complicates the identification of Pareto-optimal solutions, requiring robust algorithms capable of dynamically adapting to time-varying conditions. Addressing these challenges is critical for ensuring  efficient operation of \gls{ng} networks.
\vspace{-0.25cm}

\section{Taxonomy of Stochastic Optimization} \label{sec:Overview}
This section outlines the foundational taxonomy of \gls{so} by addressing its core principles and algorithmic diversity. First, we elucidate the fundamental characteristics that distinguish \gls{so} from deterministic approaches, emphasizing the role of randomness in modeling optimization problems (open question three). Subsequently, we classify the algorithmic landscape, reviewing the design methodologies that span from population-based metaheuristics to learning-driven frameworks (open question four).

\begin{figure}[t]
    \centering
    \includegraphics[width=0.8\linewidth]{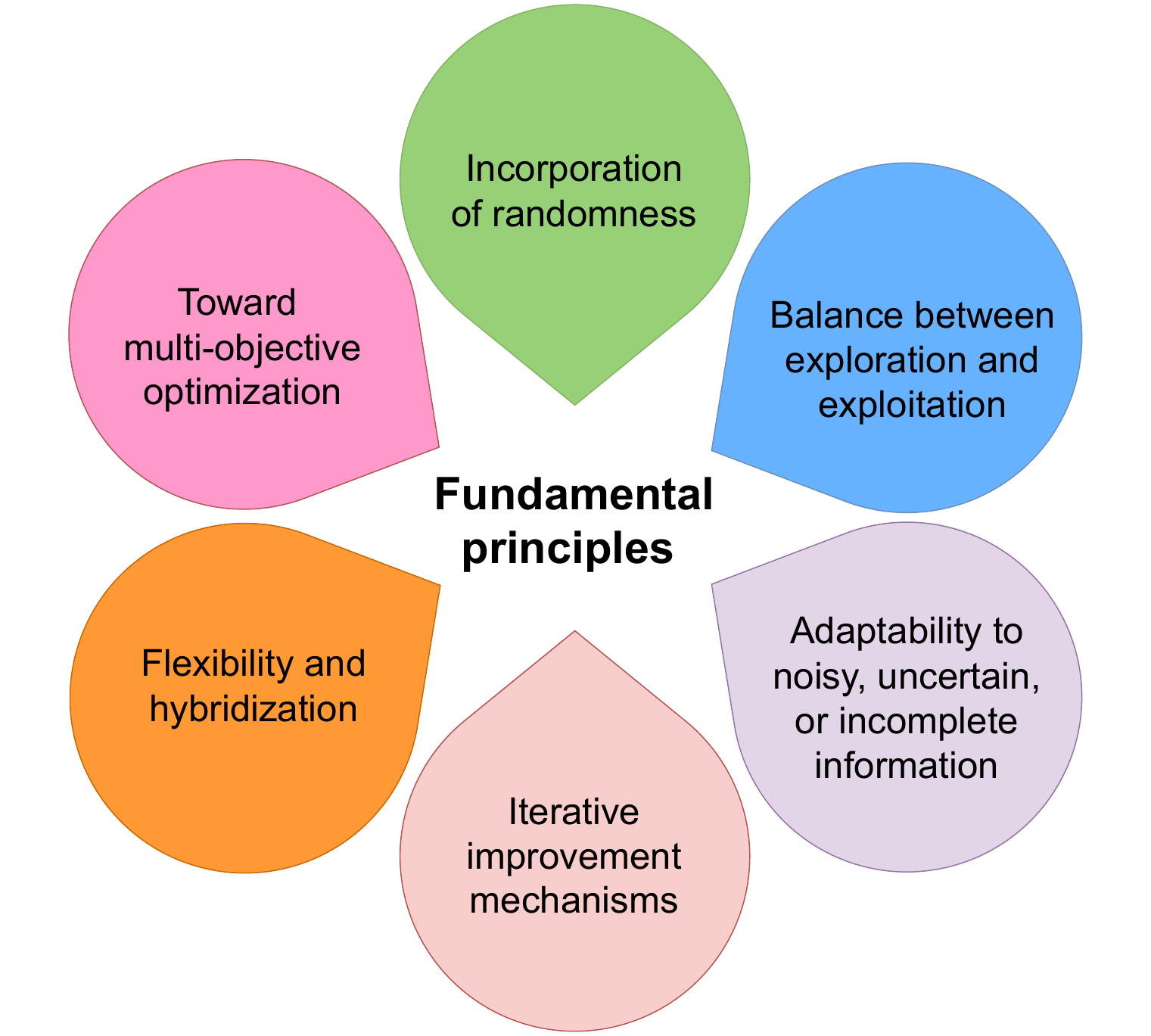}
    \caption{The fundamental principles of \gls{so}.}
    \label{fig:SOprinciple}
    \vspace{-0.5cm}
\end{figure}
\vspace{-0.5em}
\subsection*{\textbf{Open question three: What are the fundamental principles and characteristics of \gls{so}?}}

\setcounter{subsubsection}{0}
\subsubsection{Background} \gls{so} refers to a class of mathematical and computational techniques that incorporate randomness into the optimization process to solve problems characterized by uncertainty, noise, or incomplete information \cite{curtis2020adaptive}. This approach is particularly useful for complex, non-deterministic problems where traditional deterministic optimization methods pose challenges, owing to having unknown parameters or data \cite{huang2022placement_DE2022}. Fig.~\ref{fig:SOOverview} illustrates the intersection of various algorithms across different characteristics, including metaheuristics, \gls{rl}, and a range of other methods, such as \gls{SGD}, approximate algorithms, and hybrid approaches that innovatively combine multiple strategies. Specifically, metaheuristics refer to high-level strategies that proficiently guide the search process for efficiently exploring the solution space and finding optimal or near-optimal solutions \cite{qiao2024resource_GA2024}. These algorithms are generally categorized into two types: \textit{population-based}, which explore a diverse set of candidate solutions, promoting global search space exploration, and \textit{trajectory-based}, which follow a single solution path, emphasizing local refinement. By contrast, \gls{rl} is a branch of \gls{ml} focused on training an agent to make a sequence of decisions by interacting with an operational environment \cite{wang2022deep}. The agent aims to maximize cumulative rewards through trial-and-error learning. \gls{rl} algorithms are divided into three categories: value-based methods, policy-based methods, and actor-critic/hybrid methods, which are inherently stochastic due to their exploration policies and environmental uncertainties. 

\subsubsection{Key features and considerations} \gls{so} relies on the fundamental principles, shown in Fig.~\ref{fig:SOprinciple} and interpreted  as follows:
\begin{itemize}[leftmargin=*]
    \item [$i)$] \textit{Incorporation of randomness.} While deterministic algorithms produce consistent solutions from identical starting points, \gls{so} introduces randomness at various stages of the optimization process, such as solution initialization or through stochastic perturbations in iterative updates \cite{chien2025differential}. Random factors generate deliberate perturbations into the search process, allowing the optimizer to jump to hitherto unexplored areas that may contain better solutions. This mechanism is critical in complex high-dimensional landscapes, where local traps are difficult to avoid through deterministic schemes. Additionally, randomness enhances the robustness by reducing sensitivity to modeling inaccuracies and noisy observations, where deterministic methods may be impacted by imperfect information. Therefore, randomness serves as a crucial component that diversifies the search and enables the algorithm to continuously generate new candidate solutions in optimization \cite{zhu2025group_PSO2025}.     
    \item [$ii)$] \textit{Balance between exploration and exploitation.} The ability to strike a trade-off between exploration and exploitation is considered in the solution search process \cite{goudarzi2023uav}. Exploration refers to the capability of the algorithms to investigate diverse, previously unvisited regions of the solution space, increasing the likelihood of discovering globally optimal or near-optimal solutions. By contrast, exploitation focuses on intensively refining high-quality candidate solutions already identified, allowing for local improvements and faster convergence in promising regions. We emphasize the balance of factors in \gls{so}, where excessive exploration may lead to inefficiency and slow convergence, whereas a lack of strategic exploitation can cause premature convergence to local optima. \gls{so} methods incorporate stochastic mechanisms that adaptively regulate search behavior throughout the optimization process. In metaheuristic frameworks, randomized variation operators are employed to promote exploration, while selection and intensification mechanisms guide exploitation toward promising solution regions \cite{qi2023event}. In terms of \gls{rl}, typically $\varepsilon$-greedy or policy-gradient methods are utilized for explicitly managing the trade-off between exploration and exploitation through policies that strike a balance between learning new strategies and exploiting known high-reward actions \cite{mei20223d}. Similarly, other stochastic algorithms, such as stochastic gradient-based approaches and Bayesian optimization, incorporate probabilistic rules or noise to escape from local optima and encourage robust search dynamics. Therefore, the robustness and efficiency of a \gls{so} method critically depend on the ability to adaptively balance exploration and exploitation across different stages of the search under uncertainty and problem complexity.
    \item [$iii)$] \textit{Adaptability to noisy, uncertain, or incomplete information.} \gls{so} is particularly well-suited to realistic, practical contexts due to the inherent capability to handle uncertainty by harnessing adaptive mechanisms. While deterministic approaches often assume complete and accurate information, \gls{so} relies on probabilistic models, such as modeling noise, sampling from probability distributions, or evaluating solutions over repeated trials. This mitigates the impact of outliers or fluctuations in objective evaluations  \cite{curtis2020adaptive}. Some techniques, such as Monte Carlo sampling, allow the algorithm to form stable estimates of solution quality over time, thus avoiding premature convergence driven by incomplete data \cite{liu2022wireless}. In \gls{eas}, \gls{si}, and stochastic gradient methods, robustness is embedded in the algorithmic design through randomized operators, adaptive sampling, or ensemble-based evaluations. Regarding \gls{rl}, stochastic policies and reward estimators promote policy robustness by learning from diverse trajectories and variable feedback. Moreover, \gls{so} often prioritizes expected performance or risk-aware objectives, rather than optimizing for a single deterministic scenario. This makes the resultant solutions more robust to future variations or incomplete knowledge.
    \item [$iv)$] \textit{Iterative improvement mechanisms.} \gls{so} typically employs iterative improvement mechanisms for gradually approaching optimal solutions \cite{qi2023event, chien2025differential}. The search process is initialized either from a single solution or from a diverse population of candidate solutions generated via randomization or heuristic strategies to promote unbiased exploration. At each iteration, candidate solutions are evaluated using an objective function, and new solutions are produced through stochastic perturbations, recombination, or probabilistic selection. Feedback from these evaluations is then used to adaptively guide subsequent updates, enabling the search to increasingly focus on promising regions of the solution space while preserving sufficient diversity to avoid premature convergence. The cyclical process of evaluation and adaptation mimics a guided trial-and-error approach, where poorly performing solutions are either improved or discarded, while superior solutions are preserved and further exploited. This iterative refinement framework is harnessed in many well-known metaheuristic algorithms. Specifically, \gls{ga} evolves a population through selection, crossover, and mutation, balancing exploration and exploitation across generations \cite{qi2023event}. \gls{sa} iteratively perturbs a single solution, using a temperature-based probability function to accept inferior solutions early on and gradually focus on exploitation as the temperature cools \cite{he2023hybrid}. Regarding \gls{si}, \gls{pso}, and \gls{aco} utilize population-level dynamics and stochastic interactions for iteratively improving candidate solutions based on both individual and collective experiences \cite{9795684}. Additionally, feedback from environmental observation is used to adjust model parameters or policies for stochastic gradient-based methods and \gls{rl} \cite{9815075}. The iterative nature of these algorithms increases the likelihood of converging to optimal or near-optimal solutions given sufficient runtime and diversity, while allowing flexible stopping criteria based on convergence behavior, computational budgets, or application-specific requirements. 
    \item [$v)$] \textit{Flexibility and hybridization.} \gls{so} distinguish themselves through a high degree of flexibility in both algorithmic design and implementation. This flexibility arises from their problem-agnostic and modular structures, which make \gls{so} readily adaptable to domain-specific requirements. Key components of \gls{so} can be modified or extended without altering the core search paradigm \cite{qiao2024resource_GA2024}. Additionally, parameters may be tuned to fit particular problem characteristics, while domain knowledge can be injected via customized operators or initialization strategies \cite{wang2024bi_ACO2024}. Furthermore, specific constraints can be bespoke integrated seamlessly either into the solution evaluation or generation process. We emphasize an important extension of hybridization, which involves combining multiple optimization techniques for enhancing their robustness. Hybrid metaheuristics that combine \gls{eas} either with \gls{LS} or unified \gls{si} techniques may be created as memetic algorithms to solve different optimization problems \cite{van2024active,bi2020energy}. These hybrid designs are capable of balancing global exploration versus local exploitation, thereby improving robustness, convergence rate, and solution quality. Regarding \gls{rl} and \gls{ml}, hybrid techniques that combine metaheuristic and gradient-based optimization have emerged as powerful solutions for non-convex or non-differentiable problems. These methods leverage the global search capabilities of metaheuristics combined with the precision of gradient-based or policy-learning adjustment, often yielding superior results.
    \item [$vi)$] \textit{Toward multi-objective optimization.} \gls{so} algorithms are highly adaptable methods that can be extended to effectively tackle multi-objective optimization problems, which are designed to identify a set of Pareto-optimal solutions, where improving one objective necessarily leads to a trade-off with others \cite{wang2024bi_ACO2024,li2025optimizing_ACO2025}. This flexibility allows for the exploration of multiple conflicting objectives.
\end{itemize}
\subsubsection{Lessons learned} A comprehensive investigation into \gls{so} highlights the strong potential of this paradigm to address a wide range of resource allocation challenges in \gls{ng} networks. Firstly, incorporating randomness enhances the robustness of the algorithms against noise and uncertainty, while improving global exploration, hence mitigating the risk of premature convergence. Secondly, \gls{so} strikes a trade-off between exploration and exploitation. This is achieved by mechanisms such as mutation, policy variation, and stochastic sampling that are capable of successfully navigating complex solution spaces. Additionally, the adaptability of \gls{so} to environments characterized by noisy, incomplete data demonstrates its efficiency across complex problems. Iterative mechanisms such as population evolution, stochastic perturbation, and reinforcement-based updates play a key role in guiding convergence toward optimal solutions. The modular and hybrid nature of \gls{so} enables seamless integration of domain knowledge with diverse strategies for improving performance across various problem domains. Moreover, the ability of \gls{so} as a robust and principled framework to tackle multi-objective problems demonstrates its flexibility in handling conflicting objectives and generating Pareto-optimal solutions. 
\begin{figure}[t]
    \centering
\includegraphics[width=1.0\linewidth]{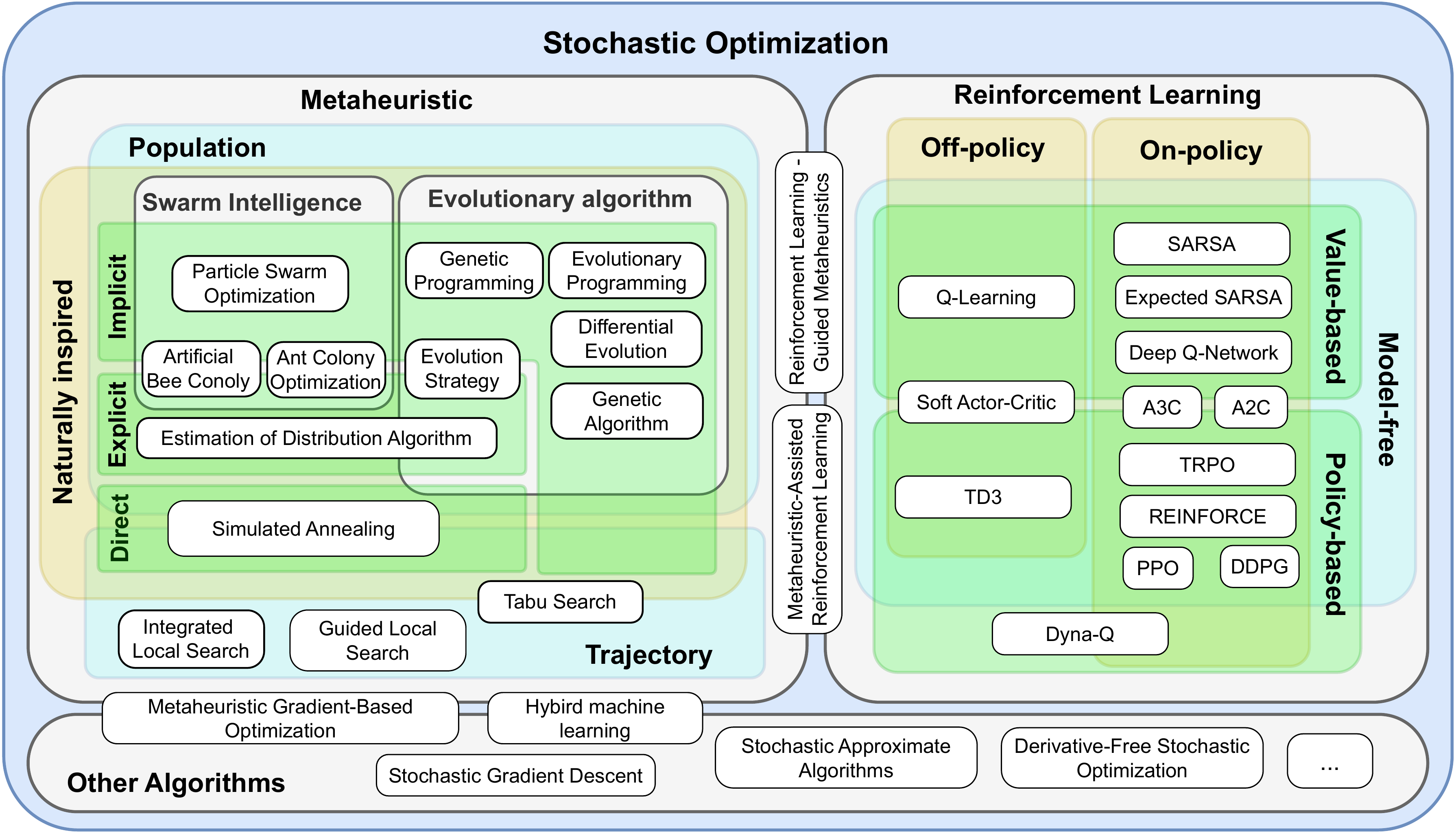}
    \caption{Classification of \gls{so}.}
    \label{fig:SOOverview}
     \vspace{-1.2em}
\end{figure}
\vspace{-1.2em}
\subsection*{\textbf{Open question four: How are the \gls{so} algorithms designed and classified?}}

\setcounter{subsubsection}{0}
\subsubsection{Background} \gls{so} algorithms are designed and classified based on diverse criteria, including search strategies, reliance on gradient information, and learning paradigms. These algorithms can be categorized into three main groups: $1)$ metaheuristics, $2)$ \gls{rl}, and $3)$ other stochastic methods, as depicted in Fig.~\ref{fig:SOOverview}. Each category encompasses diverse principles and mechanisms tailored to different optimization challenges. Metaheuristic algorithms are generally classified into population-based and trajectory-based approaches according to how they explore the search space \cite{dehghani2023coati}. Population-based metaheuristics maintain and evolve a set of candidate solutions simultaneously. This category includes \gls{eas}, which mimic natural evolution, and \gls{si} algorithms inspired by collective animal behavior observed, such as the foraging and hunting behaviors of animal groups, e.g., \gls{pso}, \gls{aco}. Trajectory-based metaheuristics, on the other hand, iteratively improve a single candidate solution by exploring its neighborhood, including  \gls{sa} \cite{he2023hybrid},  \gls{ts} \cite{yildirim2023joint}, and a suite of other algorithms. On the other hand, metaheuristic algorithms are classified based on their search type: $1)$ implicit, $2)$ explicit, and $3)$ direct search.  Implicit search methods rely on indirect mechanisms, such as probabilistic transitions or stochastic choices, to explore the solution space without explicitly computing gradients or directional moves. By contrast, explicit search algorithms generate candidate solutions using well-defined operators informed by heuristics or learned models. Finally, direct search algorithms operate by evaluating objective function values at sampled points and guiding the search solely based on these values, without requiring derivative or probabilistic models. 

Regarding \gls{rl}, the algorithms include three primary categories: value-based, policy-based, and hybrid methods. Value-based \gls{rl} learns an estimate of the expected return (value functions) for different states or state-action pairs and derives policies indirectly by choosing actions that maximize these values \cite{qi2023deep}. Policy-based methods optimize the policy directly through parameterized functions, often using gradient ascent on expected returns \cite{sohaib2023hybrid}. Hybrid methods combine value and policy learning to leverage the strengths of both approaches, such as \gls{sac}, \gls{a3c}, and \gls{a2c} \cite{pivoto2025comprehensive}. Further classification in \gls{rl} distinguishes algorithms based on their interaction with the environment: off-policy and on-policy, or model-based and model-free. On-policy algorithms learn solely from data collected by the current policy being optimized, promoting stable but sometimes sample-inefficient learning (e.g., \gls{SARSA}). Off-policy methods learn from data generated by different behavior policies, which enhances sample efficiency, albeit at the cost of increasing complexity (e.g., Q-learning). Model-based \gls{rl} incorporates an explicit or learned model of the environmental dynamics to plan or simulate future states, yielding potentially expedited learning, whereas model-free \gls{rl} relies entirely on direct interactions.

Additionally, other \gls{so} methods include randomized algorithms such as $1)$ stochastic gradient descent variants, $2)$ stochastic approximate algorithms, and $3)$ Markov Chain Monte Carlo methods, or hybrid algorithms which often combine metaheuristics and \gls{rl} by providing additional robustness in noisy or complex solution spaces. Specifically, \gls{rl} can be employed to initialize solutions or tune parameters within a metaheuristic framework, improving convergence speed and solution quality. Conversely, metaheuristics are capable of controlling hyperparameters or of guiding exploration strategies in \gls{rl} or \gls{ml} models, helping the optimizer to escape local optima or adapt dynamically to changing environments. Such synergistic designs support more flexible and resilient optimization frameworks adapted to the multifaceted challenges of real-world problems.

\subsubsection{Key features and considerations} 
Metaheuristics are high-level optimization frameworks designed to address complex problems where traditional methods are impractical due to non-linearity, discontinuity, large search spaces, or the absence of gradient information \cite{dehghani2023coati}. By employing stochastic operators to explore the solution space, metaheuristics provide an efficient alternative to exhaustive search when computational complexity is prohibitive \cite{qi2023event, van2024active}. Their effectiveness largely stems from the ability to balance exploration and exploitation, thereby improving the likelihood of attaining high-quality, near-optimal solutions.

\begin{figure}[t]
    \centering
    \includegraphics[width=0.8\linewidth]{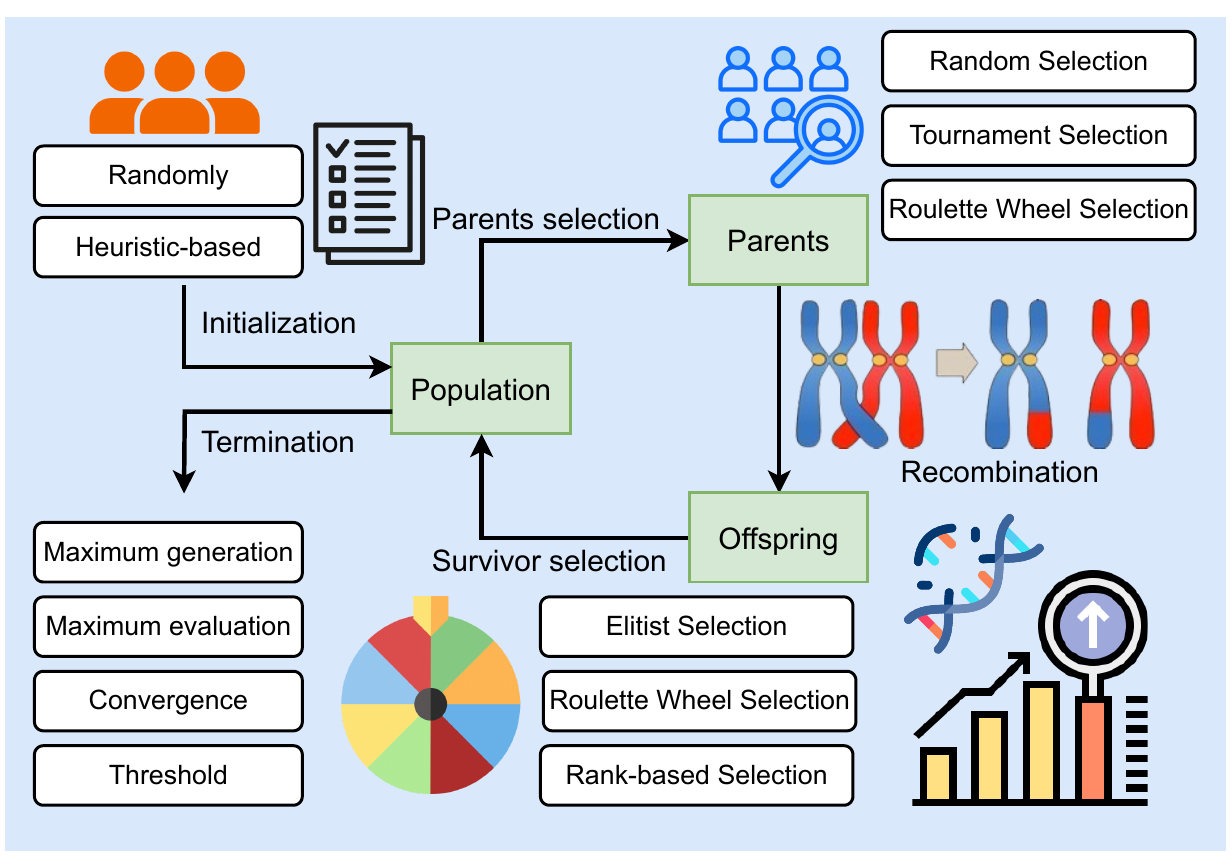}
    \caption{The framework of \gls{eas}.}
    \label{fig:EA_framework}
     \vspace{-0.5cm}
\end{figure}

The general framework of \gls{eas} is depicted in Fig.~\ref{fig:EA_framework}, consisting of two key steps: $1)$ reproduction and $2)$ natural selection. Reproduction enables the inheritance of favorable traits from parents to offspring through crossover of genes, while mutation introduces random variations that preserve genetic diversity. Natural selection ensures that fitter individuals, i.e., those better adapted to their environment, are more likely to survive and pass on their traits, whereas weaker individuals are gradually eliminated. By imitating these mechanisms, \gls{eas} iteratively improve solution quality over successive generations, making them a powerful paradigm for solving complex optimization problems. A central component of \gls{eas} is the representation of solutions. Each individual encodes a candidate solution in a bespoke structure tailored to the problem, and a collection of individuals forms the population. The encoded structure, called the genotype, is decoded into the corresponding phenotype, which can be evaluated against the problem objectives. For instance, individuals can be encoded as phase-shift matrices for \gls{ris} optimization, where decoding yields passive beamforming configurations that directly enhance system performance \cite{huang2022placement_DE2022}. Likewise, in \gls{uav}-enabled systems, individuals may represent flight schedules or trajectories, with each genome encoding task assignments or waypoint sequences for coordinated \gls{uav} operations \cite{goudarzi2023uav, pan2023joint_PSO2023}.

The parent selection mechanism determines which individuals are chosen to reproduce. Hence, parent selection is often biased toward fitter individuals to exploit promising solutions, while maintaining some degree of randomness to preserve population diversity.
New solutions are then created through genetic operators, including crossover and mutation \cite{song2022joint_GA2022}. Crossover combines information from parents to generate offspring that inherit the advantageous traits of both, while mutation introduces random modifications to preserve genetic diversity and allow exploration of new solution regions.

The next step involves survivor selection, where a strategy determines which individuals traverse into the next generation. Common approaches include elitism \cite{wang2024bi_ACO2024}, where the best individuals are preserved unchanged, generational replacement, where the entire population is replaced, or steady-state replacement, where only a subset is updated. Finally, the evolutionary process proceeds until a specific termination condition is met. Termination may be defined by a fixed number of generations, convergence of the population (e.g., minimal improvement over iterations), or achievement of a desired fitness threshold. In multi-objective optimization problems, termination can also depend on the stabilization of the Pareto front \cite{tran2025gamr}.

Leveraging the general framework of \gls{eas}, \gls{ga} inherit the core components, such as the representation of individuals and fitness evaluation, yet distinguish themselves through the design of more effective genetic operators and selection mechanisms. Various crossover strategies have been developed to address different problem structures: shuffle one-point and multi-point \cite{van2024performance}; cyclic crossover ensures that superior parental information is preserved; ordered-crossover and partially matched crossover are widely applied to permutation-based problems; and simulated binary crossover has proven effective for real-valued representations \cite{omidvar2021review}.  The mutation operator introduces stochastic variation that prevents premature convergence and maintains genetic diversity. The most common form is bit-flip based mutation for binary-encoded solutions, while Gaussian or polynomial mutations are typically employed for continuous-valued vectors. Moreover, depending on the representation, it is possible to design problem-specific crossover and mutation operators that exploit the specific structural characteristics of the encoding, such as swapping mutation and exchange-based mutation for virtual network function placement \cite{tam2024multi}. In \gls{ga}, offspring are merged with the parent population, and selection is then applied to determine the survivors for the next generation.

\begin{figure}[t]
    \centering
\includegraphics[trim=1cm 0.5cm 1.0cm 0.5cm, clip=true, width=3.4in]{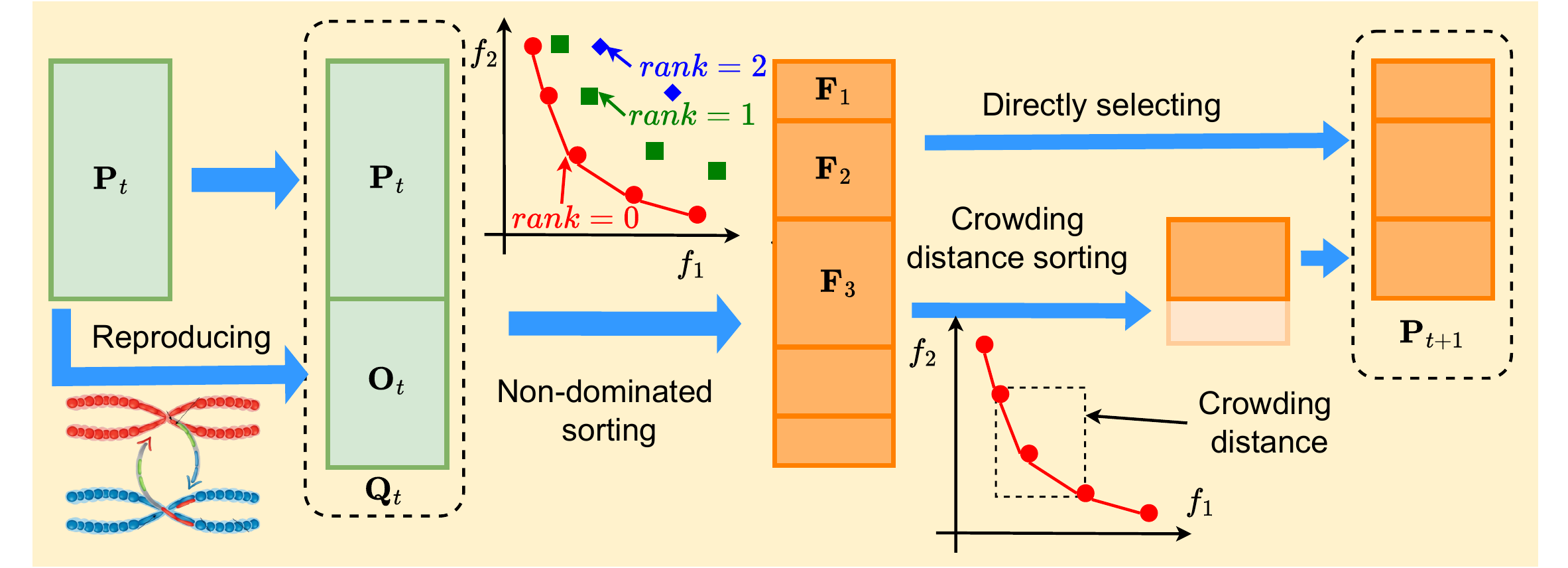}
    \caption{Flowchart of \gls{nsgaii}.}
    \label{fig:nsga2}
     \vspace{-1.2em}
\end{figure}

While single-objective \gls{ga}s have enjoyed popularity, many wireless communication problems are inherently of multi-objective nature requiring simultaneous optimization of conflicting criteria, which leads to the concept of multi-objective \gls{ga}s becoming indispensable. Pareto-based approaches, including  \gls{nsgaii} \cite{zheng2024approximation}, and \gls{nsgaiii} \cite{zheng2025systematic}, extend \gls{ga} by integrating non-dominated sorting and diversity-preserving mechanisms for efficiently approximating the Pareto front. These algorithms generate a set of trade-off solutions. As illustrated in Fig.~\ref{fig:nsga2}, the key distinction between standard \gls{ga} and  \gls{nsgaii} lies in their survivor selection strategy. Instead of evaluating individuals solely by fitness, the combined parent–offspring population is partitioned into non-dominated fronts; lower-rank fronts are preserved first, while the last admissible front is truncated using the so-called crowding distance metric to maintain solution diversity. This mechanism ensures a well-distributed Pareto front across multiple objectives. However,  \gls{nsgaii} often struggles when the number of objectives grows beyond three, since the crowding distance becomes less effective in preserving diversity in high-dimensional spaces. To address this limitation, \gls{nsgaiii} \cite{zheng2025systematic}  introduces reference points as a diversity-preserving mechanism. These reference points are pre-defined or adaptively generated along a normalized hyperplane in the objective space. During selection, if the number of individuals in the last front exceeds the population limit, candidates are chosen based on their proximity to these reference points, ensuring population diversity across all objectives. 

\gls{de} \cite{huang2022placement_DE2022} has emerged as another powerful population-based \gls{ea}, which is particularly well-suited for continuous-valued optimization problems \cite{huang2022placement_DE2022, bandyopadhyay2024quantum_DE2024}. \gls{de} generates offspring through a process of differential mutation, where new candidate solutions are created by perturbing existing individuals using scaled differences between randomly chosen members of the population. Several popular mutation strategies exist, such as DE/rand/1, DE/best/1, and DE/current-to-best/1, each striking a different balance between exploration and exploitation \cite{van2023phase_DE2023}. One of the key strengths of \gls{de} lies in its simplicity and efficiency when dealing with real-valued search spaces, since the mutation mechanism naturally adapts to the scale of the problem and encourages robust global search. Another distinguishing feature of \gls{de} is its selection mechanism, where \gls{de} employs a one-to-one comparison, where each offspring directly competes with its parent, and only the better of the two survives into the next generation. This effective greedy strategy ensures steady improvement of the population, while maintaining computational efficiency 


\gls{MODE} \cite{taha2022multi} extends the \gls{de} framework by integrating the selection mechanisms inspired by Pareto dominance and diversity preservation. \gls{MODE} evaluates solutions through non-dominated sorting, assigning ranks based on Pareto fronts and employing additional diversity measures such as crowding distance or $\epsilon$-dominance to maintain a well-distributed set of trade-off solutions. Reference point–based techniques can also be incorporated, enabling decision makers to guide the search toward preferred regions of the Pareto front, while preserving \gls{de}’s flexible mutation strategies \cite{duc2025multi}. A notable variant enhances this framework by integrating immediate parent replacement with an external population archive \cite{liao2023history}. If a trial solution dominates its parent, it replaces the parent directly to accelerate convergence, while the replaced solution is stored in the archive; otherwise, the trial solution is archived. At each generation, the current population and archive are merged and then truncated using elite- and diversity-preserving strategies inspired by \gls{nsgaii}. This design prevents the premature loss of dominated yet potentially valuable solutions, thereby balancing convergence pressure with diversity preservation and enabling more effective exploration of the Pareto front \cite{omidvar2021review}.

Beyond \gls{ga} and \gls{de}, other paradigms within the \gls{eas} family include \gls{gp}, \gls{es}, and \gls{ep}, each distinguished by their representation and variation mechanisms \cite{zhang2021two}. \gls{gp}   evolves solutions represented as tree structures, which are computer programs or symbolic expressions, allowing simultaneous adaptation of both structure and parameters. This makes \gls{gp} particularly effective for challenging tasks such as symbolic regression and automated design. By contrast, \gls{es} is developed mainly for continuous-valued optimization, and it is characterized by self-adaptive control of mutation parameters, especially step sizes. In contrast to  \gls{ga}’s emphasis on recombination, \gls{es} relies heavily on Gaussian mutations and employs $(\mu,\lambda)$ or $(\mu+\lambda)$ selection schemes to balance exploration and exploitation,  while dynamically adjusting the search. Additionally, \gls{ep}, originally motivated by \gls{ai} applications,  entirely dispenses with crossover and uses mutation as its sole variation operator. Its individuals are often modeled as finite-state machines or real-valued vectors, and evolutionary progress is achieved through stochastic perturbations combined with tournament selection.

\begin{figure}
    \centering
    \includegraphics[width=0.7\linewidth, clip=true, trim=1.8cm 1.8cm 1.8cm 0cm]{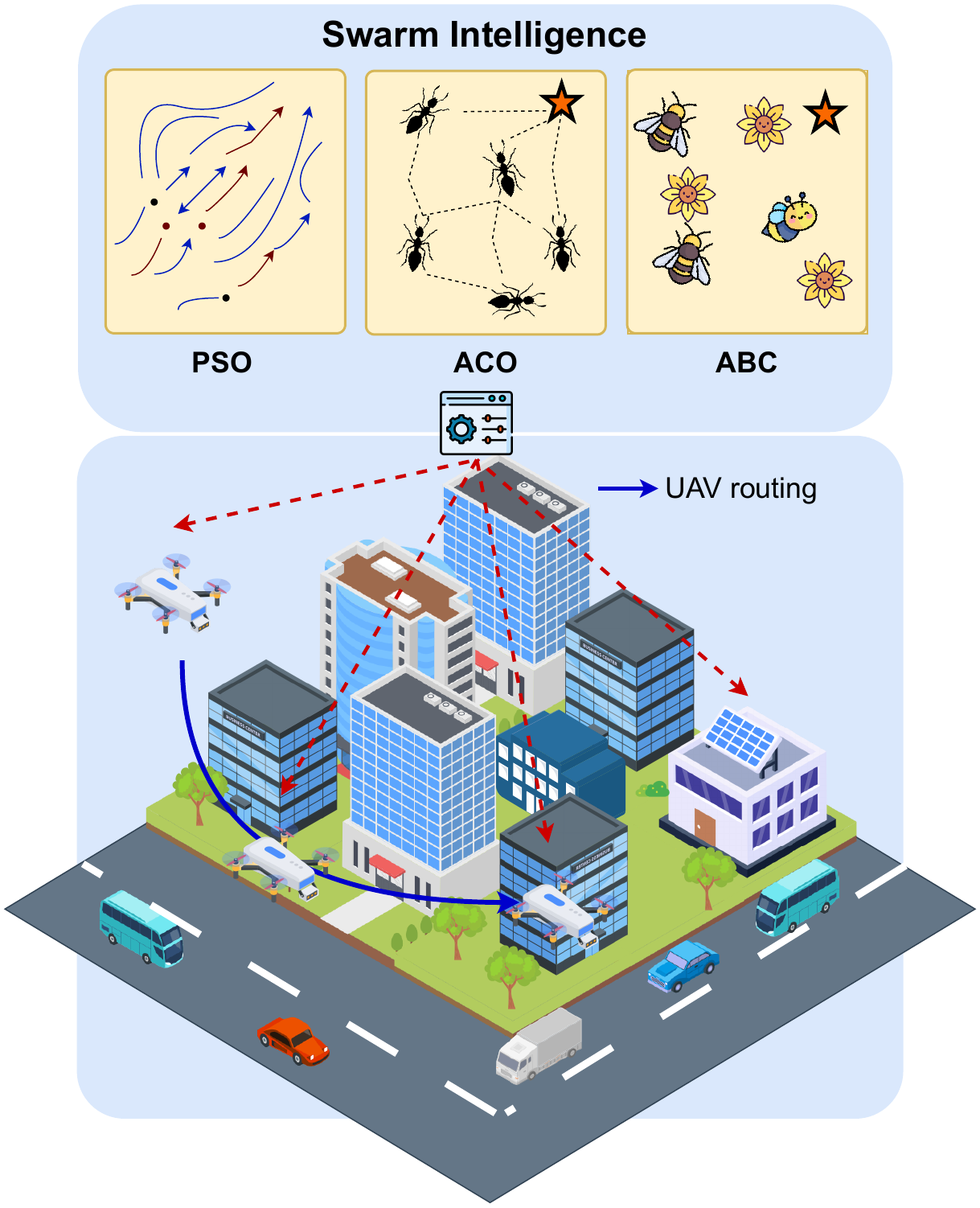}
    \caption{Applications of \gls{si}.}
    \label{fig:SI4NGnetwork}
     \vspace{-0.5cm}
\end{figure}

In parallel to \gls{eas}, the second major branch of population-based metaheuristics is \gls{si}, which is driven by the collective behavior of agents, such as particles, ants, and bees. Specifically, \gls{si} exploits self-organization and local interactions among agents to achieve global problem-solving \cite{qiao2024resource_PSO2024, li2025optimizing_ACO2025}. Agents exchange heuristic information to collaboratively guide the optimization process. Inspired by natural phenomena, such as bird flocks, ant colonies, and fish schools, \gls{si} is inherently decentralized and self-organized, with intelligence emerging from distributed interactions among agents and their environment rather than from a central controller. Operationally, swarm systems are typically composed of simple agents that follow basic behavioral rules, such as attraction or repulsion, and interact locally without global knowledge. Although each agent is individually simple, their collective interactions give rise to emergent intelligent behaviors that enable the swarm to perform complex tasks, such as discovering optimal paths or forming efficient group structures. This bottom-up paradigm underscores the power of self-organization and has been formalized into a diverse range of algorithms, such as \gls{pso} \cite{zhu2025group_PSO2025}, \gls{aco} \cite{li2025optimizing_ACO2025}, and \gls{abc}. Each of these approaches leverages collective behavior and local interactions of swarms to address a wide spectrum of optimization problems across various domains.

\gls{pso} represents candidate solutions as particles that move through the search space according to both the personal best experience and the information shared by other particles \cite{qiao2024resource_PSO2024, zhu2025group_PSO2025}. Position updates are governed by velocity equations that combine inertia, cognitive, and social components, striking a balance between exploration and exploitation. Several extensions of \gls{pso}  have been introduced to enhance its convergence behavior to reduce the risk of stagnation. Hybrid paradigms with sorted particles improve optimization performance by integrating diverse search strategies, while adaptive schemes dynamically adjust paradigm ratios and constriction coefficients across iterations to balance exploration and exploitation. Fully informed search strategies, which incorporate information from the global best solution in each generation, further mitigate premature convergence by encouraging escape from local optima \cite{zhu2025group_PSO2025}. Moreover, leader-adaptive \gls{pso}  combined with dimensionality reduction has been proposed to improve scalability in high-dimensional problems \cite{yang2024leader}. \gls{MOPSO} introduces an external archive of Pareto-optimal solutions and diversity-preserving selection strategies \cite{yuan2023particle}. Instead of converging to a single optimum, \gls{MOPSO} maintains a set of trade-off solutions, with global exemplars selected from the archive based on crowding distance, clustering, or grid-based approaches. These advances extend the applicability of \gls{pso}  and its variants to both single-objective and multi-objective optimization problems. Apart from this, \gls{aco} is inspired by the collective foraging behavior of ants, in which artificial agents construct solutions probabilistically based on their pheromone trails and heuristic information \cite{wang2024bi_ACO2024}. Pheromone levels are updated through evaporation and deposition, reinforcing components associated with higher-quality solutions \cite{li2025optimizing_ACO2025}. 

In multi-objective contexts, extensions employ either multiple pheromone matrices \cite{qian2024cooperative} or aggregation strategies to approximate the Pareto front \cite{wu2024multi}. Similarly, the \gls{abc} concept divides the population into employed, onlooker, and scout bees to search the solution space \cite{zhu2024obabc}. Employed bees refine known food sources, onlookers probabilistically select promising candidates, and scouts introduce random solutions to preserve diversity. Extensions focus on improving neighborhood search or biasing exploration toward high-quality regions, while multi-objective variants incorporate Pareto dominance to address conflicting objectives.

Trajectory-based algorithms are single-solution metaheuristics that explore the search space by iteratively transforming a current solution along a guided trajectory defined by neighborhood structures and acceptance criteria \cite{9795684,zhao2022population_ils,dat2025hsevo}. The algorithms maintain and refine only one solution at a time. Representative approaches include \gls{sa}, which employs probabilistic acceptance to escape local optima \cite{9795684}; \gls{ts}, which uses memory-based mechanisms to prevent cycling; \gls{ils}, which alternates local search with solution perturbations \cite{zhao2022population_ils}; and \gls{gls}, which dynamically penalizes frequently visited suboptimal features \cite{dat2025hsevo}. These methods are effective for combinatorial and numerical optimization problems where exploiting local structure and avoiding premature convergence are critical.



\begin{figure}
    \centering
    \includegraphics[width=0.8\linewidth]{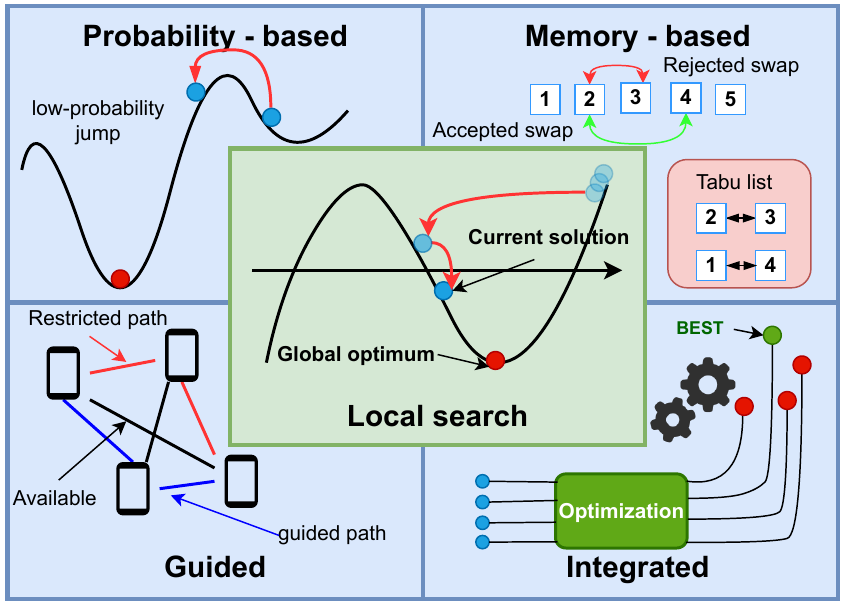}
    \caption{\gls{LS} strategues.}
    \label{fig:ls}
     \vspace{-0.5cm}
\end{figure}

Specifically, \gls{sa} modifies the \gls{LS} framework by introducing a probabilistic acceptance criterion, allowing it to escape local optima more effectively. \gls{sa}  accepts worse solutions with a probability that decreases as the algorithm progresses, controlled by a temperature parameter \cite{el2024multi_SA2024}. This modification enables \gls{sa}  to explore a broader solution space and potentially find the global optimum.  A key difference between \gls{sa}  and \gls{LS} lies in their specific neighbor selection and acceptance criteria. \gls{LS} follows a deterministic strategy,  choosing the best-improving solution, whereas \gls{sa}  uses the Metropolis criterion \cite{jeong2025quantum_SA2025} to probabilistically accept worse solutions, enhancing exploration. 

Regarding memory-based approaches, \gls{ts} guides the search process using deterministic rules augmented with adaptive memory structures. \gls{ts} overcomes key limitations such as entrapment in local optima regions through the use of memory-based constraints and historical search information. In particular, a tabu list is employed to temporarily forbid recently visited solutions or prohibited moves, thereby preventing cycling and promoting exploration of new solution regions \cite{lu2024deepeigen_TB2024}. The memory mechanisms in \gls{ts} are typically hierarchical and adaptive, encompassing short-term memory (e.g., tabu lists), long-term memory for intensification and diversification, and aspiration criteria that allow tabu restrictions to be overridden when a solution is sufficiently promising \cite{lu2024deepeigen_TB2024,yildirim2023joint}. These features enable \gls{ts} to dynamically adjust neighborhood structures based on search history and problem-specific constraints. In summary, Table.~\ref{tab:InnovativeMetaheuristic} illustrates the innovative metaheuristic strategies implemented for specific allocation of resources, probably in wireless communication. 

{
\begin{table*}[t]
\centering
\scriptsize
\caption{Innovative strategies in metaheuristic algorithms}
\label{tab:InnovativeMetaheuristic}
\begin{adjustbox}{max width=\textwidth}
\begin{tabular}{|c|c|c|p{15cm}|}
\hline
\multirow{13}{*}{\rotatebox{90}{\textbf{Metaheuristic}}} & \textbf{} & \textbf{Algorithm} & \textbf{Improvement} \\ \hline

& \multirow{2}{*}{\gls{eas}} 
& \gls{ga} & - Scalarization-based transformation of multi-objective problem into single-objective formulation \cite{nandan2021optimized_GA2021}

- Constraint-aware fitness evaluation using invalid-solution filtering and parameter-free penalty mechanisms \cite{qiao2024resource_GA2024}

- Problem-driven population initialization via conditioned random sampling \cite{song2022joint_GA2022} and cluster-aware assignment strategies \cite{kaleibar2025customized_GA2025} \\

\cline{3-4}
& & \gls{de} & - Adaptive mutation and crossover strategies with operator selection \cite{huang2022placement_DE2022} and parameter control guided by population fitness \cite{van2023phase_DE2023}

- Weighted-sum–based scalarization for handling multi-objective optimization \cite{bandyopadhyay2024quantum_DE2024}\\
\cline{2-4}
&\gls{si} & \gls{pso}  & - Distribution-aware particle initialization to improve early-stage exploration \cite{pan2023joint_PSO2023}, 

- Hybrid \gls{pso} frameworks integrating evolutionary operators for enhanced global exploration and refinement \cite{pan2023joint_PSO2023,qiao2024resource_PSO2024}

- Constraint handling through parameter-free penalty–based fitness evaluation \cite{qiao2024resource_PSO2024}

- Scalarization-based multi-objective optimization using weighted sum vector \cite{zhu2025group_PSO2025}

- Grouping–merging particle update mechanisms with diversity preservation and precision-aware perturbations \cite{zhu2025group_PSO2025} \\
\cline{3-4}
& & \gls{aco} & - Heterogeneous pheromone modeling with learning-assisted heuristics for adaptive solution construction and pheromone updating \cite{wang2024bi_ACO2024}, and using a neural network heuristics \cite{li2025optimizing_ACO2025}\\
\cline{2-4}

& \multirow{3}{*}{\gls{LS}} 
& \gls{sa}  & - Learning-assisted parameter adaptation using neural network–based surrogate modeling \cite{he2023hybrid_SA2023}

- Hybrid local search frameworks combining simulated annealing, tabu mechanisms, and cross-entropy–based neighborhood exploration \cite{el2024multi_SA2024};

- Quantum-inspired optimization via \gls{QUBO} modeling with analytically derived penalty bounds and scaling strategies. \cite{jeong2025quantum_SA2025}\\
\cline{3-4}
& & \gls{ts} & Learning-assisted initialization and probability-guided neighborhood moves with early stopping criteria \cite{lu2024deepeigen_TB2024}\\
\hline

\end{tabular}
\end{adjustbox}
\label{tab:meta_summary}
\vspace{-1.5em}
\end{table*}}

\gls{rl} have recently emerged as a promising class of \gls{so} techniques for tackling high-dynamic resource allocation problems in \gls{ng} networks. \gls{rl} methods are designed as a dynamic and adaptive framework that fits naturally within the \gls{so} family, especially in sequential decision-making problems in the face of uncertainty \cite{pivoto2025comprehensive}. In \gls{rl},  an agent interacts with an environment to learn a policy that maximizes long-term cumulative reward through trial-and-error, making it well-suited for online and adaptive optimization in time-varying wireless systems. The foundation of \gls{rl} is modeled as a \gls{MDP}, defined by a tuple $(\mathcal{S}, \mathcal{A}, P, R, \gamma)$, explicitly $\mathcal{S}$ is the set of states; $\mathcal{A}$ is the set of actions; $P(s'|s, a)$ is the probability of transitioning to state $s'$ after taking action $a$ in state $s$; $R(s, a)$ is the expected immediate reward; and $\gamma \in [0,1]$ is the discount factor reflecting the importance of future rewards.  Additionally, a policy $\pi(a|s)$ specifies the agent's behavior by defining a probability distribution over actions. The objective is to find an optimal policy $\pi^{\ast}$ that maximizes the expected return
\begin{equation}
    G_t = \sum\nolimits_{k = 0}^{\infty} \gamma^kR_{t + k + 1},
\end{equation}
where $G_t$ denotes the return at time step $t$, $R_{t+k+1}$ is the reward received $(k+1)$ steps ahead, and the discount factor $\gamma$ controls the trade-off between immediate and long-term rewards, with smaller values favoring short-term gains and larger values emphasizing future returns. Additionally, value functions are used to evaluate policy performance. The state-value function $V_{\pi}(s)$ denotes the expected return from state $s$ under policy $\pi$, while the action-value function $Q_{\pi}(s,a)$ represents the expected return obtained by taking action $a$ in state $s$ and subsequently following $\pi$. These quantities can be computed recursively using the Bellman equations as \cite{pivoto2025comprehensive}
\begin{equation}
    V_{\pi}(s)  = \mathbb{E}_{\pi} \left\{\sum\nolimits_{k = 0}^\infty \gamma^k R_{t+k+1}| S_t = s\right\},
\end{equation}
\begin{equation}
    Q_\pi(s, a) = \mathbb{E}_{\pi}\left\{ G_t|S_t = s, A_t = a\right\}.
\end{equation}
Here, $S_t$ and $A_t$ are the state and action at time step $t$, respectively. The optimal state-value function $V^{\ast}(s)$ is the maximum expected return achievable from state $s$ over all policies and satisfies the Bellman optimality equation, which expresses the state value as the maximum expected sum of the immediate reward and the discounted value of subsequent states \cite{pivoto2025comprehensive}
\begin{equation}
    V^*_{\pi} (s) = \max_{a\in \mathcal{A}} \big[ R(s, a) + \gamma \sum\nolimits_{s'\in S} P(s'|s,a)V^*(s')\big].
\end{equation}
On the other hand, the optimal action-value function $Q^*(s, a)$ estimates the maximum expected return. The relationship between $Q^{\ast}(s, a)$ and future reward is formally based on the Bellman optimality equation, which is defined as follows \cite{pivoto2025comprehensive}
\begin{equation}
    Q^*(s, a) = R(s, a) + \gamma \sum\nolimits_{s'\in S}P(s'| s, a)\max_{a'\in \mathcal{A}}Q^*(s', a'). 
\end{equation}
In terms of the value-based approach, such as those in Q-learning, \gls{SARSA} and \gls{DQN}, methods focus on estimating the action-value function, which is then used to derive the optimal policy. These methods leverage the Bellman equation to update value functions over time, gradually refining their estimation of long-term rewards. The stochastic nature of wireless environments, characterized by random state transitions due to fading channels or user mobility, is addressed by averaging over multiple experiences to obtain robust value estimates. For instance, Q-learning has been applied to dynamic spectrum access and transmit power control \cite{qlearning_tpc_naderializadeh2021resource}, where channel conditions vary probabilistically, and the agent must learn optimal strategies without explicit knowledge of the system model \cite{mei20223d}. However, value-based methods often struggle in continuous action spaces and may require discretization, which can limit performance in fine-grained control tasks such as beamforming or power allocation. Policy-based methods, including REINFORCE \cite{reinforce_wang2024deep}, \gls{td3}, \gls{DDPG} \cite{pivoto2025comprehensive}, \gls{trpo} \cite{long2025lyapunov_TRPO2025}, and \gls{PPO} \cite{ppo_guo2025secrecy}, directly parameterize and optimize the policy function, making these methods particularly effective in modeling stochastic policies, where actions are selected according to a state-dependent stochastic policy. Additionally, \gls{rl} approaches are adaptive and can be tuned according to the context of the problem, as illustrated in Table~\ref{tab:InnovativeRL}.

\begin{table*}[t]
\centering
\caption{Innovative strategies in \gls{rl}}
\label{tab:InnovativeRL}
\scriptsize
\begin{adjustbox}{max width=\textwidth}
\begin{tabular}{|c|c|c|p{15cm}|}
\hline
\multirow{12}{*}{\parbox{2mm}{\centering\rotatebox{90}{\textbf{\gls{rl}\quad\quad\quad\quad\quad}}}}
& \textbf{} & \textbf{Algorithm} & \textbf{Improvement} \\ \hline

& \multirow{4}{*}{\parbox{2mm}{\centering\rotatebox{90}{Value-based~~~~~}}}
& Q-Learning
& - Problem-aware state abstraction and latency-driven reward design leveraging channel correlation information \cite{tripathi2023optimal_QL2023}

- Enhanced temporal-difference learning with eligibility traces, multi-objective reward formulation, and adaptive learning-rate control \cite{sharvari2024improved_QL2024}

- Q-value estimation using \gls{LSTM} for fast convergence \cite{wang2025fast_QL2025} \\ \cline{3-4}

& & \gls{SARSA}
& - Multi-step temporal-difference updates with eligibility traces and deep function approximation using \gls{DNN}s and replay buffers for improved convergence stability \cite{ahsan2022reliable} \\ \cline{3-4}

& & Expected \gls{SARSA}
& - Variance-reduced value updates through expectation over action distributions \cite{pivoto2025comprehensive} \\ \cline{3-4}

& & \gls{DQN}
& - Deep value-function approximation with experience replay and target networks, including overestimation mitigation via Double \gls{DQN} \cite{9552212} \\ \cline{2-4}

& \multirow{4}{*}{\parbox{2mm}{\centering\rotatebox{90}{Policy-based~~~~~}}}
& \gls{PPO}
& - Clustering-based interaction reduction to improve scalability in multi-agent settings \cite{pan2023joint_PSO2023}

- Multi-agent with centralized training and distributed execution, incorporating global rewards and state normalization for cooperative learning \cite{sharvari2024improved_QL2024}

- Stability-aware policy optimization via Lyapunov-based decomposition and hierarchical \gls{PPO} \cite{long2025lyapunov_TRPO2025} \\ \cline{3-4}

& & \gls{DDPG}
& - Prediction-assisted continuous control with \gls{qos}- and energy-aware reward design \cite{zhao2024deep_DDPG2024}

- Predictive localization through a \gls{DNN} that forecasts the future positions of vehicular users and eavesdropper \cite{ahmad2025intelligent_DDPG2025}\\ \cline{3-4}

& & \gls{td3}
& - History-aware actor modeling with \gls{LSTM} and action-space regularization \cite{han2023two_TD32023}

- Trajectory-aware actor–critic optimization with delayed updates and target policy smoothing \cite{puspitasari2024td3_TD32024} \\ \cline{3-4}

& & Dyna-Q
& - Hybrid model-based planning and model-free \gls{rl}. \\ \cline{2-4}

& \multirow{3}{*}{\parbox{2mm}{\centering\rotatebox{90}{Actor--Critic}}}
& \gls{sac}
& - Entropy-regularized actor–critic learning with privacy-aware reward modeling \cite{shen2024sac_SAC204} 

\quad\\ \cline{3-4}

& & \gls{a3c}
& - Asynchronous parallel actor–critic learning for scalable training  \\ \cline{3-4}

& & \gls{a2c}
& - Synchronous actor–critic learning with batched gradient updates \cite{pivoto2025comprehensive} \\ \hline

\end{tabular}
\end{adjustbox}
\vspace{-0.35cm}
\end{table*}

\subsubsection{Lessons learned}   
No single \gls{so} algorithm is universally optimal; instead, performance depends strongly on problem characteristics. \gls{ga}s are particularly effective for discrete or mixed optimization problems due to their effective exploration of large search spaces via crossover and mutation. \gls{de} and \gls{pso}  are eminently well-suited for continuous optimization problems, particularly for those having non-linear objective functions and multiple local optima, and can be combined to enhance search efficiency, while also offering faster convergence. Other  \gls{si}  algorithms are effective for optimization tasks requiring cooperative behaviors, like network routing, and can be enhanced by combining them with global search strategies. Additionally, \gls{LS}-based algorithms are effective when neighborhood structures are explicitly defined, as they systematically exploit local information to refine solutions and escape poor local optima through adaptive exploration. \gls{rl} algorithms are ideal for dynamic decision-making problems, and can be integrated with \gls{ea}s for adaptability in a fluctuating environment. Thus, the algorithm orchestrations depend on the problem characteristics, and an adaptive approach can significantly improve solution quality and computational efficiency.
\vspace{-0.25cm}
\section{\gls{so} for Resource Allocation in \gls{ng} Networks} \label{sec:soa6G}
This section examines the strategic application of \gls{so} techniques in \gls{ng} networks, addressing two critical questions. First, we examine the driving forces behind the current adoption of \gls{so}, analyzing why the escalating complexity and heterogeneity of \gls{ng} environments necessitate a paradigm shift toward stochastic methods (open question five). Second, we investigate the practical efficiency of these techniques in resolving intricate resource allocation challenges posed by cutting-edge \gls{ng} technologies (open question six).

\vspace{-0.5em}
\subsection*{\textbf{Open question five: Why is now  the right time for adoption of \gls{so} in the context of \gls{ng} networks?}}

\setcounter{subsubsection}{0}
\subsubsection{Background} \gls{ng} networks demand efficient allocation of spectrum, power, time, and computing resources under highly dynamic and non-stationary conditions \cite{qi2023deep, qi2023event}. Fluctuating channel conditions, user mobility, time-varying traffic, and inter-node interference create a probabilistic environment that challenges traditional deterministic optimization methods \cite{van2024active}. \gls{so} provides a principled framework for adaptive and data-driven decision-making by incorporating randomness, probabilistic modeling, and learning from both historical and real-time observations, thereby enhancing robustness and flexibility in resource allocation. Modern advances in computational capabilities and network intelligence have made stochastic approaches increasingly practical and scalable for \gls{ng} networks. As networks evolve toward \gls{urllc}, enhanced mobile broadband, and massive machine-type communications, \gls{so} emerges as a critical tool to ensure stability, efficiency, and intelligent resource management in the \gls{ng} era.


\subsubsection{Key features and considerations} As \gls{ng} wireless networks scale in size, density, and service diversity, several factors make the adoption of computationally efficient \gls{so} not only timely but essential for attaining high performance and prompt adaptability. Tab.~\ref{tab:SO_considerations} highlights the  challenges of resource allocation in the \gls{ng} networks and the benefits of \gls{so}, which includes as:
{
\begin{table*}[t]
\centering
\scriptsize
\caption{Design considerations and the role of \gls{so} in \gls{ng} networks}
\begin{tabular}{|@{}p{3cm}|p{3.5cm}|p{6cm}|p{4.6cm}@{}|}
\hline
\textbf{Design Consideration} & \textbf{NG Network Challenge} & \textbf{Role of \gls{so}} & \textbf{Example Techniques} \\
\hline
\textbf{Scalability} & Massive device connectivity in ultra-dense deployments & \gls{so} supports scalable decision-making by decentralizing computation and approximating in high-dimensional spaces & \gls{ga},   \gls{pso},  Deep Q-Learning, Decentralized \gls{MARL} \\
\hline

\textbf{Partial Observability} & Limited or delayed feedback due to constrained control channels & \gls{so} exploits stochastic observations to derive near-optimal decisions without full state information & Thompson Sampling, Bayesian Optimization \\
\hline

\textbf{Non-convexity} & Joint optimization of user association, power control, beamforming & \gls{so} explores non-convex landscapes to identify feasible high-quality solutions beyond deterministic methods & \gls{eas}, \gls{sa}, Deep \gls{rl} with exploration strategies \\
\hline

\textbf{Real-Time Constraints} & Ultra-low latency requirements & \gls{so} supports fast convergence and online adaptation under strict latency budgets & Model-free \gls{rl},  Online Mirror Descent, Metaheuristics with early stopping \\
\hline

\textbf{Multi-Objective Trade-offs} & Conflicting \gls{kpis} & \gls{so} approximates Pareto frontiers to balance trade-offs among multiple objectives dynamically & 
\gls{nsgaii}, \gls{nsgaiii}\\
\hline

\textbf{Integration with \gls{ai} Models} & \gls{ai}-native \gls{ng} networks require runtime adaptability and resilience & \gls{so} provides a robust optimization layer for training, inference, and deployment under uncertainty & Stochastic Gradient Langevin Dynamics, Bandit Algorithms, \gls{so} for \gls{llm} pipeline tuning \\
\hline

\textbf{Dynamic Topologies} & Non-stationary networks with mobile users, \gls{uav}s, and edge nodes & gls{so} adapts policies and resource allocation to evolving network topologies & Adaptive \gls{rl},  Online Learning with Drift Detection, Contextual Bandits \\
\hline

\textbf{Limited Labelled Data / Prior Knowledge} & Sparse measurements in emerging or under-deployed regions & \gls{so} leverages prior distributions and sample-efficient learning to operate with limited or delayed feedback. & Bayesian \gls{rl},  Transfer Learning with \gls{so} priors, Gaussian Processes \\
\bottomrule
\end{tabular}
\label{tab:SO_considerations}
\vspace{-1.2em}
\end{table*}}

\textbf{Scalability} is a fundamental barrier to the adoption of \gls{so} in large-scale systems.  \gls{ng} networks are expected to support millions of devices in ultra-dense deployments, resulting in high-dimensional optimization problems with rapidly increasing complexity \cite{10169085,10922192}. The robustness of distributed \gls{so} algorithms leads to excellent convergence properties even when the network has unprecedented sizes, in contrast to centralized approaches that become computationally intractable. Advanced distributed \gls{so} frameworks leverage consensus-based mechanisms that enable network nodes to collaboratively solve global optimization problems without centralized coordination \cite{li2021distributed}. These techniques maintain robust convergence guarantees under communication delays and link failures, making them well-suited for large-scale deployments where perfect synchronization is impossible  \cite{ghasemi2023robustness}. The scalability and robustness of modern \gls{so} techniques enable the decomposition of complex network-wide optimization problems into manageable subproblems. This decomposition ensures that computational complexity grows polynomially rather than exponentially with network size, while preserving performance guarantees \cite{zhou2024decomposition}. 

\textbf{Partial observability and learning under uncertainty} pose significant challenges in \gls{ng} network optimization, where complete system state information is often available due to measurement limitations, estimation errors, and communication delays \cite{bozkus2024leveraging}. \gls{so} provides inherent robustness to partial observability by explicitly incorporating uncertainty into the optimization framework, enabling effective decision making under incomplete information. These methods maintain robust performance by modeling uncertainty explicitly and providing probabilistic guarantees on optimization outcomes. This robustness ensures network performance degradation as observability decreases, rather than catastrophic failure. \gls{ml} techniques integrated with \gls{so} enable radio networks to learn effective resource allocation strategies through exploration and exploitation under unknown system dynamics \cite{ali2024optimization}. 

\textbf{Non-convexity and convergence guarantees}: Non-convex optimization problems are fundamental in wireless networks due to unpredictable interference patterns, discrete resource allocation constraints, and complex objective functions \cite{wang2022non, chen2025anns}. Traditional optimization methods often become trapped in local optima, limiting their effectiveness in complex and time-varying wireless environments. Robust \gls{so} algorithms provide theoretical convergence guarantees even for non-convex problems by leveraging advanced techniques such as sophisticated variance reduction and momentum methods \cite{perazzone2025communication}. These algorithms can escape local optima and converge to globally optimal or near-optimal solutions with high probability, ensuring consistent performance across diverse operating conditions. Stochastic gradient descent with variance reduction techniques demonstrates robust convergence properties in non-convex settings. Its convergence rate remains stable even under noisy gradient estimates \cite{lei2024energy}. This robustness is crucial for \gls{ng} wireless applications, where gradient computations are often corrupted by channel noise or interference.

\textbf{Real-time constraints and computational efficiency}: Real-time constraints require optimization algorithms that can deliver solutions within strict time limits, often on the order of milliseconds for critical applications \cite{zhou2024predictable}. The computational robustness of \gls{so} algorithms enables near-optimal solutions even under severe time constraints, making them well-suited for time-sensitive wireless applications. Online \gls{so} algorithms adapt to time-varying network conditions in real time while maintaining robust performance guarantees \cite{9795684}. This robustness ensures stable performance even when the wireless environment deviates from design assumptions.

\textbf{Multi-objective trade-offs and Pareto optimization}: Again, the multi-objective optimization of wireless networks requires balancing conflicting objectives, such as energy efficiency, latency, throughput, and reliability \cite{zhang2022k_massiveMIMO2022}. Robust \gls{so} provides mathematical frameworks for handling multi-objective trade-offs, while ensuring that performance guarantees are maintained across all objectives. Pareto-based optimization identifies non-dominated frontiers of competing objectives while remaining robust to uncertainty in objective weights and constraints \cite{zheng2024approximation}.

\textbf{Dynamic topologies and adaptive algorithms}: Dynamic topology changes in wireless networks, driven by user mobility, node failures, and time-varying connectivity, pose significant challenges for optimization algorithms \cite{yuan2023joint}. As a remedy, robust \gls{so} algorithms can adapt to topology changes in real-time while maintaining convergence guarantees and exhibiting stable performance. 

\subsubsection{Lessons learned} Unlike deterministic methods that rely on static or average-case assumptions and are prone to local optima, \gls{so} incorporates probabilistic modeling, randomness, and data-driven adaptation to maintain stable performance under fluctuating channels, unpredictable mobility, and dynamic interference. Distributed and decomposition-based \gls{so} frameworks address scalability challenges in ultra-dense \gls{ng} deployments with polynomial complexity growth and robust convergence under imperfect synchronization. \gls{so} further mitigates partial observability through explicit uncertainty modeling and learning‐based approaches, enabling informed decision making with incomplete or delayed system information. For non‐convex optimization problems common in wireless environments, advanced \gls{so} techniques such as variance reduction and momentum-based methods facilitate prompt escape from local optima and promote near-global solutions. Moreover, the computational efficiency of online \gls{so} algorithms supports stringent real-time constraints by delivering near-optimal solutions within millisecond-level latency budgets. Finally, \gls{so}’s capability to balance multi‐objective trade‐offs through Pareto optimization, alongside its adaptability to dynamic network topologies, underscores its role as an essential tool for ensuring stability, efficiency, and intelligent resource management in \gls{ng} systems.

\vspace{-0.25cm}

\subsection*{\textbf{Open question six: How is \gls{so} applied to address the resource allocation posed by the cutting-edge technologies of \gls{ng} networks?}}

\begin{table*}[t]
\caption{Applications of \gls{so} in \gls{ng} networks}
\label{tab:application_SO}
\centering
{\scriptsize \begin{tabular}{|p{0.1\textwidth}|p{0.07\textwidth}|p{0.05\textwidth}|p{0.07\textwidth}|p{0.08\textwidth}|p{0.53\textwidth}|}
\hline
\textbf{NG-Network} & \textbf{Ref.} & \textbf{Year} & \textbf{Obj.} & \textbf{Var.} & \textbf{Algorithm Contribution} \\
\hline

\multirow{4}{*}{RIS} & \cite{huang2022placement_DE2022} & 2022 & Single & Continuous & The roulette wheel method selects one of the three mutation operators based on a weight vector, which is updated according to the fitness of the best individual \\ \cline{2-6}
& \cite{van2023phase_DE2023} & 2023 & Single & Continuous & Use multiple mutation operators and the control parameters of the mutation and crossover operators are dynamically adapted to the search behavior \\ \cline{2-6}
& \cite{el2024multi_SA2024} & 2024 & Single & Mixed & Hybrid metaheuristics combining modified tabu search, simulated annealing, and cross-entropy–based neighborhood exploration, applied to decoupled \gls{ris} optimization \\ \cline{2-6}
& \cite{bandyopadhyay2024quantum_DE2024} & 2024 & Multiple & Discrete & Convert multi-objective optimization to single-objective using a weighted-sum vector; each individual is represented by a quantum vector, from which a binary vector is generated, and each binary string is then decoded to its corresponding decimal value \\ \hline

\multirow{5}{*}{Massive \gls{mimo}} & \cite{zhang2022k_massiveMIMO2022} & 2022 & Multiple & Continuous & Propose a fitness function that is based on objectives, use K-means clustering to divide differential evolution population into three groups with different mutation operators \\ \cline{2-6}
& \cite{9964313} & 2022 & Single & Mixed & Decompose the problem into three subproblem: employ K-means clustering for \gls{uav} deployment and user pairing, \gls{pso}  for hybrid beamforming to maximize hit probability, and \gls{ga} for optimal power allocation \\ \cline{2-6}
& \cite{10169085} & 2023 & Single & Continuous & Distributed cooperative multi-agent reinforcement learning framework where each user acts as a learning agent for joint resource allocation using only local information \\ \cline{2-6}
& \cite{10289120} & 2023 & Single & Continuous & The problem is modeled as a \gls{MDP} under stochastic and time-varying channels, and the deep deterministic policy gradient algorithm is employed to optimize both the joint covariance matrix design at the base station and the passive beamforming design \\ \cline{2-6}
& \cite{10922192} & 2025 & Single & Continuous & Propose an enhanced genetic algorithm that employs multi-point randomized crossover combined with a forced local mutation mechanism around the current population optimum \\ \hline

\multirow{6}{*}{Satellite-UAV} & \cite{nguyen2023real} & 2023 & Single & Discrete & Binary‐encoded \gls{ga} with single‐point crossover, bit‐flip mutation, and roulette‐wheel selection \\ \cline{2-6}
& \cite{pan2023joint_PSO2023} & 2023 & Single & Continuous & The initialization process adopts a normal distribution, incorporates \gls{ga}’s crossover operation, \gls{de}’s mutation and natural selection mechanisms into \gls{pso}  to escape local optima \\ \cline{2-6}
& \cite{qiao2024resource_PSO2024} & 2024 & Single & Mixed & Integrate \gls{ga} to refine select particles in each \gls{pso}  generation and use a parameter-free penalty function \\ \cline{2-6}
& \cite{wang2024bi_ACO2024} & 2024 & Multiple & Discrete & Introduce heterogeneous colonies with different objective preferences, each using five pheromone matrix pairs; design feasible solution generation method for solution construction, solution division method for quality improvement, and pheromone update method for adaptive pheromone updating \\ \cline{2-6}
& \cite{li2025optimizing_ACO2025} & 2025 & Multiple & Discrete & \gls{aco} guides routing using neural network heuristics; ants explore paths via state transition, update pheromones based on cost-quality, and iteratively improve routing for Transformer-based VNF embedding \\ \cline{2-6}
& \cite{jeong2025quantum_SA2025} & 2025 & Single & Continuous & Employ \gls{QA} on \gls{QUBO} models, enhanced by deriving penalty bounds and optimizing scaling parameters with Taylor approximation and Mixed-integer linear fractional programming, avoiding heuristic tuning \\ \hline

\multirow{5}{*}{Sensing \& Comm.} & \cite{nandan2021optimized_GA2021} & 2021 & Multiple & Discrete & Convert multi-objective optimization to single-objective using a weighted-sum vector; binary-encoded \gls{ga} with single-point crossover, bit-flip mutation, and invalid-individual removal \\ \cline{2-6}
& \cite{qiao2024resource_GA2024} & 2024 & Single & Mixed & Hybrid \gls{ga}–\gls{pso}, where in each \gls{pso}  generation, part of the population is refined by \gls{ga}. The fitness is evaluated via a parameter‐free penalty function to handle constraints \\ \cline{2-6}
& \cite{song2022joint_GA2022} & 2022 & Single & Discrete & Individual is represented by a two‐dimensional concatenated matrix, with initialization performed using conditioned random search instead of purely random generation \\ \cline{2-6}
& \cite{kaleibar2025customized_GA2025} & 2025 & Single & Discrete & Generates the initial population with a higher likelihood of assigning a provider to a requester within the same cluster, and applies a tournament selection strategy with a larger tournament size \\ \hline

\multirow{3}{*}{Underwater Comm.} & \cite{9845685} & 2022 & Single & Mixed & Integrate \gls{pso},  unequal clustering and adaptive updates of cluster sizes \\ \cline{2-6}
& \cite{li2024multiobjective}  & 2024 & Multiple & Discrete & Enhance  \gls{nsgaii} with a labeled tree-based encoding scheme, steering angle prior information, and a hash-based memory search mechanism \\ \cline{2-6}
& \cite{10947185} & 2025 & Single & Mixed & Combine model-based optimization (bisection and Lagrange dual) and \gls{DDPG} for spectrum assignment \\ \hline

\multirow{4}{*}{Others} & \cite{yuan2023joint} & 2024 & Multiple & Continuous & Temperature-based perturbation, domain-specific tuning \\ \cline{2-6}
& \cite{he2023hybrid_SA2023} & 2023 & Multiple & Continuous & Neural network is trained to model geometry–S-parameter relation; \gls{sa}  optimizes geometric parameters via the ANN surrogate to broaden bandwidth while keeping the required center frequency \\ \cline{2-6}
& \cite{lu2024deepeigen_TB2024} & 2024 & Single & Continuous & DeepEigNet initialization, probability-guided moves, and an early stopping mechanism \\ \cline{2-6}
& \cite{zhu2025group_PSO2025} & 2025 & Multiple & Continuous & Convert multi-objective optimization to single-objective using a weighted-sum vector; integrate a grouping and merging strategy that enhances information sharing and diversity by allowing particles to learn within groups and periodically merge in a controlled manner; precision‐matched perturbations are introduced to maintain search momentum \\ \hline
\end{tabular}}
\vspace{-0.5cm}
\end{table*}

\subsubsection{Background} 
Although the cutting edge technologies can significantly enhance network capacity, coverage, and flexibility, they also introduce highly dynamic and stochastic resource allocation challenges. For example, \gls{ris}-assisted systems require optimization of reflection coefficients under incomplete and time-varying channel information, resulting in highly non-convex problems. Massive and Cell-Free Massive \gls{mimo} demand large-scale stochastic optimization of power control and user association to adapt to dynamic channels and user mobility. Space–air networks with \gls{uav}s and satellites face rapidly changing topologies, necessitating real-time stochastic decisions for scheduling, coverage, and handovers. In \gls{MEC}, uncertainty in wireless links, server capacity, and user mobility complicates task offloading decisions for latency-sensitive services. Underwater acoustic communications further challenge resource allocation due to severe channel variability, long delays, and energy constraints. These characteristics make \gls{so} essential for efficient resource management in highly dynamic and uncertain NG networks. Applications of \gls{so} in \gls{ng} networks are summarized in Tab.~\ref{tab:application_SO}.

\subsubsection{Key features and considerations}
\hfill

\textbf{\gls{ris}}: 
\gls{so} is primarily applied to joint active-passive resource allocation problems characterized by non-convex objective functions, unit-modulus constraints, and imperfect channel state information. A representative class of studies formulates joint beamforming and phase-shift design problems under throughput- or fairness-oriented objectives. For example, Zhi \textit{et al.} \cite{9743440} considered sum-rate and max–min rate maximization in \gls{ris}-assisted massive \gls{mimo} uplink systems, where a novel \gls{ga} is employed to search over coupled active beamforming vectors and \gls{ris} phase configurations. Regarding high-frequency scenarios, \gls{si}-based methods have been adopted to cope with severe path loss and blockage effects. 
In \cite{yildirim2022ris}, an \gls{ris}-aided \gls{mmwave} Massive \gls{mimo} uplink is considered, where \gls{pso} is employed to jointly design beamformers and \gls{ris} phase shifts. The problem is formulated as achievable-rate maximization under practical constraints on transmit power, beamforming structure, and \gls{ris} phase range, with \gls{pso} searching over the coupled active and passive beamforming variables. Specifically, the \gls{ris} phase shifts are treated as particle positions that are iteratively updated based on both their individual best performance and on the global best performance of the swarm. The presence of an \gls{ris} optimized by \gls{pso}  alleviates the deleterious disruptive effects of the wireless propagation environment. 

When network dynamics and mobility are explicitly considered, learning-based \gls{so} techniques are introduced to handle complex decision spaces. Mei \textit{et al.} \cite{mei20223d} proposed a \gls{DRL} framework for jointly optimizing \gls{uav} trajectory and \gls{ris} phase-shift in a \gls{ris}-assisted \gls{uav} communication system, aiming for minimizing the \gls{uav}’s overall propulsion energy usage across all time slots. The system involves a \gls{uav} serving ground terminals with the aid of a \gls{ris} deployed on buildings to overcome blockages. The authors employ \gls{DDQN} for discrete \gls{uav} actions and \gls{DDPG} for continuous control, both integrating \gls{ris} phase adaptation.  Reliability-oriented designs further extend \gls{so} applications beyond rate-based objectives. 
Chien \textit{et al.} \cite{van2024active} explored the optimization of communication reliability in \gls{ris}-assisted \gls{mimo} systems as a multivariate function of both the phase shifts introduced by the \gls{ris} and the beamforming vectors of the BS. To tackle this, the authors proposed a \gls{de} algorithm and invoked \gls{LS} techniques at the end of each generation, allowing effective handling of the non-convex \gls{ser} optimization problem, helps to avoid getting stuck in local optima. 

Overall, existing \gls{so}-based studies on \gls{ris}-assisted networks can be broadly categorized into metaheuristic-driven approaches for joint active–passive beamforming under unit-modulus constraints, and learning-based frameworks for dynamic scenarios involving mobility, mixed discrete–continuous decisions, and long-term performance objectives.

\textbf{Massive \gls{mimo}/Cell-Free Massive \gls{mimo}}: \gls{so} techniques are widely adopted to address large-scale and time-varying resource allocation problems involving communication, computing, and energy management. Tilahun \textit{et al.} \cite{10169085} formulated a joint communication-and-computing optimization problem to minimize the total energy consumption of users in a Cell-Free Massive \gls{mimo}-assisted \gls{MEC} network under stringent latency constraints. Both task arrivals and wireless channel states are modeled as stochastic processes. To enable scalable real-time decision making, the authors proposed a \gls{MARL} framework with centralized training and decentralized execution, allowing distributed agents to coordinate resource allocation decisions without excessive signaling overhead.
Kurma \textit{et al.} \cite{10289120} investigated \gls{RE} optimization in \gls{ris}-aided full-duplex \gls{mimo} systems comprising a multi-antenna \gls{bs}, uplink and downlink users, and distributed \gls{ris}s. The optimization objective is to maximize \gls{RE}, defined as a weighted combination of \gls{SE} and \gls{EE}, subject to power constraints and unit-modulus \gls{ris} phase-shift constraints. The non-convex joint optimization problem is addressed by a policy-gradient-based \gls{DRL} framework with an actor–critic architecture and experience replay under dynamic channel conditions. As a further extension, Chen \textit{et al.} \cite{10922192} studied polarized hybrid beamforming in reconfigurable antenna sub-arrays for \gls{mmwave} multi-user Massive \gls{mimo} systems. The beamforming design is formulated as a masking matrix selection problem, where the polarization distribution, digital beamforming, and analog beamforming, implemented via phase shifters, are jointly optimized to maximize the sum rate. To solve this problem, the authors compared random mask sampling, a conventional \gls{ga}, and an enhanced \gls{ga} incorporating targeted perturbation strategies, highlighting the role of evolutionary search in exploring large combinatorial beamforming spaces. Finally, Chien \textit{et al.} \cite{11080325} employed augmented \gls{de} for \gls{ris} phase optimization in Cell-Free Massive \gls{mimo} channel estimation. By injecting targeted perturbations into elite candidates, the method minimizes normalized mean squared error while mitigating premature convergence.


\gls{so} in Massive and Cell-Free Massive \gls{mimo} systems is predominantly driven by scalability and coordination requirements. Learning-based approaches, particularly multi-agent and policy-gradient \gls{rl}, are employed to enable distributed decision making across antennas, users, and edge resources, while \gls{ea} and \gls{si}-based methods are leveraged to handle high-dimensional, mixed-integer, and gradient-intractable beamforming and configuration problems.

\textbf{Airborne communication}:  
Optimizing resource usage in space-air communication involves challenging path planning problems characterized by complex 3D environments, dynamic obstacles, and energy constraints. In \cite{9795684}, Yu \textit{et al.} formulated the \gls{uav} trajectory planning task as an NP-hard optimization problem to minimize a weighted path cost capturing the propagation distance, obstacle avoidance, yaw and pitch angles, and altitude constraints. To address the non-convex and multi-objective nature of this problem, the authors employed a hybrid \gls{pso} framework that integrates \gls{sa} for escaping local optima and a dimensional learning strategy to improve convergence behavior. This aims for a balance between exploration versus exploitation for high-quality \gls{uav} trajectory planning. Regarding a network-layer perspective, airborne communication  also face resource allocation challenges at higher protocol layers, particularly in content delivery and caching under dynamic demand and limited  opportunities. The research \cite{guo2024deep} investigated a three-layer airborne communication architecture integrating satellites, terrestrial base stations, and aircraft-assisted caching. The content placement problem is formulated to minimize end-to-end delivery latency while accounting for time-varying content popularity, heterogeneous cache capacities, and constrained communication durations. To address this problem, the authors modeled content placement and delivery as a Markov decision process and employed a \gls{DRL}–based approach to learn adaptive caching policies in dynamic airborne environments.

\gls{so} in airborne communication networks is mainly applied at two levels: trajectory-level optimization to handle mobility and environmental uncertainty, and network-layer resource management to adapt content delivery and caching decisions under dynamic demand and connectivity constraints.

\textbf{Satellite communication}:
\gls{so} methods are utilized in optimizing constellation layouts of \gls{ng} satellites, dynamically steering beams, and judiciously allocating both spectrum and power under uncertain link conditions to maximize coverage and throughput. Song \textit{et al.} \cite{10241983} investigated task scheduling for a group of electromagnetic exploration satellites by formulating the problem as a mixed-integer optimization task that maximizes detection profit. To solve this problem, a \gls{ga} is combined with a learning-based adaptation mechanism, where a recurrent neural network dynamically adjusts crossover and mutation parameters under the guidance of a policy-gradient \gls{rl} strategy. This hybrid design enables adaptive search behavior in evolving large-scale satellite scheduling scenarios. Beyond scheduling, learning-based \gls{so} frameworks have been adopted for joint resource allocation and caching in integrated terrestrial–satellite systems. In \cite{10003098}, Li \textit{et al.} proposed a \gls{MARL} framework based on \gls{MADDPG} to jointly optimize user association, power control, and cache placement in satellite-assisted \gls{NOMA} networks. The problem is formulated as a constrained \gls{EE} maximization task, where different agents coordinate to manage communication and caching decisions.  In addition to throughput-oriented objectives, information freshness has emerged as an important performance metric in satellite-enabled \gls{IoT} networks. 


\gls{so} in satellite communication primarily addresses mixed-integer scheduling and long-term resource coordination in dynamic, large-scale, and heterogeneous environments, enabling adaptive decision making under time-varying traffic demands, diverse link conditions, and complex system architectures.

\textbf{\gls{ISAC}}: 
\gls{so} is employed to manage tightly coupled sensing–communication resource allocation under mobility, uncertainty, and dynamic traffic conditions. Typical challenges include joint bandwidth,  power allocation, infrastructure activation, and latency–sensing trade-offs in highly dynamic environments. Zhang \textit{et al.} \cite{9920954} studied \gls{ISAC} optimization in vehicle-to-infrastructure systems by maximizing the aggregate radar sensing rate across multiple \gls{bs}s and time segments. The problem involves joint base station activation and bandwidth allocation under stochastic network conditions. The authors, then, adopted a \gls{rl} approach using a double \gls{DQN}, which simultaneously learns the optimal bandwidth distribution and \gls{bs} selection. The approach enables adaptive decision-making across sensing and communication dimensions. In Internet of Vehicles scenarios, Liu \textit{et al.} \cite{10716739} considered \gls{ISAC}-enabled multi-\gls{bs} systems under spectrum scarcity. The resource allocation problem is decomposed into two stages, where a \gls{DDPG} framework dynamically allocates transmission power and channels based on vehicle locations to maximize \gls{SE}, followed by a \gls{DQN} that assigns bandwidth to further coordinate communication performance while satisfying sensing constraints and adapting to varying traffic loads. \gls{ISAC} optimization has also been investigated in high-mobility scenarios such as \gls{uav}-assisted high-speed rail systems. Qiao \textit{et al.} \cite{10742627} formulated a mixed-integer nonlinear optimization problem to maximize the fair sum rate of passengers under radar sensing constraints. Then, a hybrid \gls{pso}–\gls{ga} approach is employed, combining the convergence capabilities of \gls{pso} with the strong global search capability of \gls{ga}. In the \gls{pso} phase, particles navigate the search space by adjusting the positions based on their own personal bests and the swarm’s global best. 

\textbf{Mobile edge computing}:  \gls{so} methods have demonstrated strong potential in efficiently managing task distribution and resource scheduling amidst erratically fluctuating unpredictable demands. Goudarzi \textit{et al.} \cite{goudarzi2023uav} investigated a \gls{uav}-assisted \gls{MEC} architecture consisting of a cloud center, mobile edge servers mounted on \gls{uav}s, and mobile users coordinated via an \gls{sdn} controller.  To overcome issues such as limited accessibility and high energy consumption, the authors proposed a cooperative \gls{eas} that jointly optimizes task offloading and bandwidth allocation under energy and communication constraints. Multi-objective task offloading has also been extensively studied in \gls{MEC} systems. Wang \textit{et al.} \cite{wang2021multiobjectives} proposed a comprehensive multiobjective optimization framework for task offloading, power allocation, and resource scheduling in multi-user, multi-server \gls{MEC} systems.  Each offloading decision must balance three conflicting objectives: minimizing response time, energy consumption, and usage cost, while operating under constraints such as limited server resources and device power budgets. To solve this, the authors developed an improved multiobjective \gls{ea} based on decomposition, which encodes decisions into chromosomes and employs potent constraint-handling strategies, crossover/mutation operations, and decision-making methods for solution selection, such as simple additive weighting or multiple criteria-based decision-making. Partial computation offloading has been considered to enhance \gls{EE}. Bi \textit{et al.} \cite{bi2020energy} designed an \gls{EE} partial computation offloading strategy for \gls{MEC} supporting smart mobile devices, aiming for minimizing the total energy consumption by jointly optimizing offloading ratios, central processing unit speeds, bandwidth allocations, and transmission power, subject to latency, resource, and energy constraints. Then, the authors proposed a novel hybrid metaheuristic algorithm, namely \gls{GSP}. Specifically, the \gls{ga}'s crossover and mutation operations are applied for producing superior particles and for enhancing the \gls{pso}'s global search capability. Additionally, the \gls{sa}'s Metropolis acceptance rule is used when updating particle positions, allowing the algorithm to escape local optima.

\begin{figure}[t]
    \centering
    \includegraphics[width=0.95\linewidth]{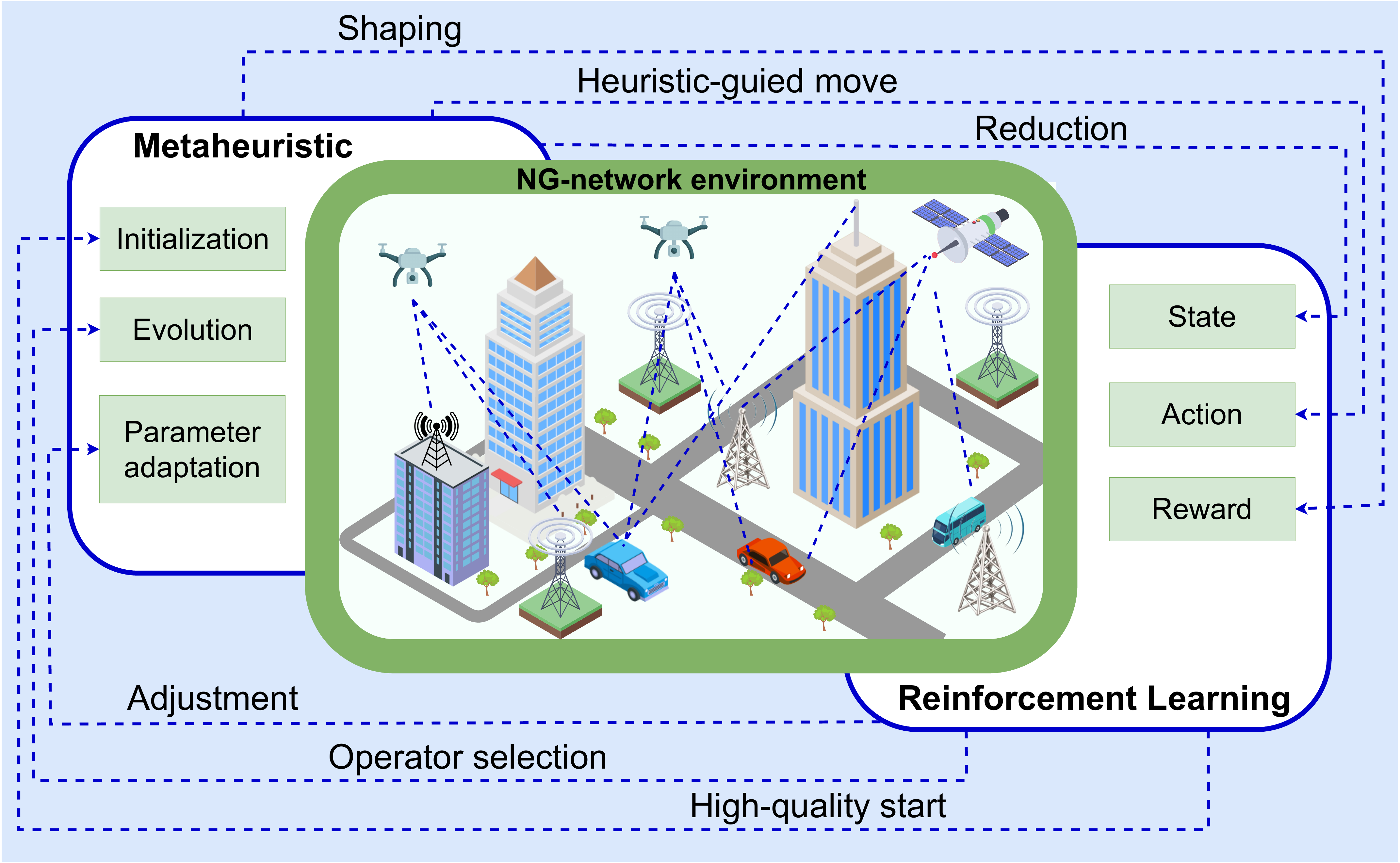}
    \caption{Hybridization of metaheuristics and \gls{rl}.}
    \label{fig:RL for MH}
    \vspace{-1.5em}
\end{figure}
\begin{figure*}[t] 
    \centering
    \includegraphics[width=\textwidth, trim=0cm 3.8cm 0cm 3.5cm, clip]{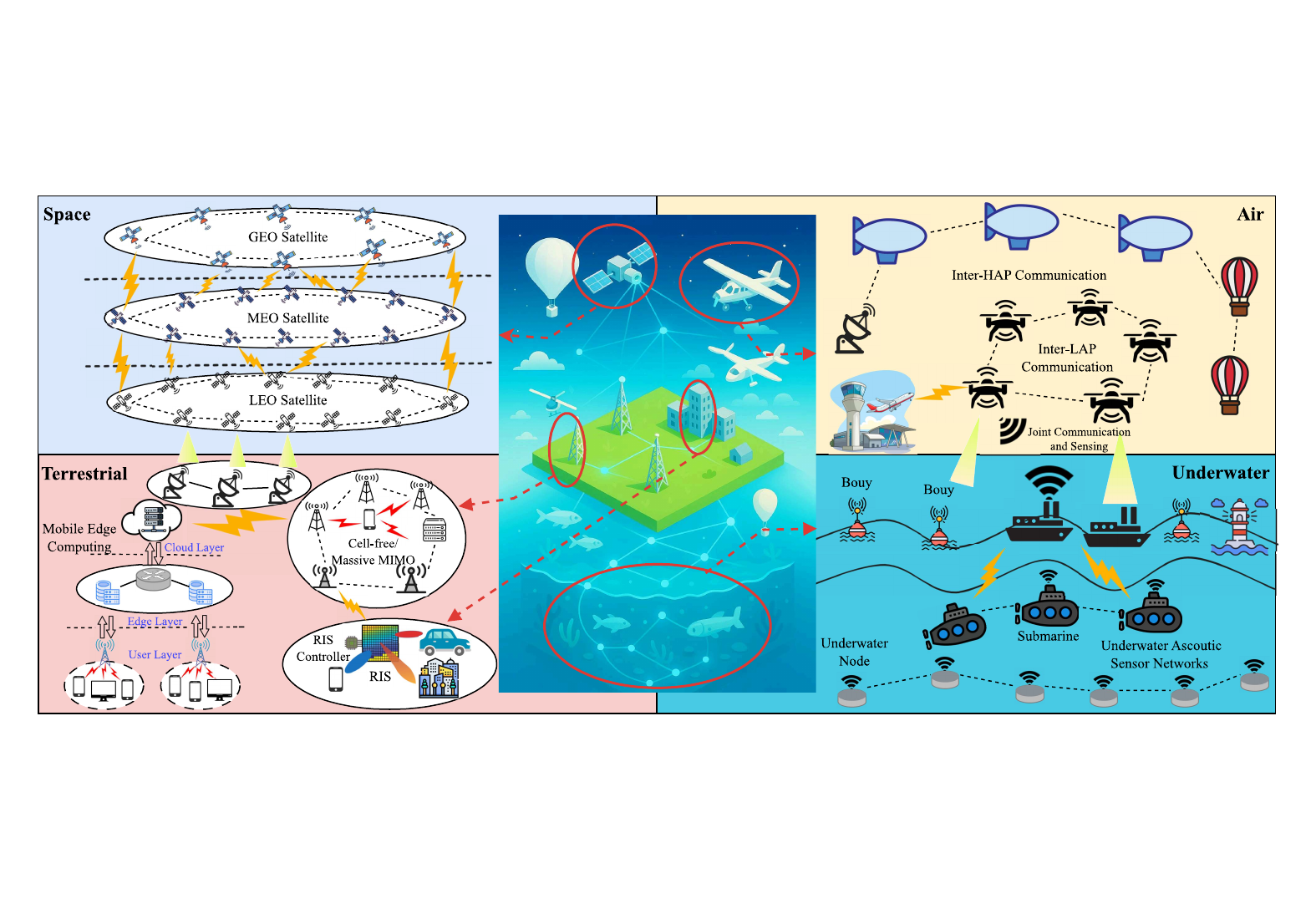} 
    \caption{Cutting-edge technologies integrated in \gls{ng} wireless networks classified into four layers consisting of space, air, terrestrial, and underwater communications.}
    \label{fig:ngnetwork}
    \vspace{-0.8cm}
\end{figure*}

\textbf{Underwater acoustic communications:} \gls{so} plays a critical role in overcoming severe energy constraints, hostile time-variant channel conditions, and multi‑hop routing challenges. 
Hou \textit{et al.} \cite{9845685} investigated energy-aware clustering and routing in three-dimensional \gls{UASNs} by formulating a network lifetime maximization problem that balances energy consumption among sensor nodes. The optimization focuses on joint cluster-head selection, cluster size configuration, and routing path planning, with objective functions incorporating residual energy, communication distance, and load distribution. A \gls{pso}-based approach is adopted to iteratively explore feasible clustering and routing configurations under dynamic topology conditions. 

Joint spectrum and power allocation problems have been addressed in underwater environments. Tang \textit{et al.} \cite{10947185} presented a comprehensive solution for resource allocation in \gls{UASNs}, where sensor nodes transmit data to a central sink over harsh underwater acoustic channels. The goal is to maximize the minimum rate across all users, subject to constraints on power allocation and partial spectrum sharing. As the absence of closed-form spectrum assignment models and the presence of non-convex interference coupling, the authors combined a \gls{ga} for coarse spectrum allocation with a \gls{DRL} approach based on \gls{DDPG} to refine power and spectrum decisions in a data-driven manner. The \gls{ga} is initially applied for approximating the global optimum, while the \gls{DRL}-based approach is employed to efficiently explore the entire solution space and achieve the optimal resource allocation. Energy harvesting and mobility further increase the complexity of underwater resource management. Han \textit{et al.} \cite{9877737} studied a cooperative uplink \gls{UASNs} with mobile relays powered by energy harvesting. The problem of jointly optimizing relay selection and transmission power is formulated to maximize long-term uplink capacity under battery and power constraints. The main challenge lies in the coupled discrete-continuous nature of the problem, aggravated by the stochastic nature of energy harvesting, node mobility, and limited channel information. To overcome these challenges, the authors leveraged a divide-and-conquer \gls{DRL}-based approach. \gls{DQN} for relay selection and \gls{DDPG} for power allocation, along with a reconstructed state space and an adaptive reward mechanism that balances short-term throughput with long-term energy sustainability.

\begin{figure*}[t]
    \centering
    \begin{minipage}[t]{0.48\textwidth}
        \centering
        \begin{subfigure}[b]{0.48\linewidth}
            \centering
            \includegraphics[width=\linewidth, trim = 3cm 8cm 4cm 7cm, clip = true]{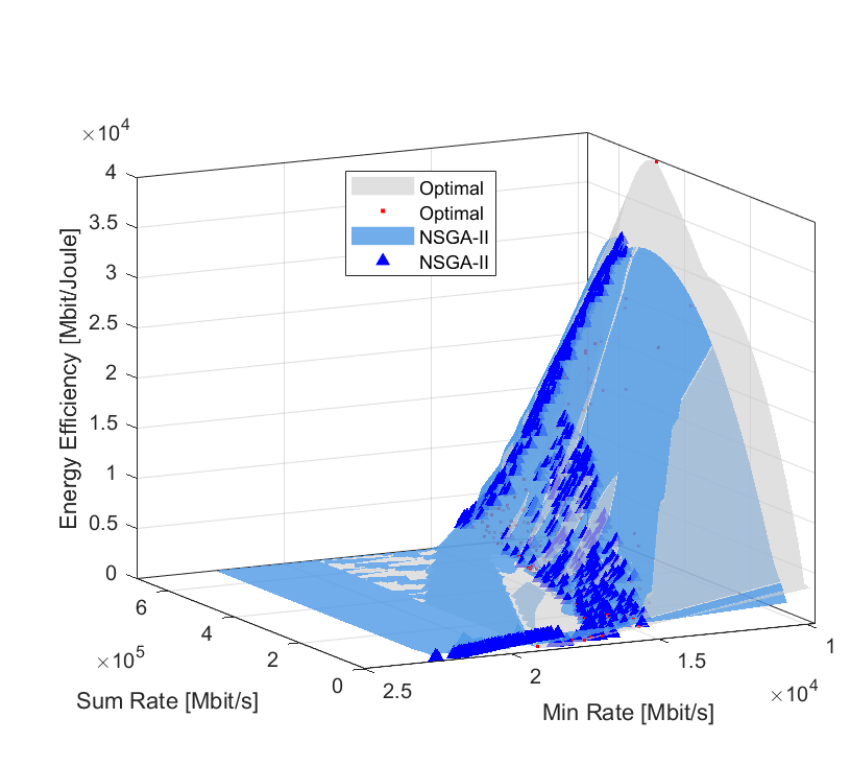}
            \caption{3D Pareto front}
            \label{fig:surface}
        \end{subfigure}
        \hfill
        \begin{subfigure}[b]{0.48\linewidth}
            \centering
            \includegraphics[width=\linewidth, trim = 3cm 8cm 4cm 7cm, clip = true]{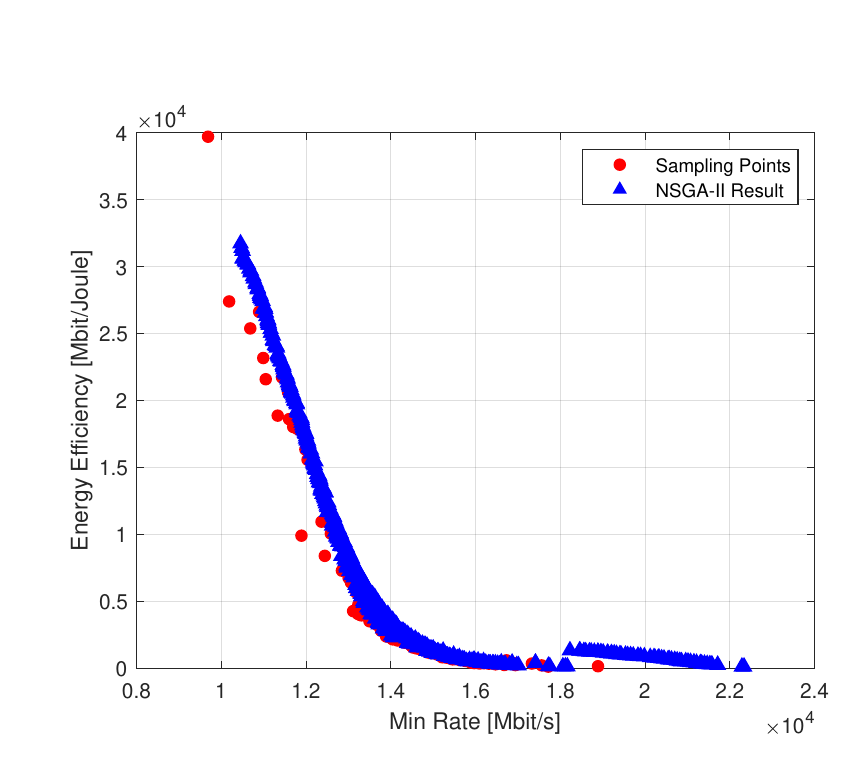}
            \caption{2D Pareto front}
            \label{fig:Pareto}
        \end{subfigure}
        
        \caption{Pareto front obtained by \gls{nsgaii} for the joint design of \gls{ris} elements and the number of antennas at the \gls{bs}, considering the three different objectives.}
        \label{fig:Pareto2hinh}
    \end{minipage}%
    \hfill 
    \begin{minipage}[t]{0.48\textwidth}
        \centering
        \begin{subfigure}[b]{0.48\linewidth}
            \centering
            \includegraphics[width=\linewidth, trim = 3cm 8cm 4cm 7cm, clip = true]{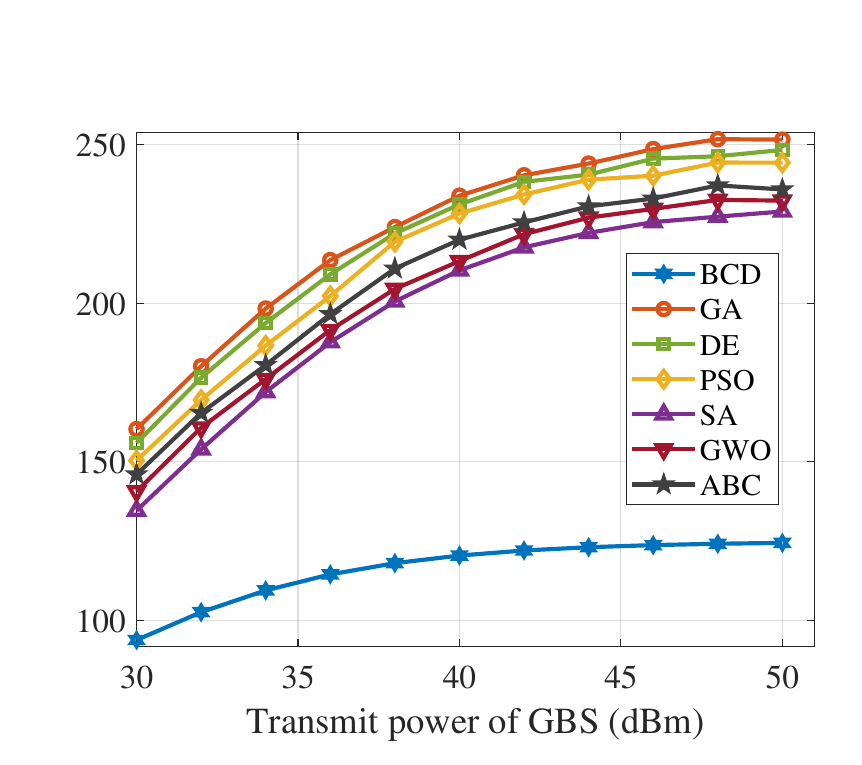}
            \caption{}
            \label{fig:achievableRate}
        \end{subfigure}
        \hfill
        \begin{subfigure}[b]{0.48\linewidth}
            \centering
            \includegraphics[width=\linewidth, trim = 3cm 8cm 4cm 7cm, clip = true]{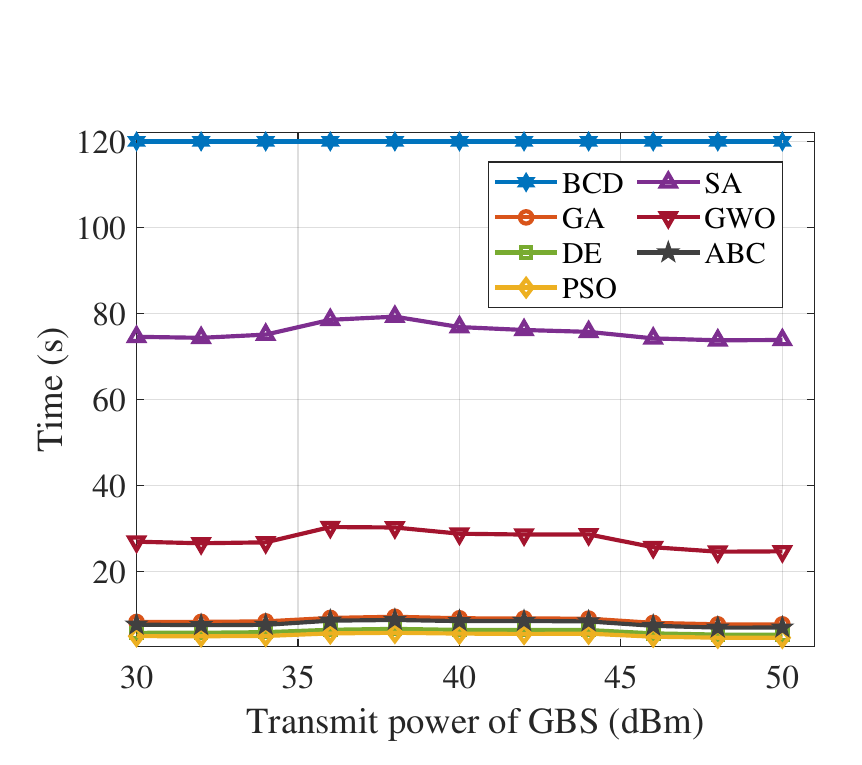}
            \caption{}
            \label{fig:time}
        \end{subfigure}
        \caption{Stochastic versus deterministic optimization for joint trajectory and dynamic time-splitting ratio in a two-\gls{uav}-assisted system.}
        \label{fig:uavrouting}
    \end{minipage}
    \vspace{-1.5em}
\end{figure*}

\textit{Performance results:}
In Fig.~\ref{fig:Pareto2hinh}, we consider a Massive \gls{mimo} system in which a \gls{bs} having up to 100 antennas serves 50 users uniformly distributed within an area of $0.5 \times 0.5~\text{km}^2$. To enhance communication performance, a \gls{ris} having 256 elements is installed in the coverage area and the channel information is modeled according to~\cite{van2024active}. A multi-objective optimization problem  jointly designs the number of active \gls{bs} antennas and the configuration of \gls{ris} elements for simultaneously maximizing three objectives: sum rate, minimum rate, and \gls{EE}.  Fig.~\ref{fig:surface} demonstrates the convergence performance of the  \gls{nsgaii} algorithm, configured with a population size of 100 and 100 generations. The  Pareto front obtained closely approximates the true front after approximately 10,000 fitness evaluations.  The 2D Pareto front between \gls{EE} and minimum rate illustrates the robustness and diversity preservation capability of  \gls{nsgaii} in capturing trade-offs among the objectives.

In Fig.~\ref{fig:uavrouting}, we consider a network where a ground \gls{bs} communicates with \gls{uav}s equipped with backscatter devices that relay information to end users. We formulate a novel optimization problem for jointly determining the dynamic time-splitting ratio and flight trajectory of the \gls{uav}s under imperfect channel state information, while considering key factors such as \gls{uav} flight duration, altitude, and transmission power. The performance comparison between a  deterministic \gls{BCD}-based algorithm and various \gls{so} algorithms, including \gls{ga}, \gls{de}, \gls{pso},  \gls{sa}, \gls{gwo}, and \gls{abc}, under different transmit power levels of the GBS. Quantitatively, the achievable rate consistently increases with the GBS transmit power for all methods, where the \gls{so} algorithms substantially outperform the \gls{BCD} benchmark. Specifically, the \gls{so} techniques achieve rate improvements ranging from 73.79\% to 95.89\%, while the computational time is reduced by as much as  15.78\% to 95.47\%.  Additionally, \gls{ga} exhibits the highest achievable rate. \gls{sa} and \gls{gwo} require longer execution times compared to other \gls{so} methods. In contrast, \gls{de} and \gls{pso} present a favorable trade-off between solution quality and computation cost.
\vspace{-0.1cm}
\subsubsection{Lessons learned} Since cutting-edge technologies like Massive/Cell-Free \gls{mimo},  joint communication and sensing, and \gls{ris} might introduce complex SE expressions under dynamic and uncertain environments, \gls{so} proves valuable by offering flexibility in adapting to real-time changes. 
It is capable of outperforming static or deterministic methods in handling unpredictable channel conditions, mobility, and interference. In a nutshell, while \gls{so} is attractive for complex and uncertain environments introduced by \gls{ng} technologies, its continued development must focus on scalability, adaptability, and real-world applicability. These insights are crucial for designing high-performance \gls{ng} wireless systems that are resilient and efficient.

\vspace{-0.5em}

\vspace{-0.15cm}
\section{Integration of \gls{so} with Future Technologies} \label{sec:Integrated}

\subsection*{\textbf{Open question seven: How can \gls{so} collaborate with emerging technologies, e.g. generative \gls{ai}, in resource allocation for \gls{ng} networks?}}

\subsubsection{Background}

The transition to \gls{6g} is underpinned by the convergence of Distributed Systems \cite{10922192}, Generative \gls{ai} \cite{Khoramnejad10855897, Vu10737408}, and Automation frameworks \cite{hamidi20215g}.
\begin{figure}[t]
    \centering
\includegraphics[width=0.7\linewidth, clip=true]{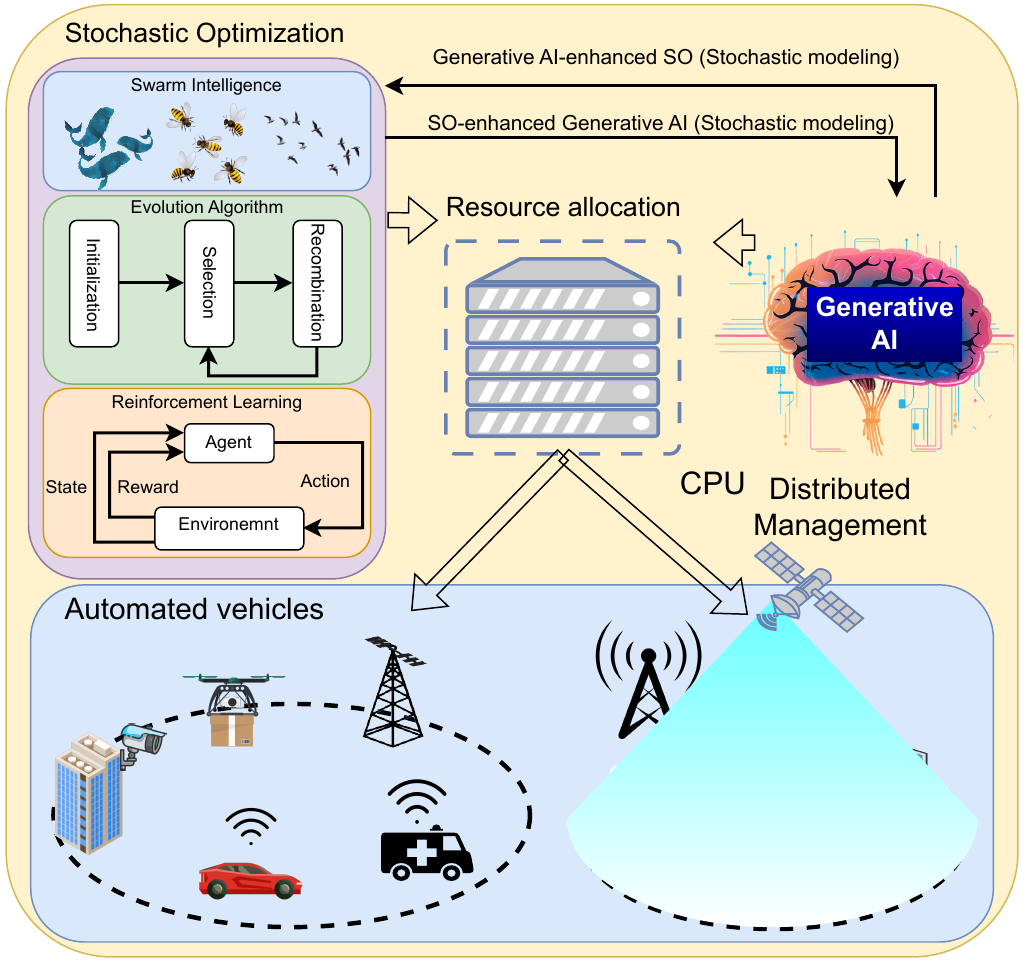}
    \caption{SO combining with the emerging technologies in \gls{ng} wireless networks.}
    \label{fig:SO_future}
    \vspace{-1.5em}
\end{figure}
While distributed architectures enhance scalability \cite{duan2022distributed}, their efficacy is compromised by inherent network stochasticity, rendering centralized deterministic controls inadequate. Concurrently, Generative \gls{ai} models, encompassing \gls{llm}s \cite{Liu11071329}, diffusion models \cite{Wen11205961}, and \gls{gan}s \cite{Khoramnejad10855897}, drive intelligent operations \cite{Vu10737408, zhou2024large} but face severe integration challenges at the edge, including energy constraints and non-stationary data distributions. Meanwhile, network automation is critical for managing complexity via frameworks like intent-based networking and zero-touch management \cite{Coronado9913206}. By translating high-level user intents (e.g., maximizing \gls{QoE}) into autonomous decisions, these systems enhance scalability across multi-service environments. However, their efficacy is challenged by the stochastic, partially observable nature of \gls{ng} networks, where ambiguous intents and delayed feedback complicate real-time control. Across these technological domains, the dominance of stochasticity renders deterministic optimization fundamentally inadequate. Conversely, \gls{so} provides the essential mathematical scaffolding to navigate uncertainty (see Fig.~\ref{fig:SO_future}). By enabling robust decision-making under noisy, partial information and facilitating model-free learning in non-stationary environments, \gls{so} acts as a foundational enabler for distributed systems, \gls{llm}-driven intelligence, and autonomous orchestration in \gls{ng} networks.

\subsubsection{Key features and considerations}
\hfill

\textbf{Distributed implementation:} Distributed systems are exemplified by technologies such as \gls{MEC}, \gls{FL}, cell-free massive \gls{mimo},  and \gls{ris} clusters, just to name a few. Operating in inherently decentralized architectures, these systems must handle dynamic constraints, which often break convexity assumptions and limit the feasibility of centralized optimization approaches. Therefore, decision-making is typically based on asynchronous, partial, and non-stationary feedback, which is reliably managed by population-based \gls{so} methods.
\begin{itemize}[leftmargin=*]
        \item \textit{\gls{si}}  represented by, for example, \gls{pso} and  \gls{aco} is inherently decentralized, making them well-suited for distributed optimization. These algorithms rely on local interactions and limited information exchange, allowing agents (e.g., edge devices, \gls{uav}s, \gls{ris} elements) to independently adapt based on their own local observations and shared best-known solutions. This results in low communication overhead and high scalability, critical for large-scale dynamic networks.  

        \item \textit{\gls{ea}s} as gradient-free and population-based methods inherently support parallel computation by evolving candidate solutions locally on distributed nodes (e.g., island models). This architecture accommodates heterogeneous environments with varying node capabilities and link qualities. For instance, Li \textit{et al.} \cite{Li9733794} introduced a fully distributed \gls{de} framework utilizing a three-layer architecture for parallel fitness evaluation and adaptive migration, significantly accelerating convergence in resource allocation tasks.

        \item \textit{\gls{rl}} facilitates autonomous, asynchronous decision-making in distributed systems by optimizing long-term objectives via localized environmental interactions. This capability is critical for tasks constrained by partial observability, such as distributed caching \cite{Zhou10123387} or federated learning \cite{seid2023multiagent}. Furthermore, \gls{rl} enables model-free adaptation to stochastic dynamics \cite{Urmonov9869639}, while advanced techniques like mean-field approximation and parameter sharing support scalable cooperation in continuous action spaces \cite{xu2024joint}.

        \item \textit{Other \gls{so} Frameworks:} Stochastic frameworks, such as gradient descent and dual decomposition, align with distributed architectures by partitioning global objectives into decoupled local subproblems governed by message passing or shared dual variables. Their robust convergence properties under noisy gradients render them indispensable for managing computation-latency trade-offs in \gls{MEC} and federated learning, thereby facilitating real-time adaptive control.
                
    \end{itemize}

\textbf{\gls{so} with generative \gls{ai}:} Integrating Generative \gls{ai} models, encompassing \gls{llm}s, diffusion models, and \gls{gan}s, into network optimization pipelines requires scalable, uncertainty-resilient decision-making mechanisms. Therein, \gls{so} provides a natural framework for enabling sample-efficient search, robust adaptation, and multi-objective trade-off handling. The interplay between \gls{so} and Generative \gls{ai}-based intelligence unfolds along the following key dimensions:

\begin{itemize}[leftmargin=*]
    \item \textit{\gls{si}} in the convergence with generative \gls{ai} establishes a dual-functional framework. Classical algorithms, such as \gls{pso}, function as external optimizers that refine generative outputs to satisfy stringent \gls{qos} constraints like latency and energy \cite{zhu2025swarmintelligenceenhancedreasoning}. Conversely, Generative \gls{ai} agents (most notably \gls{llm}s) can themselves be employed within a swarm-like architecture, forming distributed decision-making entities capable of interpreting local context and coordinating via prompt-driven protocols. In this paradigm, \gls{si} principles provide a coordination scaffold for distributed \gls{llm} agents, enabling decentralized negotiation and emergent policy synthesis within complex multi-agent environments \cite{jimenezromero2025multiagentsystemspoweredlarge}. 
    
    \item \textit{\gls{ea}s} combining with generative \gls{ai} results in a powerful synergy between generative capability (e.g., distribution modeling in \gls{gan}s \cite{Vu10737408}) and global search \cite{Wu10767756}. Meanwhile, \gls{eas}-enhanced Generative \gls{ai} utilizes \gls{eas} for optimizing the inputs (prompts/noise vectors), parameters, or inference trajectories of models, particularly when model internals are inaccessible (black-box) or when optimizing over multiple objectives. For example, Liu \textit{et al.} \cite{Liu10611913} introduced an \gls{LMEA}, utilizing \gls{llm}s to execute crossover and mutation operations via prompt engineering. Reciprocally, Generative \gls{ai}-enhanced \gls{eas} refers to using models to assist the evolutionary search process. While \gls{gan}s could generate high-quality initial populations or surrogate models to accelerate convergence \cite{Vu10737408, Khoramnejad10855897}, \gls{llm}s offer a semantic advantage by interpreting challenging problem contexts \cite{Tian_Han_Wu_Guo_Zhou_Zhang_Li_Wei_Zhang_2025}. These bidirectional synergies facilitate a more adaptive approach to \gls{so}, leveraging both structured search and generative reasoning.
    
    \item \textit{\gls{rl}} integrating with generative \gls{ai} constitutes a robust closed-loop framework. Generative models augment \gls{rl} agents by synthesizing training trajectories or distilling unstructured states, exemplified by \gls{llm}-guided policy updates in O-\gls{ran} slicing \cite{lotfi2025prompttunedllmaugmenteddrldynamic}. Conversely, \gls{rl} optimizes the deployment of generative models. For instance, \gls{rl} is used for determining split points in edge hierarchies to minimize latency and energy consumption \cite{chen2024adaptivelayersplittingwireless}. Overall, this combination unifies structured policy learning with flexible generative capabilities, paving the way for autonomous \gls{ng} network control.

    \item \textit{Other \gls{so} frameworks:} Beyond \gls{llm}s, diffusion models and \gls{vae}s \cite{Wen11205961, Huynh10490142} synergize with zero-order optimization and \gls{bo} to address black-box uncertainty. Specifically, deep generative simulators provide robust modeling of temporal-spatial channel randomness \cite{Liu10515203}. For instance, robust optimization frameworks utilizing deep generative simulators (e.g., diffusion models) provide a principled way of modeling and handling temporal-spatial randomness in wireless channels \cite{Zhang10937314}. Meanwhile, \gls{bo} ensures sample-efficient hyperparameter tuning, exemplified by prompt-guided \gls{llm} placement in edge environments \cite{liu2024largelanguagemodelsenhance}.
    
\end{itemize}

\textbf{SO in automation:} As \gls{ng} networks evolve toward full autonomy, \gls{so} provides the essential algorithmic scaffolding for self-configuration, self-optimization, and self-healing amidst high-dimensional uncertainty. By facilitating both reactive adaptation and proactive intent management under stochastic traffic and topology changes, \gls{so} empowers the multi-layer automation stack.

\begin{itemize}[leftmargin=*]
    \item \textit{\gls{si}} algorithms, such as \gls{pso} and \gls{aco}, enable adaptive behavior through local interactions, facilitating scalable decision-making under dynamically fluctuating and uncertain conditions. Conforth and Meng \cite{Conforth4668289} proposed a \gls{si}-based \gls{rl} framework for training \gls{ann}, combining  \gls{aco} and \gls{pso} for determining an \gls{ann} topology and the associated connection weights. The method is validated on a robot localization task, demonstrating the effectiveness of the \gls{si}-\gls{rl} synergy for distributed control applications.
    
    \item \textit{\gls{ea}s} support automation in \gls{ng} networks by evolving adaptive policies and configurations without gradient dependence through population-based search and adaptive operators. Through mechanisms like island-model evolution, they facilitate automated intent-to-policy translation and distributed decision-making. These capabilities make \gls{eas} natural fit for zero-touch management, allowing self-optimize and adapt to dynamic, uncertain environments. For example, Li \textit{et al.} \cite{Li9733794} present an automated distributed differential evolution framework. Therein, evolutionary search is performed by multiple \gls{de} populations, while automation is introduced through performance-driven fitness evaluation allocation. This allows the system to autonomously manage computational resources and accelerate convergence.

    \item \textit{\gls{rl}} has emerged as a key pillar for closed-loop automation, enabling systems to learn optimal policies via interaction with the network environment. In \gls{ng} networks, \gls{rl} supports autonomous decision-making for routing, spectrum allocation, energy management, and more. \gls{MARL} further enables coordination across heterogeneous nodes, while online \gls{rl} techniques allow continuous adaptation in real deployments where conditions shift frequently. Farzanullah  \textit{et al.} in \cite{Farzanullah9771770}   employ \gls{MARL} for resource allocation in factory automation networks to meet \gls{urllc} requirements, for enabling decentralized decision-making, enhancing communication reliability, and supporting real-time industrial automation.
    \item \textit{Other \gls{so} frameworks:} Complementary techniques such as Bayesian optimization, zero-order optimization, and generative models are vital for black-box automation where gradients are unavailable. \gls{bo} excels in sample-efficient hyperparameter tuning, exemplified by its role in guiding multi-agent deep reinforcement learning for automated, energy-efficient \gls{uav} coordination \cite{gong2022bayesianoptimizationenhanceddeep}. Similarly, zero-order optimization facilitates online configuration in noisy, model-free environments, proving particularly effective for gradient-independent tasks as adaptive load balancing and real-time sensor management \cite{liu2020primerzerothorderoptimizationsignal}.
\end{itemize}
\begin{table*}
\centering
\scriptsize
\caption{Integration of \gls{so} with distributed systems, generative \gls{ai}, and automation tasks: Summary of key approaches, their features, and applications.}
\label{Table: SO with future techs}
\begin{tabular}{| p{0.07\textwidth}| p{0.06\textwidth}| p{0.42\textwidth} | 
    p{0.38\textwidth} |}
\hline
\textbf{Technology} & \textbf{Approach} & \textbf{Key Features} & \textbf{Synergies \& Applications} \\
\hline

Distributed systems and edge intelligence & \gls{si} & Decentralized coordination; local feedback adaptation; scalable and robust to noise. & Distributed resource allocation \cite{Mukherjee8834771}. \\
\cline{2-4}
& \gls{eas} & Parallelizable; robust to mobility and fading; mixed-variable optimization. & Distributed \gls{de} resource allocation \cite{Li9733794}. \\
\cline{2-4}
& \gls{rl} & Asynchronous local decision-making; handles non-stationary dynamics. & Cooperative edge caching \cite{Zhou10123387}, \gls{FL} \cite{seid2023multiagent}, \gls{uav} pathing \cite{Nie9563249}. \\
\cline{2-4}
& Other \gls{so} & \gls{SGD}, dual decomposition, stochastic approximation. & \gls{MEC}/\gls{FL} adaptation under energy and latency. \\
\hline
Generative \gls{ai} & \gls{si} & Optimizing generative outputs (e.g., \gls{llm}/diffusion) or coordinating agents. & Reasoning enhancement \cite{zhu2025swarmintelligenceenhancedreasoning}, swarm \gls{llm}s \cite{jimenezromero2025multiagentsystemspoweredlarge}. \\
\cline{2-4}
& \gls{eas} & Generative population initialization (\gls{gan}s); \gls{llm}-enhanced operators. & \gls{gan}-assisted optimization \cite{Vu10737408, Khoramnejad10855897}, \gls{LMEA} \cite{Wu10767756}, \gls{eas}-guided slicing \cite{Tian_Han_Wu_Guo_Zhou_Zhang_Li_Wei_Zhang_2025}. \\
\cline{2-4}
& \gls{rl} & GenAI-based state abstraction/augmentation; \gls{rl} for model deployment. & Prompt-augmented \gls{rl} policy \cite{lotfi2025prompttunedllmaugmenteddrldynamic}, \gls{llm} split computing \cite{chen2024adaptivelayersplittingwireless}. \\
\cline{2-4}
& Other \gls{so} & Generative simulators (Diffusion/VAE); robust zero-order solvers. & Diffusion-based channel modeling \cite{Wen11205961, Liu10515203}, VAE-based design \cite{Huynh10490142}, \gls{llm}-enhanced \gls{bo} \cite{liu2024largelanguagemodelsenhance}. \\
\hline
Automation & \gls{si} & Agent coordination; dynamic adaptive behavior. & \gls{si}-enhanced model training \cite{Conforth4668289}. \\
\cline{2-4}
& \gls{eas} & Long-term planning and policy evolution. & Adaptive resource allocation \cite{Li9733794}. \\
\cline{2-4}
& \gls{rl} & Closed-loop control, \gls{MARL}, online learning. & \gls{MARL} in \gls{urllc} factory settings \cite{Farzanullah9771770}. \\
\cline{2-4}
& Other \gls{so} & \gls{bo}, zero-order tuning, generative models. & \gls{uav} offloading w/ \gls{bo} \cite{gong2022bayesianoptimizationenhanceddeep}, sensor management \cite{liu2020primerzerothorderoptimizationsignal}. \\
\hline
\end{tabular}
\vspace{-1.5em}
\end{table*}

Table~\ref{Table: SO with future techs} provides an overview of how various \gls{so} approaches are applied across distributed systems, Generative \gls{ai} models, and automation in \gls{ng} networks, highlighting their key features, tasks, and real-world applications.

\subsubsection{Lessons learned}
The integration of \gls{so} with distributed architecture, Generative \gls{ai}, and automation offers critical insights into the design of self-evolving \gls{ng} networks. \gls{so} enhances distributed systems by enabling decentralized decision-making, robust adaptation to dynamic conditions via \gls{si} and \gls{ea}s. Furthermore, a bidirectional synergy characterizes the interplay between \gls{so} and Generative \gls{ai}. While models encompassing \gls{llm}s, diffusion models, and \gls{gan}s enhance \gls{so} through semantic reasoning, latent state abstraction, and synthetic data augmentation. Reciprocally, \gls{so} grounds these generative outputs, acting as a rigorous feedback mechanism to optimize prompts, refine black-box inference trajectories, and enforce physical constraints. For network automation, \gls{so} and \gls{rl} enable self-optimization and self-healing in response to changing conditions, enhancing efficiency and service quality. However, challenges remain, such as computational and energy constraints at the edge and partial observability in dynamic environments. Real-time adaptation and scalability are difficult with the limited resources of distributed systems. Addressing these necessitates a pivot toward lightweight, model-free optimization and federated learning paradigms to reconcile algorithmic complexity with distributed resource limitations.

\vspace{-0.5em}
\subsection*{\textbf{Open question eight. How can \gls{QC} boost the capabilities of \gls{so} in \gls{ng} networks?}}
\setcounter{subsubsection}{0}
\subsubsection{Background}
As \gls{ng} network problems grow in complexity, driven by multi-agent coordination, large-scale device populations, and multi-objective trade-offs, classical \gls{so} algorithms increasingly suffer from four major limitations including $i)$ \textit{exponential computational complexity}: The combinatorial nature of problems like \gls{ris} configuration, task off-loading, or trajectory control leads to search spaces that grow exponentially with the number of variables, making global optimization intractable for classical \gls{cpu}/\gls{gpu}; $ii)$ \textit{convergence inefficiency}: Population-based algorithms and deep reinforcement learning require large iteration counts and extensive sampling to reach acceptable performance, often violating latency constraints in practice; $iii)$ \textit{sample inefficiency under uncertainty}: In partially observed, non-stationary environments, classical \gls{so} suffers from limited exploration capability and difficulty in adapting to unseen conditions or environmental shifts; and $iv)$ \textit{energy and hardware inefficiency}: Classical processors consume significant energy when scaling \gls{so} across large network scenarios, raising concerns about sustainability, especially at the edge.

These limitations necessitate a paradigm shift toward \gls{QC} \cite{Coccia9800933, Duong9849061}. By leveraging superposition, entanglement, and tunneling, \gls{QC} facilitates intrinsic parallel exploration, thereby reducing the computational complexity of \gls{so} beyond the reach of classical hardware acceleration.


\subsubsection{Key features and considerations}
We now examine how \gls{QC} augments each class of \gls{so} techniques, from population-based metaheuristics and \gls{rl} to black-box and sampling-based frameworks, and discuss implementation trade-offs at the hardware and algorithmic level.

\begin{figure}
    \centering
    \includegraphics[trim=0cm 0cm 0cm 0cm, clip=true, width=0.8\linewidth]{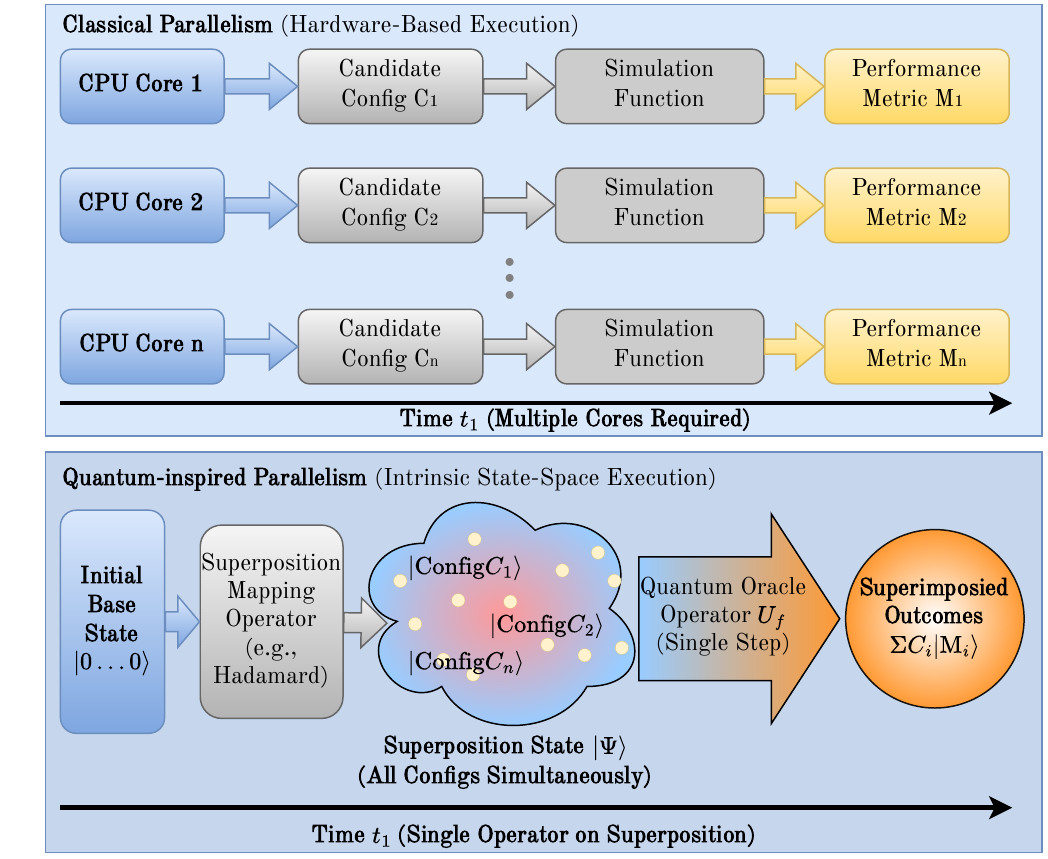}
    \caption{Conceptual comparison between classical hardware-based parallelism (top) and quantum-inspired intrinsic parallelism via superposition mapping (bottom).}
\label{fig:classical_vs_quantum_parallel}
\vspace{-1.5em}
\end{figure}
\begin{figure}
    \centering
    \includegraphics[width=0.7\linewidth]{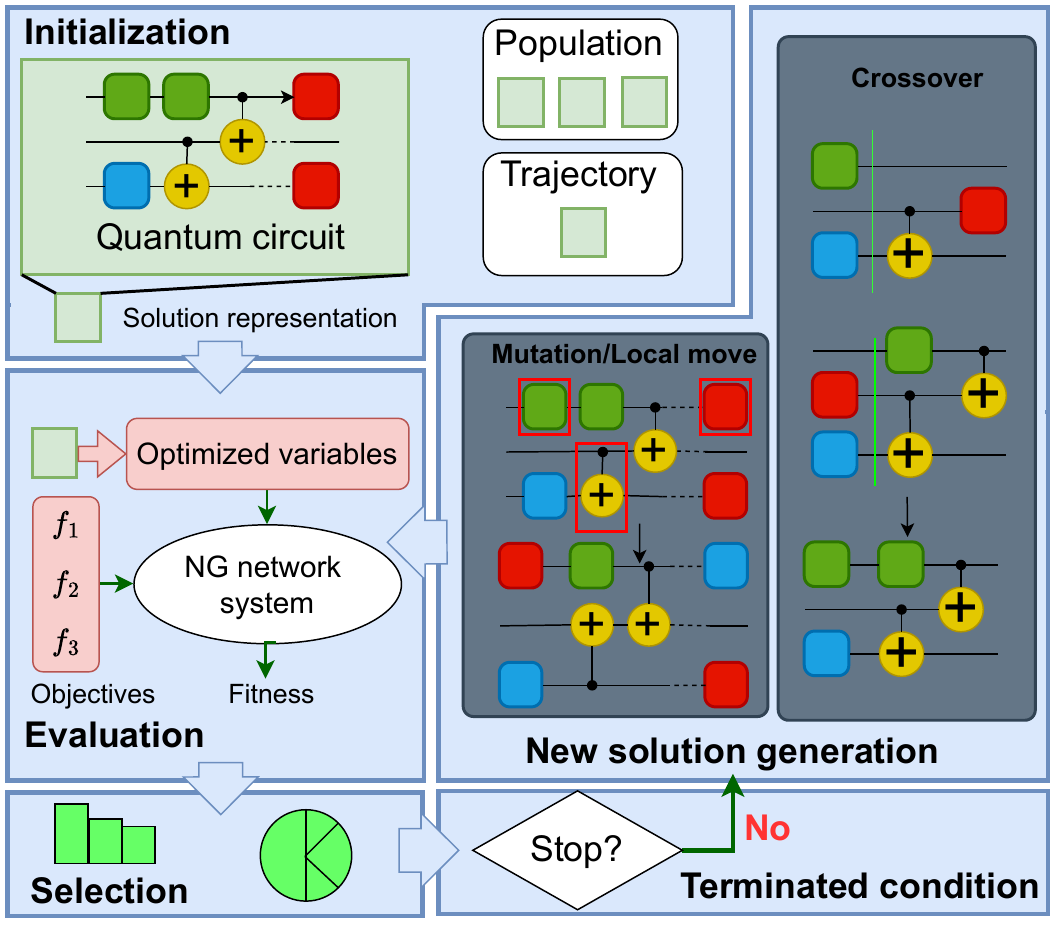}
    \caption{\gls{so} based on quantum computation.}
    \label{fig:Quantum_meta}
    \vspace{-1.5em}
\end{figure}

\textbf{\gls{QIM}:} Quantum-inspired metaheuristics integrate quantum principles into classical algorithms such as \gls{ga}, \gls{pso},  or \gls{aco}. As illustrated in Fig.~\ref{fig:Quantum_meta}, the \gls{QIM}-based stochastic optimization employs probabilistic qubit representations during initialization, thereby maximizing diversity. Subsequently, quantum-inspired operators, such as rotation gates or phase shifts, dynamically update these states, enabling interference-like behavior and escape from local optima. Therefore, this integration accelerates convergence and improves performance in high-dimensional, multi-objective wireless optimization tasks. To justify the intrinsic amalgamation of quantum principles within our framework, Fig.~\ref{fig:classical_vs_quantum_parallel} illustrates the fundamental operational distinction. Classical algorithms (top panel) rely on serial processing or hardware-based parallelism. In contrast, the low panel demonstrates the proposed quantum-inspired mechanism acting as a quantum coprocessor. Specifically, a superposition mapping operator (e.g., Hadamard) transforms classical information into a single superimposed state ($|\Psi\rangle$) \cite{RevModPhys15004}. This exploits intrinsic state-space parallelism, evaluating the entire solution space via a quantum oracle ($U_f$) in a single step, thereby accelerating the search for high-dimensional \gls{ng} network configurations \cite{RevModPhys15004}. The detailed features are  
\begin{itemize}[leftmargin=*]
    \item \textit{Quantum encoding for richer search representation:} In classical metaheuristics (e.g., \gls{ga}, \gls{pso}), solution representations are fixed and explicit, limiting each candidate to a single deterministic point in the search space devoid of probabilistic superposition. Conversely, quantum-inspired metaheuristics utilize qubit encodings to represent decision variables as state superpositions. This facilitates a probabilistic mapping of the entire search space, enabling the simultaneous exploration of multiple potential solutions. For instance, quantum-behaved \gls{pso} \cite{Fu5941032} utilizes phase angle encoding and quantum attractor-based stochastic sampling to represent particle positions, thereby facilitating broader probabilistic exploration.
    
    \item \textit{Quantum-inspired operators}: While classical operators (e.g., crossover) rely on static heuristics, quantum-inspired frameworks employ adaptive mechanisms like rotation gates and phase shifts. These operators update probability amplitudes via success-driven adjustments, leveraging quantum interference to dynamically guide the search. \gls{QIGA}, for instance, apply quantum operators such as rotation gates, Hadamard transformations, and quantum mutation (e.g., Pauli-X gates) to evolve qubit-based solutions in a probabilistic manner. This enables gradient-free, adaptive search across complex, high-dimensional spaces. Saad \textit{et al.} \cite{Saad9364969} demonstrated that \gls{QIGA} leverages superposition and gate-based updates to optimize resource-constrained scheduling efficiently. This approach is particularly suited for \gls{ng} network tasks like offloading and routing, coping effectively with non-differentiable or simulation-dependent objectives.
    
    \item \textit{Interference and quantum tunneling for escaping local optima:} A major challenge in classical metaheuristics is premature convergence to local optima, especially in non-convex optimization landscapes. Quantum-inspired metaheuristics integrate quantum tunneling and wave interference principles to enable escape from local traps. In practice, this can be implemented via probabilistic reinitialization schemes or non-local update operators that allow solutions to ``jump'' between regions of the search space. Notably, \gls{QA} frameworks, such as those realized by D-Wave systems \cite{Jeong10907925}, use quantum tunneling to traverse energy barriers between local and global optima in a potential landscape, which can be emulated by quantum-inspired algorithms through probabilistic perturbations or hybrid annealing schedules. 

    \item \textit{Parallel evaluation via quantum coprocessor paradigm:} Unlike classical optimizers that evaluate solutions sequentially or rely on hardware duplication, \gls{QC} leverages superposition to encode and process a large set of candidate states simultaneously. This reflects a hybrid quantum-classical workflow suitable for the \gls{NISQ} era, where the quantum module acts as a quantum coprocessor. As demonstrated in recent frameworks \cite{Jeong10907925}, this collaboration is bidirectional: classical algorithms are employed to fine-tune the parameters (e.g., rotation angles $\theta$) of the quantum circuit, while the quantum coprocessor executes the computationally intensive evaluation of multiple configurations in parallel via superposition. For instance, quantum annealers (e.g., D-Wave) exploit this property to evaluate global energy states for \gls{QUBO} problems, achieving near-optimal sum-rate performance with significantly reduced running time. 
\end{itemize}

\textbf{\gls{qrl}:} \gls{rl} is frequently used for dynamic spectrum access, mobility management, energy-efficient control, and intent-based orchestration. However, classical \gls{rl} suffers from slow convergence, poor sample efficiency, and local policy entrapment under highly stochastic and multi-objective environments. In \gls{ng} networks, \gls{qrl} offers meaningful improvements:
\begin{itemize}[leftmargin=*]
    \item \textit{Quantum encoding for richer policy and state representation:} Quantum encoding fundamentally enhances environmental representation in reinforcement learning by utilizing qubits, which, unlike binary bits, exist in superposition to capture continuous probabilistic configurations \cite{Silvirianti10153404}. This capacity enables the efficient encoding of exponentially larger state-action spaces using fewer physical units. Furthermore, entangled qubit configurations efficiently model complex dependencies, such as spatial-temporal correlations. This provides a compact, information-rich representation for high-dimensional \gls{ng} network states, enabling agents to evaluate diverse policies with greater nuance \cite{Ryu11005393}.
    
    \item \textit{Quantum-inspired operators for adaptive learning:} Quantum reinforcement learning replaces neural networks with parameterized gates and \gls{VQCs} as function approximators \cite{Wei10752359, Silvirianti10153404}. By leveraging trainable rotation and entanglement, these operators achieve high expressivity with significantly fewer parameters, yielding compact models robust to high-dimensional control tasks. In practical \gls{ng} network scenarios, such as \gls{MEC} \cite{Wei10752359} or joint sensing and communication systems \cite{Paul10680116}, these quantum-enhanced update mechanisms allow agents to rapidly adjust their strategies in response with improved stability and convergence behavior.

    \item \textit{Quantum superposition for parallel policy evaluation:} Quantum superposition enables the simultaneous evaluation of multiple policy trajectories within a single state \cite{Ansere10318071, Kim10461080}. This intrinsic parallelism significantly enhances search efficiency, allowing broader exploration per iteration without additional computational overhead. In scenarios like dense \gls{IoT} coordination, this parallelism significantly accelerates learning and sample efficiency. Consequently, \gls{qrl} converges to optimal strategies faster than classical counterparts, satisfying the critical temporal constraints of \gls{ng} networks.

    \item \textit{Reduction in training parameters and model complexity:} \gls{qrl} exploits the exponential expressiveness of Hilbert spaces to approximate complex policies via compact parameterized gates \cite{Ryu11005393}. This drastic reduction in parameterization minimizes memory and computational overhead, thereby accelerating policy optimization \cite{Kim10461080}. Such efficiency is particularly advantageous in resource-constrained settings \cite{Seon10994421, Silvirianti10153404, Wei10752359}, where classical deep \gls{rl} models may be impractical due to hardware limitations. By aligning learning performance with stringent resource constraints, \gls{qrl} offers a promising framework for scalable deployment in \gls{ng} network environments.
\end{itemize}
To fully leverage the potential of quantum reinforcement learning, several architectural variants, each with distinct strengths, have been proposed, as summarized in Table~\ref{Table: Quantum Architectures for NGNs}.
\begin{table*}[t]
\centering
\scriptsize
\caption{Comparison of quantum-specific architectures driving performance in \gls{ng} networks}
\label{Table: Quantum Architectures for NGNs}
\begin{tabular}{| p{0.2\textwidth} | p{0.22\textwidth} | p{0.22\textwidth} | p{0.25\textwidth} |}
\hline
\textbf{Architecture} & \textbf{Structure} & \textbf{Strength} & \textbf{Representative Application} \\
\hline

\textbf{Hybrid quantum-classical non-sequential architecture} & Classical feedforward layers for input/output, with multiple non-sequential \gls{VQCs} in the middle layer. & Enables finer control over qubits, reduces entanglement overhead, improves generalization and scalability in quantum-classical systems. & Joint task offloading and resource allocation in \gls{MEC} environments \cite{Wei10752359, Ansere10318071}, \gls{leo} satellite routing \cite{Seon10994421}. \\
\hline

\textbf{Layerwise quantum deep reinforcement learning} & Deep quantum network with RX/RY/CZ gates per layer and local loss calculation at each quantum layer instead of a single global loss. & Avoids barren plateaus, supports deeper quantum networks, enables faster convergence and stable training with less memory overhead. & \gls{uav} trajectory and power allocation optimization under energy-efficiency constraints \cite{Silvirianti10153404}. \\
\hline

\textbf{Multi-agent quantum reinforcement learning} & Decentralized agents each with independent quantum policy networks using \gls{VQCs}, trained from local observations and rewards. & Scalable multi-agent coordination; lowers communication overhead; resilient to partial observability; suitable for large-scale systems. & Irregular repetition slotted ALOHA-\gls{NOMA} adaptive transmission \cite{Ryu11005393}. \\
\hline

\textbf{Quantum actor-critic with quantum policy circuits} & Actor and/or critic networks modeled with parameterized quantum circuits; can use shared encoders and hybrid classical output layers. & Effective for continuous action control; enables expressive policy approximation with fewer parameters and enhanced convergence speed. & Joint Direction of Arrival (DoA) estimation and task offloading in \gls{ISAC}-based \gls{6g} systems \cite{Paul10680116}. \\
\hline

\end{tabular}
\vspace{-1.5em}
\end{table*}

\textbf{Hardware and algorithmic considerations:} To employ quantum-enhanced \gls{so} in practice, both hardware feasibility and algorithmic robustness must be considered. These considerations determine whether a technique is realistic in near-term hybrid systems or requires fault-tolerant \gls{QC}.

\begin{itemize}[leftmargin=*]
    \item \textit{Encoding constraints and embedding overheads:} Quantum hardware imposes strict constraints on how problems must be encoded. For example, quantum annealers (e.g., D-Wave) require problems to be formulated as \gls{QUBO} or Ising models. By contrast, gate-based quantum computers necessitate variational formulations with \gls{PQCs}. Additionally, embedding overhead arises when logical variables must be mapped onto physical qubits with limited connectivity. Complex problems such as user association, joint power-bandwidth allocation, or \gls{ris} phase control often require thousands of variables, far exceeding current hardware capabilities unless aggressive dimensionality reduction or clustering is applied. For instance, Dinh \textit{et al.} \cite{Dinh10720152} applied \gls{QA} to satellite beam placement, employing Hamiltonian reduction to decompose the problem into sub-instances compatible with D-Wave hardware limits.
    
    \item \textit{Hybrid quantum-classical integration:} Given the \gls{NISQ} nature of today's quantum processors, most real-world applications require hybrid workflows where quantum routines augment classical optimization components. A common structure involves using quantum modules, such as \gls{QA} for search, sampling, or policy generation, while the objective function evaluation and reward feedback occur in classical simulators or digital twins \cite{Fan9964012}.

    \item \textit{Noise, decoherence, and error tolerance:} Quantum algorithms for \gls{so} must tackle hardware-induced quantity noise \cite{kanno2021noise, Kandala_2017}, such as decoherence, crosstalk, and gate infidelity, which can degrade performance, particularly in feedback-intensive wireless applications. To mitigate these effects, error-aware strategies, including noisy gradient training and measurement error mitigation, enhance robustness. Additionally, redundant circuit encodings can average out stochastic failures. Minimizing quantum-classical communication latency is also critical for meeting the real-time demands of \gls{ng} network controls.

    \item \textit{Quantum acceleration benchmarks:} Justifying \gls{QC} integration necessitates rigorous benchmarking against classical baselines (e.g., \gls{de}, \gls{pso}) across algorithmic and hardware dimensions. Critical metrics include convergence rates, multi-objective Pareto coverage, and robustness to dynamic network conditions, while hardware evaluations must quantify the wall-clock latency and power efficiency of QPUs relative to \gls{cpu}s/\gls{gpu}s. In \gls{uav} scheduling, Jeong \textit{et al.} \cite{Jeong10907925} demonstrated that \gls{QA} surpasses classical clustering in convergence and throughput across diverse user densities, thereby providing essential empirical validation of quantum advantage in wireless scenarios.
\end{itemize}

\subsubsection{Lessons learned}
\gls{QC} holds transformative potential for \gls{so} in \gls{ng} networks, yet realizing its full potential requires careful alignment between algorithm design, hardware limitations, and network-specific requirements, with the key lessons:
\begin{itemize}[leftmargin=*]
    \item Quantum-inspired algorithms are immediately valuable even without access to fault-tolerant quantum hardware (e.g., quantum-inspired genetic algorithm, quantum-behaved \gls{pso}). This provides measurable gains in search diversity, convergence, and scalability across high-dimensional \gls{ng} network problems. These methods offer a practical bridge between classical and \gls{QC} paradigms.

    \item \gls{qrl} offers compact and expressive control models via exploiting quantum encoding and variational circuits to reduce the number of training parameters and improve policy adaptability, which are keys for dynamic and constrained wireless environments. Their strength lies not only in expressivity but also in sample efficiency and robustness to partial observability.

    \item The convenient hybrid quantum-classical workflow concept emerges as the most feasible approach in the \gls{NISQ} era. Embedding quantum subroutines into classical optimization pipelines, regardless whether for initialization, sampling, or policy approximation, because it can offer speedups without full quantum stack deployment.

    \item Effective quantum-enhanced \gls{so} mandates \gls{QUBO}, Ising, or \gls{PQCs} encodings, yet embedding overhead and limited connectivity necessitate dimensionality reduction or structured decomposition to ensure hardware compatibility.

    \item For evaluation, demonstrating quantum advantage involves more than improved solution quality. Realistic benchmarking must include energy efficiency, wall-clock latency, adaptability under uncertainty, and Pareto front coverage. Early deployments should leverage quantum simulators, bespoke hybrid components, and integration with \gls{llm}s and/or digital twins to support explainable and intent-driven network automation.

    \item Given the susceptibility of quantum processors to impairments, error-aware circuit design, e.g., low-depth \gls{PQCs}, redundant encoding, and quantum error mitigation with or without error correction, are essential for reliable performance, especially in feedback-driven applications like real-time beamformingor resource slicing.
\end{itemize}
These insights indicate that \gls{QC}, even in the current \gls{NISQ} era, can meaningfully enhance \gls{so} for \gls{ng} networks, especially when leveraged through hybrid, modular, and error-tolerant strategies tailored to wireless-domain constraints.
   
\vspace{-0.25cm}
\section{DESIGN GUIDELINES, FUTURE RESEARCH DIRECTIONS,  
AND CONCLUSIONS} \label{sec:conclusions}

This section provides the design guidelines of ultra-dense deployments, multi-layer heterogeneous \gls{ng} networks spanning deeply from space to underwater communications. They are expected to support massive connectivity, extreme service demands, and integration of sensing, communications, and computing functions. Then, we present a suite of potential research directions and draw tangible conclusions.
\vspace{-0.5em}
\subsection{Generic Design Guidelines}
\begin{itemize}[leftmargin=*]
\item \textbf{Scalable System Design \& Problem Formulation}: Effective formulation constitutes the mathematical foundation for scalability in \gls{ng} networks. It requires defining variables across continuous (e.g., beamforming \cite{van2024active}), discrete (e.g., user association \cite{pan2023joint_PSO2023}), and mixed domains \cite{nguyen2023fairness}. However, to tackle the curse of dimensionality in ultra-dense deployments, formulation must move beyond monolithic models by adopting \textit{decomposition strategies} that partition large-scale problems into parallelizable subtasks \cite{liu2022fair_decomposition}. Furthermore, efficient problem representation is critical; mapping multi-modal data (e.g., channel states, vision) into compact embeddings ensures tractability. Crucially, where exact global optimality is computationally infeasible, formulations should prioritize robust feasibility via risk-aware approaches (e.g., chance-constrained optimization \cite{huang2020stochastic}), providing reliability bounds within the strict energy budgets of edge hardware.

\item \textbf{Resource-Efficient Algorithmic Design}: Designing algorithms for \gls{ng} networks mandates balancing the exploration-exploitation trade-off with stringent hardware and latency constraints. \par \textit{(i) Initialization and Representation:} As problem instances scale, naive random sampling risks premature convergence. Design must leverage domain-specific heuristics—such as using \gls{sa} or \gls{pso} to generate robust initial populations \cite{sadana2025survey}, to guide optimization toward feasible regions rapidly. \par \textit{(ii) Hardware-Aware Execution:} Given that super-linear complexity growth taxes battery-powered devices, algorithms must incorporate \textbf{Green AI principles} (e.g., model compression) to minimize thermal stress and processing overhead \cite{Pasricha9017997}. Large-scale designs should further utilize decentralized multi-agent coordination with lightweight consensus to mitigate central bottlenecks \cite{tang2024dynamic}. \par \textit{(iii) Adaptation:} Exploitation refines candidates via local search or gradient fine-tuning, while self-adaptive mechanisms (e.g., \gls{bo}) dynamically calibrate parameters to handle non-stationary dynamics without exhaustive manual tuning \cite{sun2021learning}.

\item \textbf{Multi-dimensional Performance Evaluation}:Evaluating \gls{so} in ultra-dense networks requires going beyond simple sum-rate maximization to assess practical scalability. The evaluation framework must capture the tension between solution quality and computational cost: \textit{(i) Optimality-Latency Trade-off:} In \gls{urllc} contexts, algorithms should be evaluated on their ability to rapidly converge to robust, near-optimal operating points rather than solely penalized for missing elusive global optima \cite{fei2016survey, teng2018resource}. \textit{(ii) Computational Efficiency:} Metrics must explicitly quantify the runtime, memory footprint, and energy cost of the optimization process itself, ensuring the overhead does not negate system gains \cite{Shuvo9985008}. \textit{(iii) Stability and Robustness:} Consistency under stochastic disturbances is vital for risk-aware designs. Finally, for multi-objective tasks, Pareto front visualization and specialized metrics (Table~\ref{tab:multiobjective_metrics}) are necessary to analyze how well the algorithm navigates conflicting goals under high-dimensional pressure.

\item \textbf{Performance evaluation}: 
Selecting an evaluation framework is critical for assessing \gls{so} efficacy in \gls{ng} networks. Performance analysis must leverage realistic scenarios and metrics that capture solution quality, efficiency, stability, and scalability. The primary evaluation dimensions include:
\textit{(i)} Solution quality representing how close the  result obtained is to the optimal or reference solution, typically measured through metrics such as achievable sum-rate, system utility, or energy efficiency; \textit{(ii)} Convergence speed, quantified by the number of iterations or total runtime required to reach a near-optimal solution;
\textit{(iii)} Stability and robustness, which evaluate the consistency of results under varying initializations, stochastic disturbances, or environmental dynamics; and
\textit{(iv)} Computational complexity, reflecting algorithmic scalability in terms of runtime, memory, and communication overhead. Because \gls{so} is inherently non-deterministic, both the average performance and variance across multiple runs should be evaluated. Many \gls{ng} network optimization tasks involve multiple conflicting goals. Evaluating algorithmic performance in such contexts requires specialized metrics that assess both convergence to the Pareto-optimal front and the diversity of solutions. Table~\ref{tab:multiobjective_metrics} presents some common multi-objective metrics. Additionally, the visualization of Pareto fronts can further assist in analyzing trade-offs, allowing for the intuitive comparison of algorithmic performance under different optimization priorities.

\end{itemize}

\begin{table*}[t]
\centering
\scriptsize
\caption{Multi-objective performance metrics}
\begin{tabularx}{0.98\textwidth}{|l|X|X|}
\hline
\textbf{Metric} & \textbf{Description} & \textbf{Feature} \\ \hline
Hypervolume \cite{li2024multiobjective,tran2025gamr,duc2025multi} & Volume dominated by Pareto set & Indicates convergence and diversity \\ \hline
Inverted Generational Distance \cite{tam2024multi,tran2025gamr} & Average distance from reference Pareto front to obtained front & Evaluates both convergence and coverage of solutions \\ \hline
Generational Distance \cite{wang2024bi_ACO2024,xing2025multi} & Mean distance from obtained front to reference front & Measures closeness to Pareto-optimal front \\ \hline
Spacing   \cite{chaudhry2021multi} & Standard deviation of distances between neighboring solutions & Assesses uniform distribution of Pareto points \\ \hline
Spread  \cite{wang2024bi_ACO2024} & Diversity ratio across extremes and intermediate solutions & Quantifies how solutions are distributed along Pareto front \\ \hline
Epsilon indicator  \cite{goudos2018multi} & Minimum shift to dominate another front & Provides pairwise comparison between Pareto fronts \\ \hline
Scalarization  \cite{nandan2021optimized_GA2021,bandyopadhyay2024quantum_DE2024,zhu2025group_PSO2025} & Linear combination of objectives & Simplifies comparison via single-objective equivalence \\ \hline
\end{tabularx}
\label{tab:multiobjective_metrics}
\vspace{-1.5em}
\end{table*}

\vspace{-0.5em}
\subsection{Future Research Directions}
Drawing on the insights gained from each of the above open questions, we distill the key lessons learned and outline promising avenues for future research aimed at advancing the seamless integration of \gls{so} with \gls{ng} networks.
\begin{itemize}[leftmargin=*]
    \item \textbf{Uncertainty-aware and ergodic optimization:} A fundamental challenge is that the inherent stochasticity of the \gls{ng} networks erodes the reliability of deterministic allocation schemes. Hence, future research should develop \gls{so} frameworks that explicitly account for uncertainty, either by leveraging robust risk-aware formulations such as chance-constrained or distributionally robust optimization, or by employing ergodic evaluations over multiple coherence intervals. By incorporating probabilistic models and sample-based methods, resource allocation can be designed to provide reliability guarantees even in the face of highly variable hostile environments. This capability positions  \gls{so} as a key enabler for robust resource management in large-scale deployments.
    \item \textbf{Multi-objective designs for stochastic environments:} The second critical challenge arises from the need to simultaneously optimize conflicting objectives such as spectral efficiency, energy consumption, latency, and fairness. In practice, striking compelling trade-offs becomes even more of a challenge when channel dynamics and device mobility shift the optimal balance over time, causing static single-objective solutions to degrade rapidly. Multi-objective \gls{so} offers a promising framework, yet the identification of robust Pareto-optimal solutions remains difficult in these non-stationary conditions. Future research should thus focus on algorithms capable of tracking evolving Pareto fronts, learning operator preferences on the fly, and adapting priorities across multiple time scales. Such methods will facilitate the creation of  \gls{ng} networks to deliver differentiated service guarantees, while maintaining efficiency and fairness that static or single-metric optimization cannot provide.
    \item \textbf{Gradient-free and black-box optimization:} Another pressing limitation in \gls{ng} networks is the prevalence of non-convex, noisy, and simulator-only optimization problems, where no gradients are available, or they are unreliable. Resource allocation tasks such as scheduling, beam selection, or hardware-in-the-loop optimization fall directly into this category. \gls{so} provides a natural solution by offering gradient-free methods that rely on population-based search, probabilistic sampling, or reinforcement-based exploration to discover near-optimal policies without requiring analytical tractability. Future research should refine these methods to improve sample efficiency, reduce computational overhead, and enable online adaptation. By doing so, black-box \gls{so} will become indispensable for optimizing real-world systems, where no closed-form models exist, allowing deployment-ready algorithms to operate directly in noisy, incomplete, and non-differentiable environments.
    \item \textbf{Scalability and real-time feasibility:} As networks evolve toward ultra-dense scenarios, scalability emerges as both an algorithmic and architectural bottleneck. The dimensionality of optimization problems escalates rapidly, while real-time constraints require millisecond-level decision making under strict energy budgets. Centralized solvers, even when theoretically optimal, are impractical in such settings. Future research should focus on decomposition-based \gls{so}, decentralized multi-agent coordination, and adaptive learning strategies that can break large-scale problems into smaller components, while maintaining global coordination. Furthermore, lightweight communication protocols and hardware-aware algorithm design are needed for ensuring tractability even on resource-constrained devices. Achieving scalable and energy-efficient stochastic optimization will unlock the ability to deliver near-optimal solutions in real time, thereby enabling \gls{ng} networks to support city-scale deployments without sacrificing services.
    \item \textbf{Integration with distributed systems, generative \gls{ai}, and automation:}  The paradigm shift toward distributed,
automated and intelligence-oriented networks lead to
complex optimization problems.
On one hand, distributed environments are characterized by non-IID data, partial observability, and limited computational capacity at the edge. Simultaneously,
Generative \gls{ai} models (e.g., \gls{llm}s) and intent-based
automation offer novel interfaces for translating high-level
service-oriented specifications into executable policies. Future work should
investigate federated and model-free SO methods that
enhance decentralized decision-making while preserving
privacy, as well as hybrid pipelines where Generative
\gls{ai} carries out the parsing of ambiguous intents into broad optimization
constraints. The family of \gls{so} algorithms is capable of refining these outputs into
feasible solutions. When synergized with digital twins
and closed-loop automation, these techniques will lead to
self-optimizing and self-healing networks that operate
reliably in the face of uncertainty. This integration opens the path
toward zero-touch network management, where human
intervention is minimized, efficiency is maximized, and
service quality is maintained across heterogeneous
environments.
    \item \textbf{Quantum-enhanced SO:} Finally, \gls{QC} presents a transformative yet hitherto underexplored opportunity for \gls{so} in next-generation networks. While \gls{NISQ} hardware remains limited in qubit count and connectivity, quantum-inspired algorithms already demonstrate tangible benefits in diversity, convergence, and scalability. In parallel, hybrid quantum–classical workflows allow the embedding of quantum subroutines into classical pipelines, enabling \gls{so} to leverage potent tunneling and superposition effects without requiring fully fault-tolerant devices. Future research should focus on overcoming embedding overhead, reducing circuit depth, and integrating quantum error-mitigation and error-correction coded techniques, while also establishing realistic benchmarking metrics that include latency, energy efficiency, and robustness to uncertainty. The pursuit of quantum-enhanced \gls{so} promises not only faster exploration of large combinatorial spaces but also new performance frontiers that classical methods cannot readily reach, thereby positioning it as a long-term strategic direction for \gls{ng} network optimization.
\end{itemize}

\begin{figure}[t]
        \centering
        \includegraphics[width=0.9\linewidth, trim = 0cm 0.75cm 0cm 1.25cm, clip = true ]{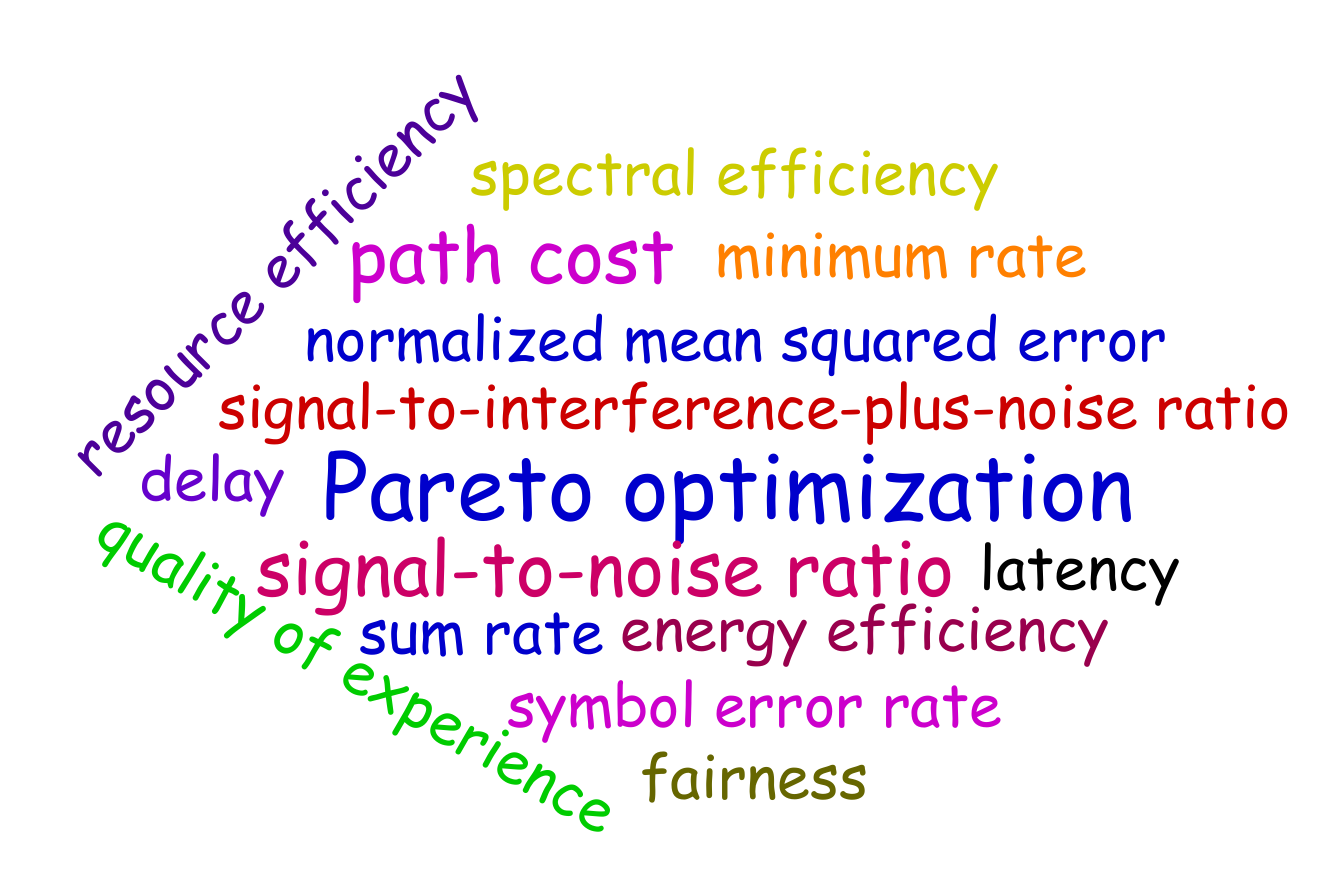}
        \caption{Common objective functions in \gls{ng} networks.}
        \label{fig:objective_function}
        \vspace{-0.25cm}
\end{figure}
\begin{figure}[t]
        \centering
        \includegraphics[width=0.9\linewidth, trim = 0cm 0cm 0cm 0cm, clip = true]{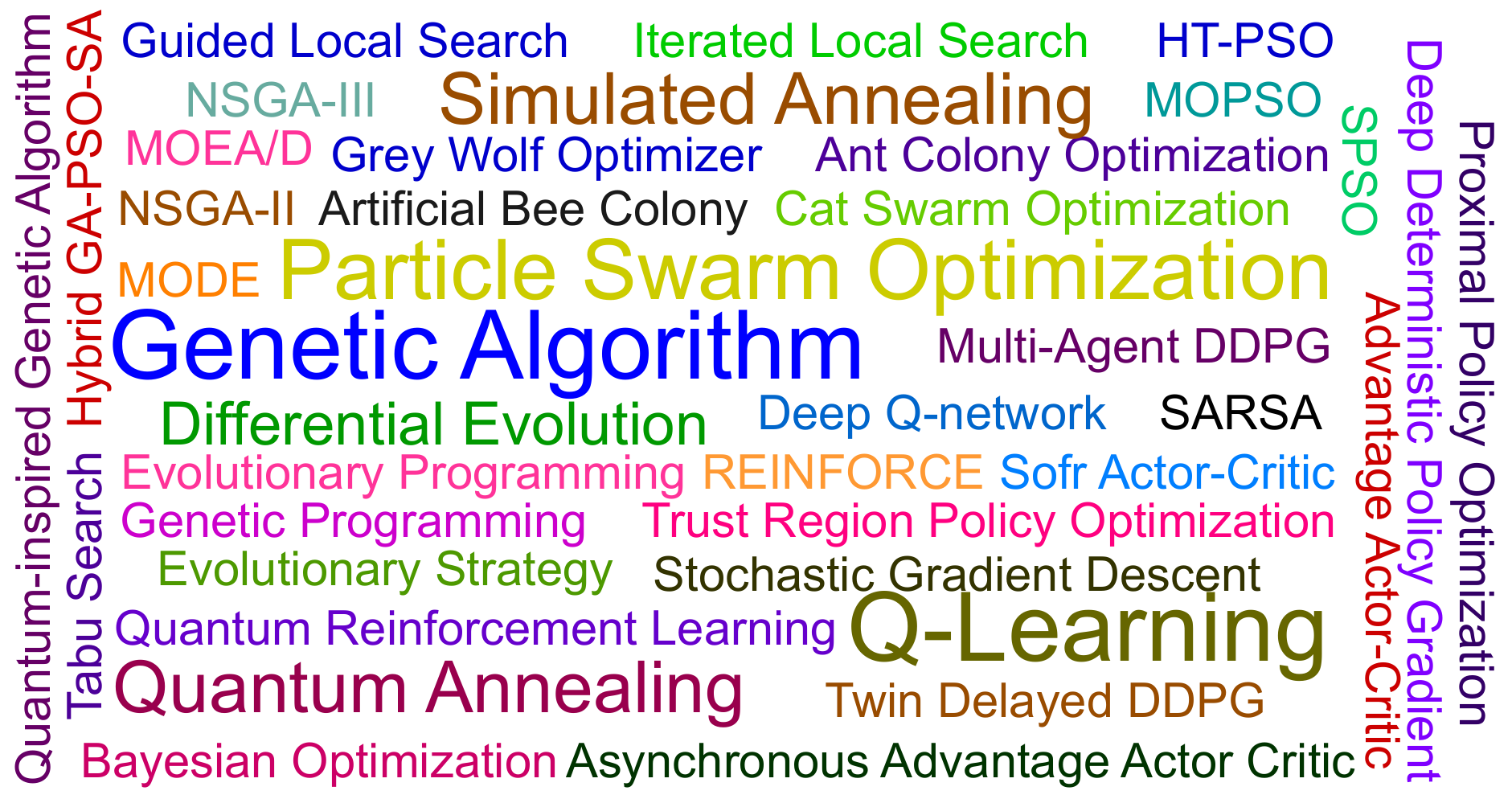}
        \caption{Family of \gls{so} algorithms.}
        \label{fig:Algorithms}
        \vspace{-0.5cm}
\end{figure}
\vspace{-0.5cm}
\subsection{Summary and Conclusions}
This tutorial \& survey provided a comprehensive exploration of \gls{so} as a unifying methodology for resource allocation in \gls{ng} networks. We commenced by revisiting the fundamental models and algorithmic frameworks that differentiate \gls{so} from classical deterministic optimization, emphasizing their suitability for high-dimensional, dynamic, and uncertainty-prone environments. Building on this foundation, we examined how \gls{so} supports both single and multi-objective trade-offs across diverse \gls{ng} systems, spanning key technologies such as massive \gls{mimo},  \gls{ris}, THz communications, satellite-aerial-terrestrial integration, and mobile edge computing. In Section~\ref{sec:Integrated}, we further highlighted the integration of \gls{so} with distributed network architectures, enabling adaptive and scalable decision-making under partial or delayed information. The discussion also extended to hybrid designs that combine \gls{so} with emerging tools like large language models and automation, providing novel avenues for tackling unstructured and complex decision spaces. Finally, we also explored the role of \gls{QC} as a natural extension of SO, underscoring its potential to accelerate convergence, enhance solution diversity, and address otherwise intractable optimization tasks. Taken together, these insights position \gls{so} not merely as a convenient mathematical framework, but as a foundational enabler for the design, deployment, and evolution of intelligent, resilient, and trustworthy \gls{ng} networks.

We have critically appraised the whole suite of objective functions seen
in Fig.~\ref{fig:objective_function}, which are often relied upon in the \gls{so} of \gls{ng}
wireless networks, while the family of \gls{so} algorithms considered was
portrayed in Fig.~\ref{fig:Algorithms}. The parting message gleaned is that the
choice of the objective function - such as for example the sum-rate or
min-max objective function - is more influential in determining the overall system
performance than that of the \gls{so} algorithms. However, not all the \gls{so} algorithms
lend themselves to conveniently determining the optimum of the objective function.
\vspace{-0.2cm}
\bibliographystyle{IEEEtran}
\bibliography{IEEE}
\end{document}